\documentclass[10pt,onecolumn]{article}
\usepackage[utf8]{inputenc}
\usepackage[T1]{fontenc}
\usepackage[table,xcdraw]{xcolor}
\usepackage{amsmath}
\usepackage{amssymb}
\usepackage{amsfonts}
\usepackage{colortbl}
\usepackage{longtable}
\usepackage{graphicx}
\usepackage{epstopdf}
\DeclareGraphicsExtensions{.pdf,.png,.jpg,.jpeg}
\usepackage{hyperref}
\hypersetup{colorlinks=true, linkcolor=blue, citecolor=blue, urlcolor=blue}
\usepackage{caption}
\usepackage{subcaption}
\usepackage{float}
\usepackage{esint}
\usepackage{bm}
\usepackage{soul}
\usepackage{tikz}
\usetikzlibrary{shapes,arrows,positioning,fit,backgrounds,calc,shapes.geometric}
\usepackage{dcolumn}
\usepackage{array}
\usepackage{multirow}
\usepackage{booktabs}
\RequirePackage[numbers,sort&compress]{natbib}

\usepackage[margin=1.8cm]{geometry}

\newcommand{\be}{\begin{equation}}
\newcommand{\ee}{\end{equation}}
\allowdisplaybreaks

\begin{document}

\onecolumn

\begin{center}
\LARGE{\bf Shadow, Hawking radiation, and thermodynamics of Letelier black hole in electromagnetic universe with Barrow entropy}
\end{center}

\vspace{0.4cm}
\begin{center}
{\bf \.{I}zzet Sakall{\i}}$^{1}$\footnote{izzet.sakalli@emu.edu.tr; ORCID: 0000-0001-7827-9476},
{\bf Erdem Sucu}$^{1}$\footnote{erdemsc07@gmail.com; ORCID: 0009-0000-3619-1492},
{\bf Ahmad Al-Badawi}$^{2}$\footnote{ahmadbadawi@ahu.edu.jo; ORCID: 0000-0002-7633-3048},
{\bf Faizuddin Ahmed}$^{3}$\footnote{faizuddinahmed15@gmail.com; ORCID: 0000-0003-2196-9622}
\end{center}

\vspace{0.2cm}
\begin{center}
{\small
$^{1}${\it Physics Department, Eastern Mediterranean University, Famagusta, 99628 North Cyprus, via Mersin 10, T\"{u}rkiye}\\[0.1cm]
$^{2}${\it Department of Physics, Al-Hussein Bin Talal University, 71111, Ma'an, Jordan}\\[0.1cm]
$^{3}${\it Department of Physics, The Assam Royal Global University, Guwahati, 781035, Assam, India}
}
\end{center}

\vspace{0.4cm}

\begin{abstract}
\noindent
We present a theoretical investigation of a static, spherically symmetric Letelier black hole immersed in an electromagnetic universe, characterized by the cloud of strings parameter $\alpha$ and the electromagnetic universe parameter $a$. The spacetime structure is established by deriving the event horizon radii, curvature invariants, and isometric embedding diagrams that visualize the horizon expansion with increasing string density. The Hawking temperature and Bekenstein-Hawking entropy are computed, revealing monotonic suppression of thermal emission as $\alpha$ increases. The photon sphere and shadow radius are obtained analytically, demonstrating that both parameters enlarge the apparent black hole silhouette compared to the Schwarzschild geometry. We extend the shadow analysis to homogeneous and inhomogeneous plasma environments, quantifying the frequency-dependent reductions in observed shadow size. The optical appearance is decomposed into direct emission, lensed emission, and photon ring contributions following Event Horizon Telescope methodology. The Hawking radiation spectra and energy emission rates are computed for scalar, electromagnetic, and Dirac fields, with greybody factors calculated using the rigorous bounds method. The thermodynamic phase structure is analyzed within the Barrow entropy framework, where the heat capacity exhibits second-order phase transitions and the Joule-Thomson expansion reveals cooling-heating inversion curves governed by the model parameters. The stability under fermionic perturbations is established through supersymmetric quantum mechanics, confirming that the positive chirality effective potential remains strictly positive outside the horizon across the parameter space. Our results provide theoretical predictions testable by current shadow observations and future gravitational wave measurements.
\end{abstract}

\vspace{0.2cm}

\vspace{0.4cm}

\tableofcontents

\vspace{0.6cm}

\twocolumn

\section{Introduction}\label{isec1}

{\color{black}

The study of BHs has occupied a central position in gravitational physics since the pioneering work of Schwarzschild, who obtained the first exact solution to Einstein's field equations describing the exterior geometry of a spherically symmetric, static mass distribution \cite{isz01}. Over the past century, BH physics has evolved from a theoretical curiosity into an active area of observational astronomy, driven by the detection of gravitational waves (GWs) from binary BH mergers by the LIGO-Virgo-KAGRA collaboration \cite{isz02,isz03} and the imaging of supermassive BH shadows by the Event Horizon Telescope (EHT) \cite{isz04,isz05}. These groundbreaking observations have opened new windows for testing general relativity (GR) in the strong-field regime and searching for deviations that might signal the presence of new physics beyond the standard paradigm \cite{isz06,isz07}.

Among the various extensions of the Schwarzschild geometry, BH solutions embedded in matter fields and topological defect configurations have attracted considerable attention. Letelier introduced a class of spacetimes describing BHs surrounded by a CoS, which represents a collection of one-dimensional topological defects formed during symmetry-breaking phase transitions in the early universe \cite{Letelier1979}. The CoS generates an anisotropic energy-momentum tensor with the distinctive equation of state $p_r = -\rho$, where the radial pressure equals the negative of the energy density \cite{isz09,isz10}. This configuration produces a deficit solid angle at spatial infinity, modifying the asymptotic structure of the spacetime in a manner analogous to global monopoles and cosmic strings \cite{isz11,isz12,isz12x1,isz12x2,isz12x3,isz12x4}. The Letelier solution has been extensively studied in the context of thermodynamics \cite{isz13}, quasinormal modes (QNMs) \cite{isz14}, gravitational lensing \cite{isz15}, and particle dynamics \cite{isz16}, revealing that the CoS parameter introduces measurable deviations from the standard Schwarzschild predictions.

Parallel developments in cosmology have established the importance of electromagnetic (EM) backgrounds and quintessence fields in modeling the accelerated expansion of the universe \cite{isz17,isz18}. The EMU model, originally proposed by Bonnor and later extended by various authors, describes spacetimes permeated by a background EM field that modifies the effective gravitational potential experienced by test particles \cite{isz19,isz20}. When a BH is immersed in such an EMU, the metric function acquires additional terms that mimic the behavior of a Reissner-Nordström (RN) geometry with an effective charge parameter determined by the EM background strength \cite{isz21}. The combination of CoS and EMU effects produces a two-parameter family of BH solutions that interpolates continuously between the Schwarzschild, RN, and pure Letelier spacetimes, offering a rich phenomenology for studying how multiple modifications interact in strong gravitational fields \cite{isz22,isz23,isz23x1,isz23x2,isz23x3,isz23x4,isz23x5}.

The thermodynamic properties of BHs have been a subject of intense investigation since Bekenstein and Hawking established the foundations of BH thermodynamics in the 1970s \cite{isz24,isz25}. The discovery that BHs radiate thermally with a temperature proportional to the surface gravity opened profound connections between gravity, quantum mechanics, and statistical physics \cite{isz26}. Recent developments have extended the thermodynamic description to include modified entropy formulations, among which the Barrow entropy has emerged as a particularly interesting proposal \cite{isz27}. Barrow suggested that quantum-gravitational effects may deform the event horizon (EH) geometry, leading to an effectively fractal surface characterized by a deformation parameter $\Delta \in [0,1]$ \cite{isz28}. This modification has been applied to cosmological settings \cite{isz29,isz29x1,isz30} and BH thermodynamics \cite{isz31}, revealing novel phase transition structures and modified thermodynamic relations that could encode signatures of quantum gravity.

The optical appearance of BHs, particularly the shadow cast by the photon sphere (PS) against a luminous background, has become a primary observable for testing gravity theories \cite{isz32,isz33}. The EHT observations of M87* and Sgr~A* have provided shadow size measurements with approximately 10\% precision, enabling constraints on deviations from the Kerr geometry \cite{isz34,isz35}. The shadow radius depends sensitively on the near-horizon geometry and the properties of null geodesics, making it a powerful probe of BH parameters and potential modifications to GR \cite{isz36}. Plasma effects in the accretion environment can further modify the apparent shadow size through frequency-dependent refraction, requiring careful modeling of both the spacetime geometry and the surrounding medium \cite{isz37,isz38,isz38x1,isz38x2,WOS:001632119100001AHMED,sucu2025quantumRoyal,sucu2025scalar,sucu2025astrophysical}. Moreover, the Hawking radiation (HR) spectrum and greybody factors provide additional channels for probing BH physics. The greybody factors quantify the transmission probability of HR through the effective potential barrier surrounding the BH and depend on the spin of the emitted particles \cite{isz39,isz40}. For fermionic fields, the Dirac equation in curved spacetime leads to a pair of supersymmetric (SUSY) partner potentials whose structure determines both the emission characteristics and the dynamical stability of the BH under perturbations \cite{isz41,isz42}. The stability analysis is particularly important for establishing whether a given BH solution represents a physically viable configuration that can exist in nature without spontaneous decay \cite{isz43}.

Motivated by these developments, we investigate in this paper the Letelier BH immersed in an EMU, examining its geometric, thermodynamic, optical, and radiative properties. Our aims are threefold. First, we seek to establish the complete spacetime structure, including the EH configuration, curvature invariants, and asymptotic behavior, for the two-parameter family of solutions characterized by the CoS parameter $\alpha$ and the EMU parameter $a$. Second, we aim to compute the observable signatures relevant for current and future BH observations, including the shadow radius, photon ring (PR) structure, and HR spectra for different spin fields. Third, we explore the thermodynamic phase structure within the Barrow entropy framework, analyzing the heat capacity and Joule-Thomson expansion (JTE) to identify phase transitions and cooling-heating regimes. The combination of CoS and EMU effects produces distinctive phenomenology that distinguishes this solution from both the standard Schwarzschild geometry and its individual modifications. The CoS parameter $\alpha$ enters the metric function as a constant shift, reducing the asymptotic value of the lapse function and producing a deficit angle geometry at spatial infinity. The EMU parameter $a$ introduces an effective charge-like term that scales as $1/r^2$, mimicking the RN behavior while arising from the background EM field rather than an intrinsic BH charge. The interplay between these two effects modifies the EH structure, PS location, Hawking temperature, and effective potentials for field perturbations in ways that can be disentangled through careful parameter studies.

Our analysis employs a combination of analytical derivations and numerical computations. The EH radii and thermodynamic quantities are obtained in closed form, allowing direct comparison with known limiting cases. The shadow radius and greybody factors are computed using established techniques, including the Visser bounds method for transmission coefficients. The Barrow entropy framework is applied to study the heat capacity and JTE coefficient, revealing connections between the phase transition structure and inversion curves. The stability under Dirac perturbations is established using the SUSY quantum mechanics formalism, exploiting the factorized structure of the effective potentials. From an observational perspective, the results presented in this work provide theoretical predictions that can be compared with EHT shadow measurements and future GW ringdown observations. The shadow radius varies by up to 48\% across the parameter space examined, a deviation well within the sensitivity of current and next-generation instruments. The HR suppression caused by the CoS could explain the survival of primordial BHs in certain mass ranges, with implications for dark matter searches and the diffuse gamma-ray background. The modified phase structure within the Barrow entropy framework offers potential connections to quantum gravity phenomenology that could be tested through precision thermodynamic measurements \cite{isz44,isz45}.

The paper is organized as follows. In Sec.~\ref{isec2}, we establish the spacetime structure and horizon properties of the Letelier BH in EMU, deriving the metric function, EH radii, curvature invariants, and thermodynamic quantities including the Hawking temperature and entropy. Section~\ref{isec3} presents the optical properties, including the PS and shadow analysis in vacuum and plasma environments, the emission decomposition into direct, lensed, and PR contributions, and the photon trajectory structure. The HR properties and energy emission rates are computed in Sec.~\ref{isec4}, where we analyze the spectra for scalar, EM, and Dirac fields and discuss the implications for BH evaporation. Section~\ref{isec5} explores the phase structure and JTE within the Barrow entropy framework, identifying second-order phase transitions and inversion curves. The Dirac field perturbations and stability analysis are presented in Sec.~\ref{isec6}, where we derive the effective potentials, compute greybody factors, and establish dynamical stability using the SUSY formalism. Finally, Sec.~\ref{isec7} summarizes our conclusions and outlines directions for future research.

\section{Spacetime Structure and Horizon Properties of the Letelier BH in EMU}
\label{isec2}

This section establishes the geometric foundation of the Letelier BH embedded in an EMU. We derive the EH structure, examine the behavior of the metric function across the parameter space, compute the curvature invariants, and visualize the spatial geometry through isometric embedding diagrams. The interplay between the CoS parameter $\alpha$ and the EMU parameter $a$ produces a two-parameter family of BH solutions that interpolates continuously between the Schwarzschild, RN, and pure Letelier spacetimes.

\subsection{Metric Structure and Parameter Space}

We consider a static, spherically symmetric BH solution describing a Letelier spacetime surrounded by a CoS and immersed in an EMU. The line element takes the standard form \cite{Letelier1979,isz13}
\begin{equation}
ds^2 = -f(r)\, dt^2 + \frac{dr^2}{f(r)} + r^2 \left( d\theta^2 + \sin^2\theta\, d\phi^2 \right),
\label{eq:metric}
\end{equation}
where the lapse function $f(r)$ encapsulates all gravitational modifications \cite{Al-Badawi:2025kbi}:
\begin{equation}
f(r) = 1 - \alpha - \frac{2M}{r} + \frac{M^2}{r^2}(1 - a^2).
\label{eq:lapse}
\end{equation}
Here $M$ denotes the BH mass as measured by an asymptotic observer, $\alpha \in (0,1)$ is the CoS parameter characterizing the energy density of the string cloud permeating the spacetime, and $a \in [0,1]$ is the EMU parameter that quantifies the strength of the electromagnetic background. The CoS arises from a network of one-dimensional topological defects formed during symmetry-breaking phase transitions in the early universe \cite{isz11,sec2is04}. Such string configurations generate an anisotropic energy-momentum tensor with the distinctive equation of state $p_r = -\rho$, where the radial pressure equals the negative of the energy density \cite{NunesdosSantos:2025alw,sec2is05}.

The metric function (\ref{eq:lapse}) encompasses several well-known spacetimes as special limits. Setting $\alpha = 0$ and $a = 1$ yields $f(r) = 1 - 2M/r$, recovering the standard Schwarzschild solution that describes an isolated, uncharged, non-rotating BH in vacuum \cite{isz01}. For $a = 1$ with $\alpha \neq 0$, we obtain $f(r) = 1 - \alpha - 2M/r$, which describes the Letelier spacetime with a deficit solid angle $\delta\Omega = 4\pi\alpha$ due to the CoS \cite{Letelier1979}. This deficit angle removes a wedge from the solid angle at spatial infinity, producing a conical geometry analogous to cosmic string spacetimes \cite{isz11,sec2is06}. Setting $\alpha = 0$ with $a \neq 1$ gives $f(r) = 1 - 2M/r + M^2(1-a^2)/r^2$, representing a RN-like BH where the effective charge parameter is $Q_{\rm eff}^2 = M^2(1-a^2)$. The complete solution with both $\alpha \neq 0$ and $a \neq 1$ captures the combined effects of topological defects and electromagnetic backgrounds, offering a fertile ground to study how multiple modifications to the Schwarzschild geometry interact.

\subsection{EH Structure and Existence Conditions}

The EH radii are determined by the roots of $f(r_h) = 0$. Rearranging Eq.~(\ref{eq:lapse}), we obtain the quadratic equation
\begin{equation}
(1-\alpha)r_h^2 - 2Mr_h + M^2(1-a^2) = 0.
\label{eq:horizon_quadratic}
\end{equation}
Applying the quadratic formula yields the outer ($r_+$) and inner ($r_-$) horizon radii:
\begin{equation}
r_{\pm} = \frac{M\left(1 \pm \sqrt{1-(1-\alpha)(1-a^2)}\right)}{1-\alpha}.
\label{eq:horizons}
\end{equation}
For real-valued horizons, the discriminant must be non-negative, imposing the constraint $(1-\alpha)(1-a^2) \leq 1$. This condition is automatically satisfied throughout the allowed parameter range $\alpha \in (0,1)$ and $a \in [0,1]$, ensuring that the BH possesses at least one horizon. The equality defines the extremal limit where both horizons merge at $r_{\rm ext} = M/(1-\alpha)$, corresponding to a degenerate horizon with vanishing surface gravity \cite{sec2is07}.

In the pure Letelier limit ($a = 1$), the $1/r^2$ term vanishes, and the horizons simplify to $r_+ = 2M/(1-\alpha)$ and $r_- = 0$. The outer horizon expands monotonically with increasing $\alpha$, reaching $r_+ = 4M$ at $\alpha = 0.5$ and diverging as $\alpha \to 1$. In the pure EMU limit ($\alpha = 0$), we recover $r_\pm = M(1 \pm a)$, which matches the RN horizon structure upon identifying $Q_{\rm eff} = M\sqrt{1-a^2}$. The extremal configuration occurs at $a = 0$, where both horizons coalesce at $r_+ = r_- = M$.

Table~\ref{tab:horizons} presents representative horizon radii for various combinations of $\alpha$ and $a$. The rows marked with $\bigstar$ correspond to the 3D embedding configurations displayed in Fig.~\ref{fig:embedding_LetEMU}. Several trends emerge from the tabulated values. As $\alpha$ increases at fixed $a$, the outer horizon $r_+$ expands monotonically, reflecting the gravitational dilution caused by the CoS. The inner horizon $r_-$ decreases with increasing $\alpha$, eventually disappearing in the single-horizon configurations at $a = 1$. For fixed $\alpha$, increasing $a$ from 0 to 1 causes the two horizons to separate, transitioning from the near-extremal regime to the single-horizon Letelier geometry.

\begin{table*}[ht!]
\centering
\setlength{\tabcolsep}{10pt}
\renewcommand{\arraystretch}{1.6}
\begin{tabular}{cc|cc|l}
\hline\hline
\rowcolor{orange!50}
$\alpha$ & $a$ & $r_+/M$ & $r_-/M$ & Configuration \\
\hline
0.00 & 0.0 & 1.0000 & 1.0000 & Extremal BH \\
\rowcolor{yellow!20}
0.00 & 0.5 & 1.8660 & 0.1340 & Non-extremal BH $\bigstar$ \\
\rowcolor{yellow!20}
0.00 & 1.0 & 2.0000 & -- & Schwarzschild BH $\bigstar$ \\
0.10 & 0.5 & 1.7445 & 0.4777 & Non-extremal BH \\
\rowcolor{yellow!20}
0.10 & 1.0 & 2.2222 & -- & Letelier BH $\bigstar$ \\
\rowcolor{yellow!20}
0.20 & 0.5 & 2.0406 & 0.4594 & Non-extremal BH $\bigstar$ \\
0.20 & 1.0 & 2.5000 & -- & Single horizon BH \\
0.30 & 0.5 & 2.4132 & 0.4440 & Non-extremal BH \\
\rowcolor{yellow!20}
0.50 & 0.5 & 3.5811 & 0.4189 & Non-extremal BH $\bigstar$ \\
\rowcolor{yellow!20}
0.50 & 1.0 & 4.0000 & -- & Single horizon BH $\bigstar$ \\
0.70 & 0.5 & 6.2678 & 0.3989 & Non-extremal BH \\
0.90 & 1.0 & 20.0000 & -- & Single horizon BH \\
\hline\hline
\end{tabular}
\caption{Representative EH radii for the Letelier BH in EMU with $M=1$. Rows marked with $\bigstar$ correspond to the 3D embedding configurations in Fig.~\ref{fig:embedding_LetEMU}. The dash (--) indicates configurations with a single horizon where $r_- = 0$.}
\label{tab:horizons}
\end{table*}
\subsection{Lapse Function Behavior}

The behavior of the lapse function $f(r)$ determines the causal structure of the spacetime and the nature of particle orbits. Figure~\ref{fig:metric} illustrates $f(r)$ for various parameter choices. Panel (a) shows the variation with the CoS parameter $\alpha$ at fixed $a = 0.5$. As $\alpha$ increases, two effects become apparent: the asymptotic value $f(\infty) = 1 - \alpha$ decreases, reflecting the deficit angle geometry, and the outer EH radius $r_+$ shifts outward. For $\alpha = 0.5$, the horizon is located at $r_+ \approx 3.6M$, nearly twice the Schwarzschild value. Panel (b) displays the variation with the EMU parameter $a$ at fixed $\alpha = 0.1$. As $a$ approaches unity, the RN-like $1/r^2$ term diminishes, recovering the single-horizon Letelier geometry. For small $a$, the metric function exhibits a local maximum between the two horizons, characteristic of the RN-type causal structure with a Cauchy horizon at $r_-$. Panel (c) presents special limiting cases including the Schwarzschild, pure Letelier, RN-like, and extremal configurations. The intersections with $f(r) = 0$ (dashed gray line) identify the horizon locations.

\begin{figure*}[ht!]
\centering
\begin{subfigure}[b]{0.48\textwidth}
\centering
\includegraphics[width=\textwidth]{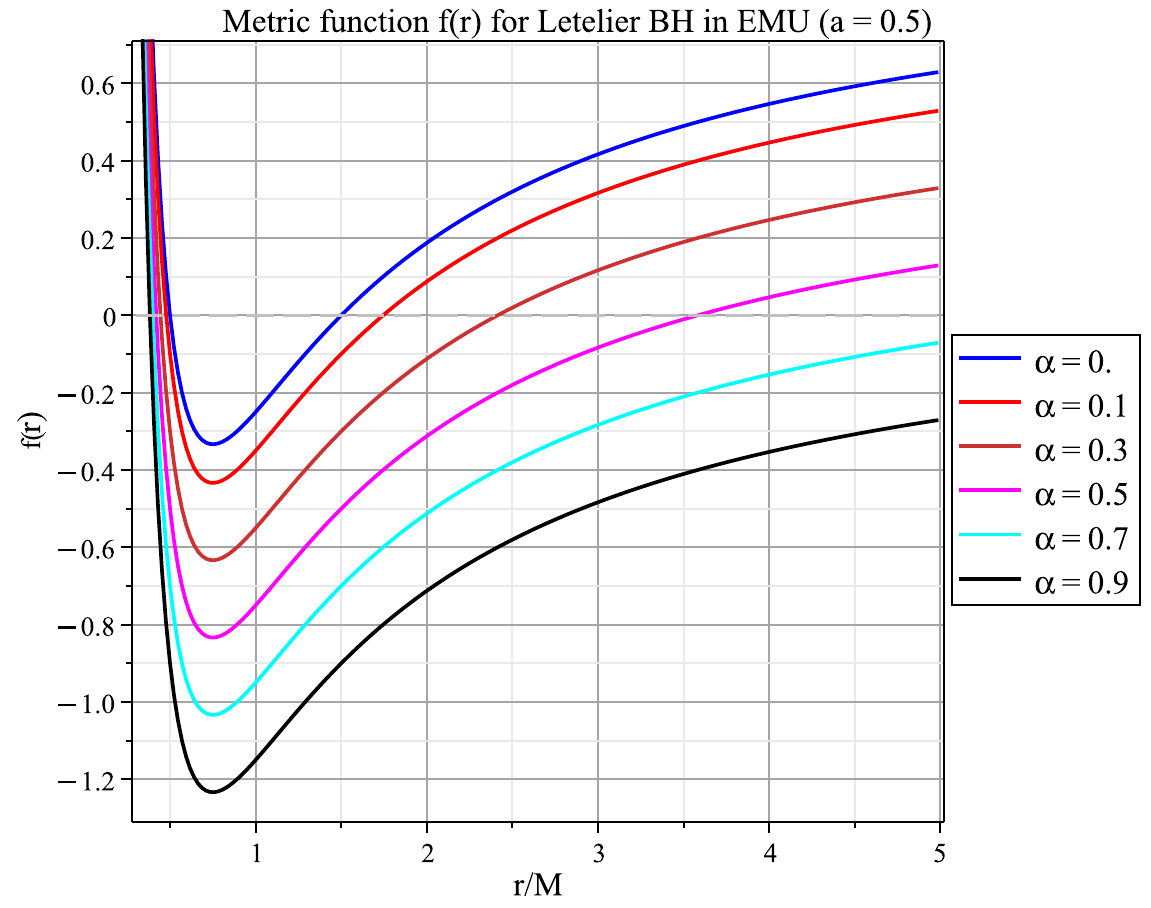}
\caption{}
\label{fig:metric_alpha}
\end{subfigure}
\hfill
\begin{subfigure}[b]{0.48\textwidth}
\centering
\includegraphics[width=\textwidth]{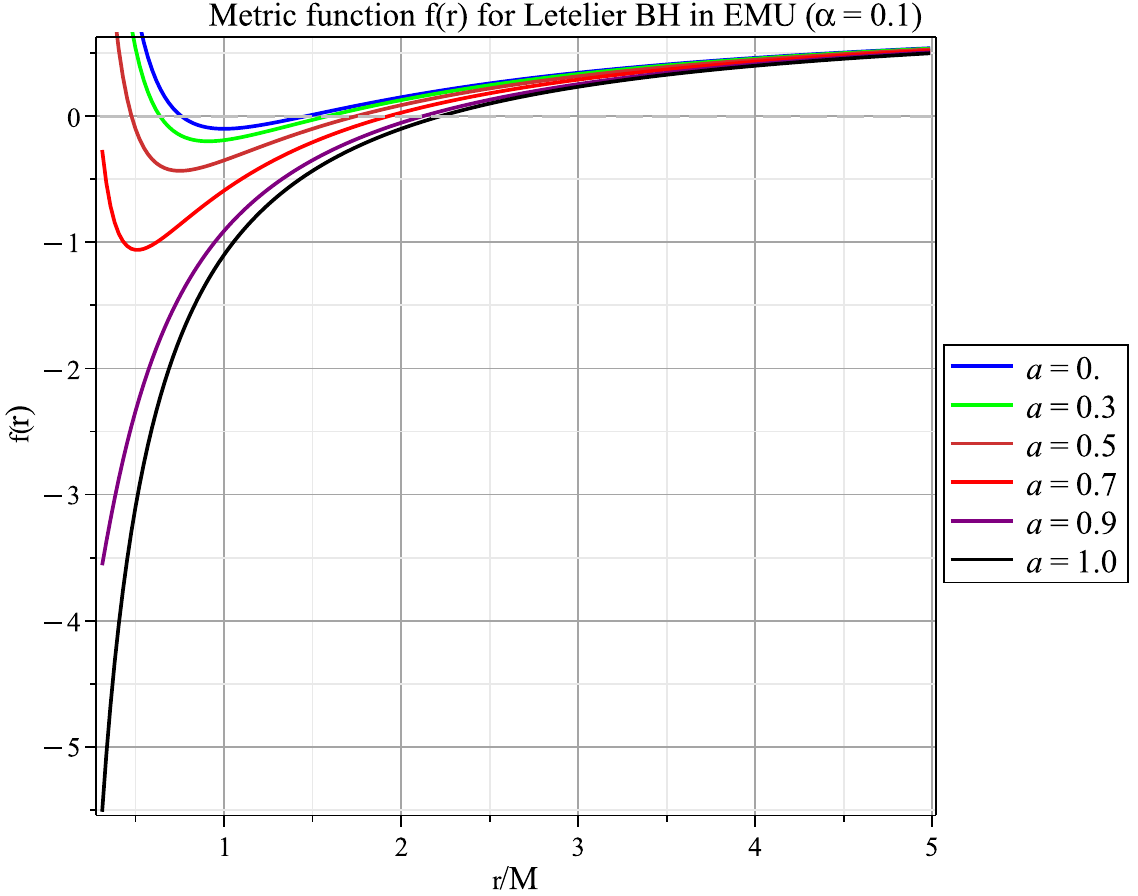}
\caption{}
\label{fig:metric_a}
\end{subfigure}
\\[0.4cm]
\begin{subfigure}[b]{0.75\textwidth}
\centering
\includegraphics[width=\textwidth]{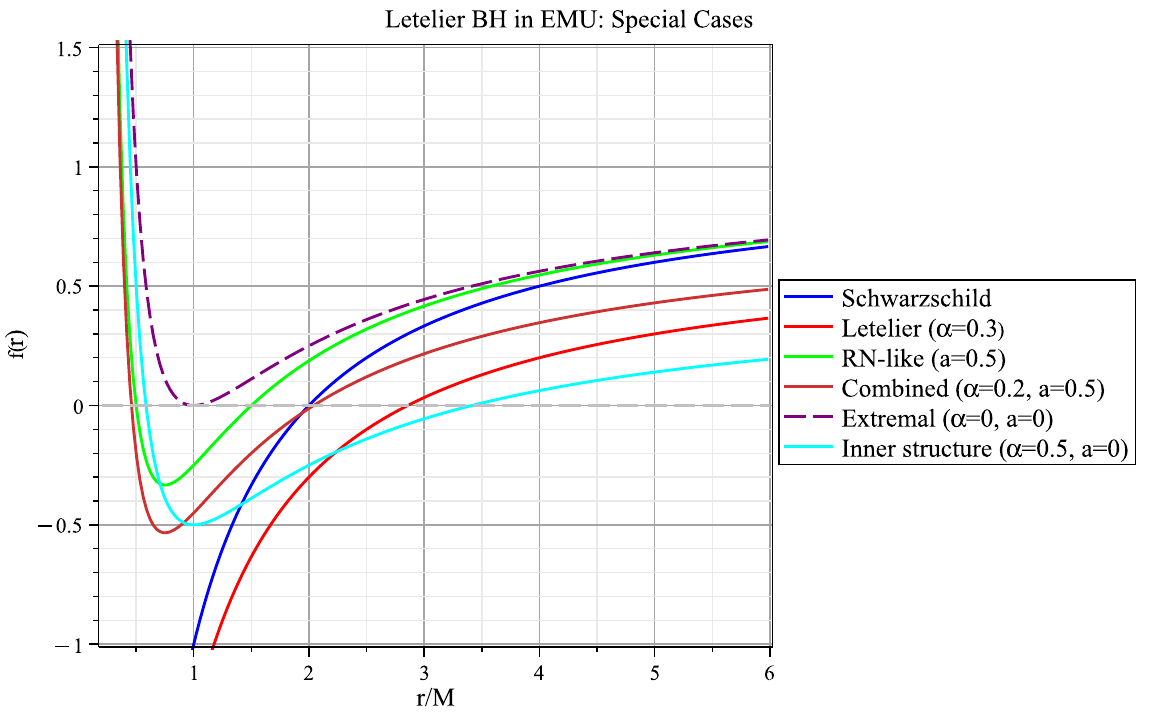}
\caption{}
\label{fig:metric_special}
\end{subfigure}
\caption{Lapse function $f(r)$ for the Letelier BH in EMU with $M=1$. (a) Variation with CoS parameter $\alpha \in \{0.0, 0.1, 0.3, 0.5, 0.7, 0.9\}$ at fixed $a = 0.5$. (b) Variation with EMU parameter $a \in \{0.0, 0.3, 0.5, 0.7, 0.9, 1.0\}$ at fixed $\alpha = 0.1$. (c) Special limiting cases including Schwarzschild, Letelier, RN-like, and extremal configurations. The dashed gray line marks $f(r) = 0$, with intersections indicating horizon locations.}
\label{fig:metric}
\end{figure*}

\subsection{Asymptotic Structure and Curvature Invariants}

At large distances, the metric function expands as
\begin{equation}
f(r) \approx (1-\alpha) - \frac{2M}{r} + \mathcal{O}(r^{-2}),
\label{eq:asymptotic_f}
\end{equation}
indicating that the spacetime approaches a deficit angle geometry characterized by $\alpha$ rather than flat Minkowski space \cite{isz12}. The solid angle at spatial infinity is reduced to $4\pi(1-\alpha)$, producing a globally conical structure that distinguishes the Letelier BH from standard asymptotically flat solutions.

The curvature invariants quantify the gravitational field strength in a coordinate-independent manner. The Ricci scalar evaluates to
\begin{equation}
R = \frac{2\alpha}{r^2},
\label{eq:Ricci_scalar}
\end{equation}
which vanishes in the absence of the CoS ($\alpha = 0$) but diverges as $r \to 0$ for any non-zero $\alpha$. The non-vanishing Ricci scalar reflects the energy-momentum contribution from the CoS, which acts as an anisotropic fluid with equation of state $p_r = -\rho$ \cite{Letelier1979,sec2is09}. This equation of state characterizes Nambu-Goto strings and produces a traceless stress tensor in the string-orthogonal directions.

The Kretschmann scalar $K = R_{\mu\nu\rho\sigma}R^{\mu\nu\rho\sigma}$ takes the form
\begin{equation}
K = \frac{48M^2}{r^6} - \frac{96M^3(1-a^2)}{r^7} + \frac{56M^4(1-a^2)^2}{r^8},
\label{eq:Kretschmann}
\end{equation}
which diverges as $r \to 0$, confirming the presence of a curvature singularity at the origin \cite{sec2is10}. The leading $r^{-6}$ term matches the Schwarzschild contribution, while the subleading terms arise from the EMU modification. The Kretschmann scalar is independent of $\alpha$, indicating that the CoS contributes only to the Ricci curvature and not to the Weyl tensor.

\subsection{Isometric Embedding Diagrams}

To visualize the spatial geometry outside the EH, we construct isometric embedding diagrams following the standard procedure \cite{sec2is11}. For a constant-time ($t = \text{const}$), equatorial ($\theta = \pi/2$) slice, the induced 2D metric reads
\begin{equation}
d\sigma^2 = \frac{dr^2}{f(r)} + r^2 d\phi^2.
\label{eq:induced_2D}
\end{equation}
This metric can be embedded in 3D Euclidean space $(r, \phi, Z)$ provided the embedding height $Z(r)$ satisfies
\begin{equation}
\frac{dZ}{dr} = \sqrt{\frac{1}{f(r)} - 1}.
\label{eq:embedding_eq}
\end{equation}
The embedding is valid for $f(r) < 1$, which holds outside the EH throughout the parameter space.

Figure~\ref{fig:embedding_LetEMU} displays 3D embedding diagrams for six representative configurations. Panel (i) shows the Schwarzschild limit ($\alpha = 0$, $a = 1$) with the familiar funnel shape and horizon at $r_+ = 2M$. Panel (ii) presents the RN-like case ($\alpha = 0$, $a = 0.5$), where the effective charge term modifies the near-horizon geometry, producing a narrower throat. Panel (iii) illustrates the pure Letelier BH ($\alpha = 0.1$, $a = 1$) with an expanded horizon at $r_+ = 2.22M$. Panels (iv)--(vi) demonstrate the combined effects: as $\alpha$ increases, the EH expands dramatically, reaching $r_+ = 3.58M$ at $(\alpha = 0.5, a = 0.5)$ and $r_+ = 4.00M$ at $(\alpha = 0.5, a = 1)$. The red ring in each diagram marks the EH location $r_+$, and the black spiral represents a timelike geodesic spiraling toward the singularity.

\begin{figure*}[ht!]
    \centering
        \includegraphics[width=0.3\textwidth]{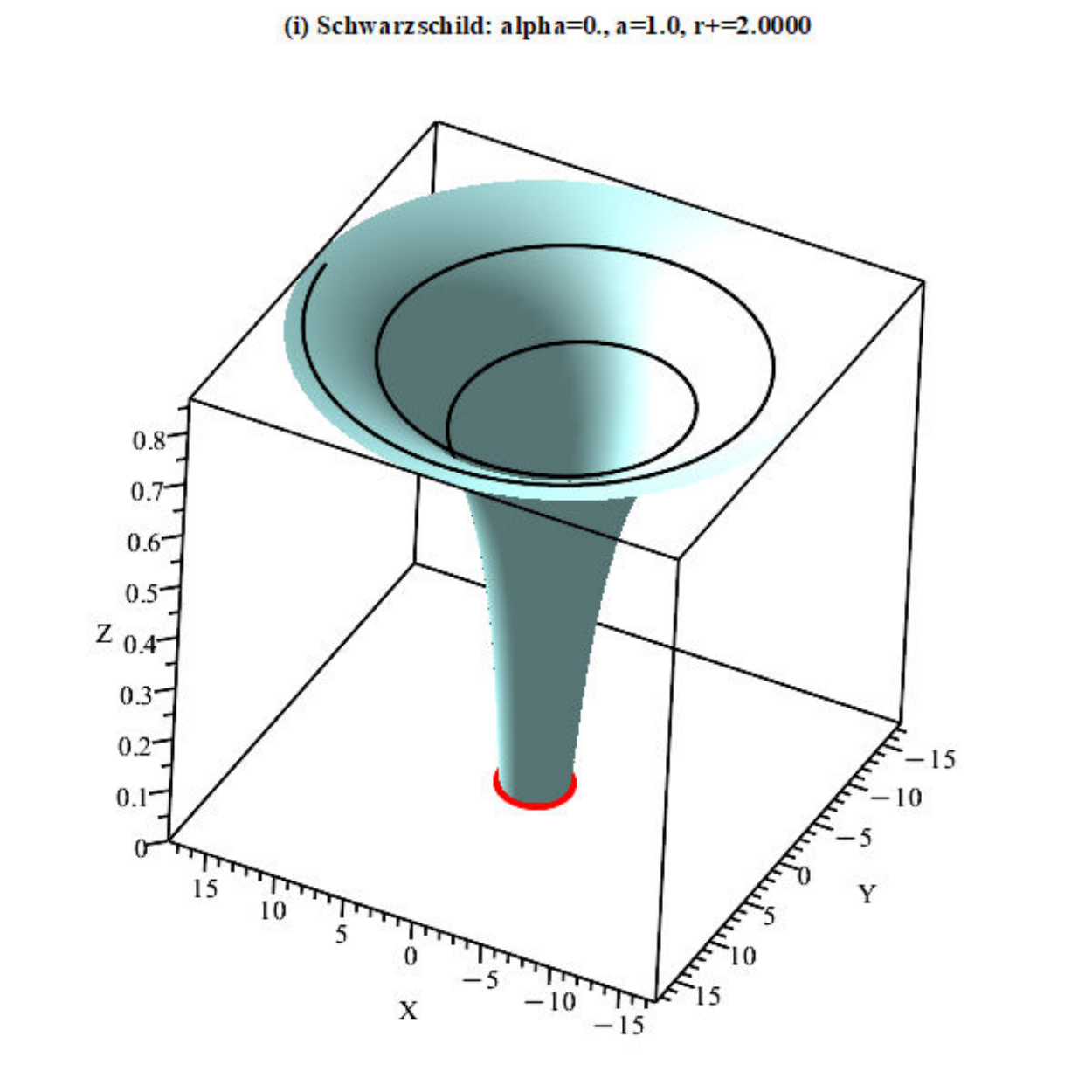}
        \includegraphics[width=0.3\textwidth]{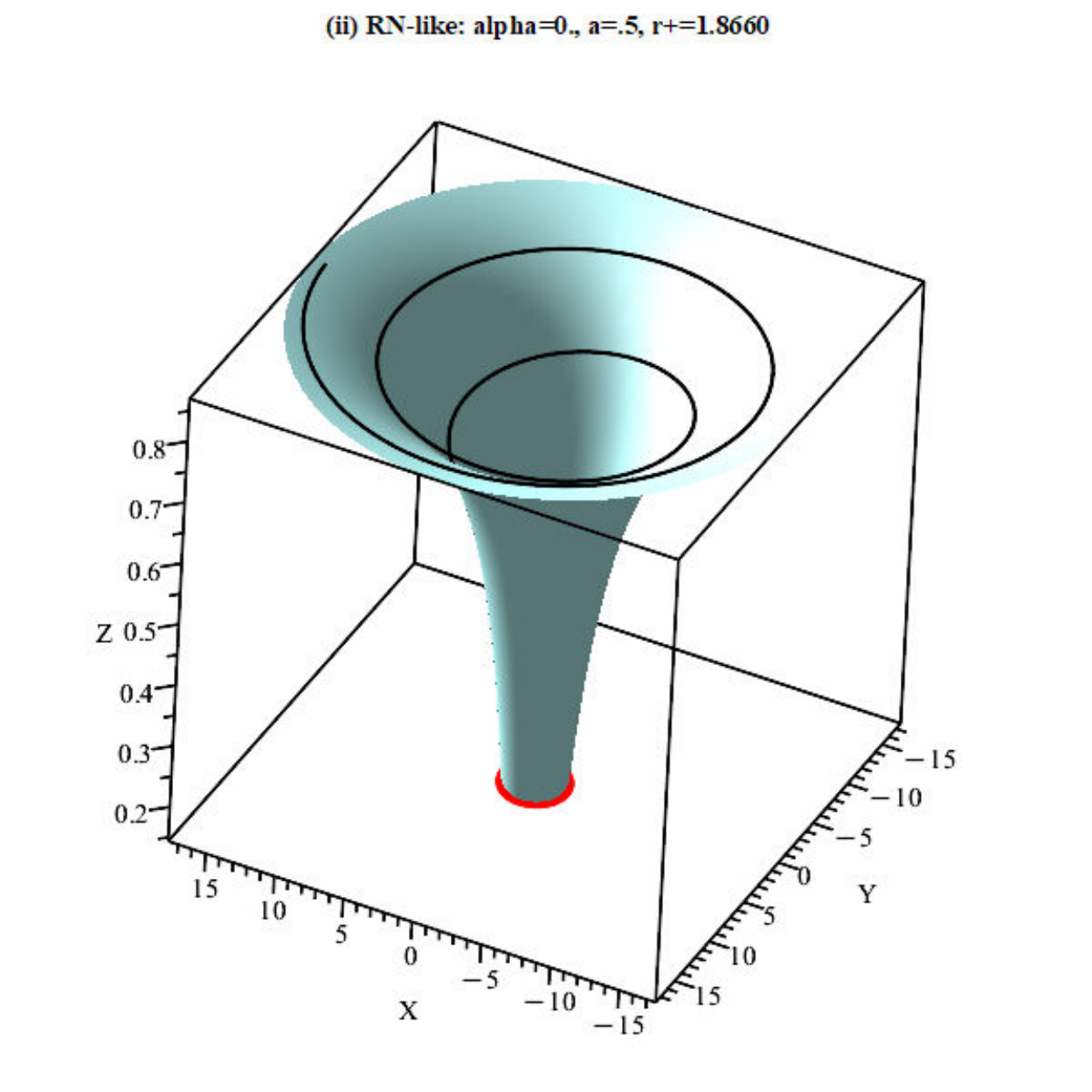}
        \includegraphics[width=0.3\textwidth]{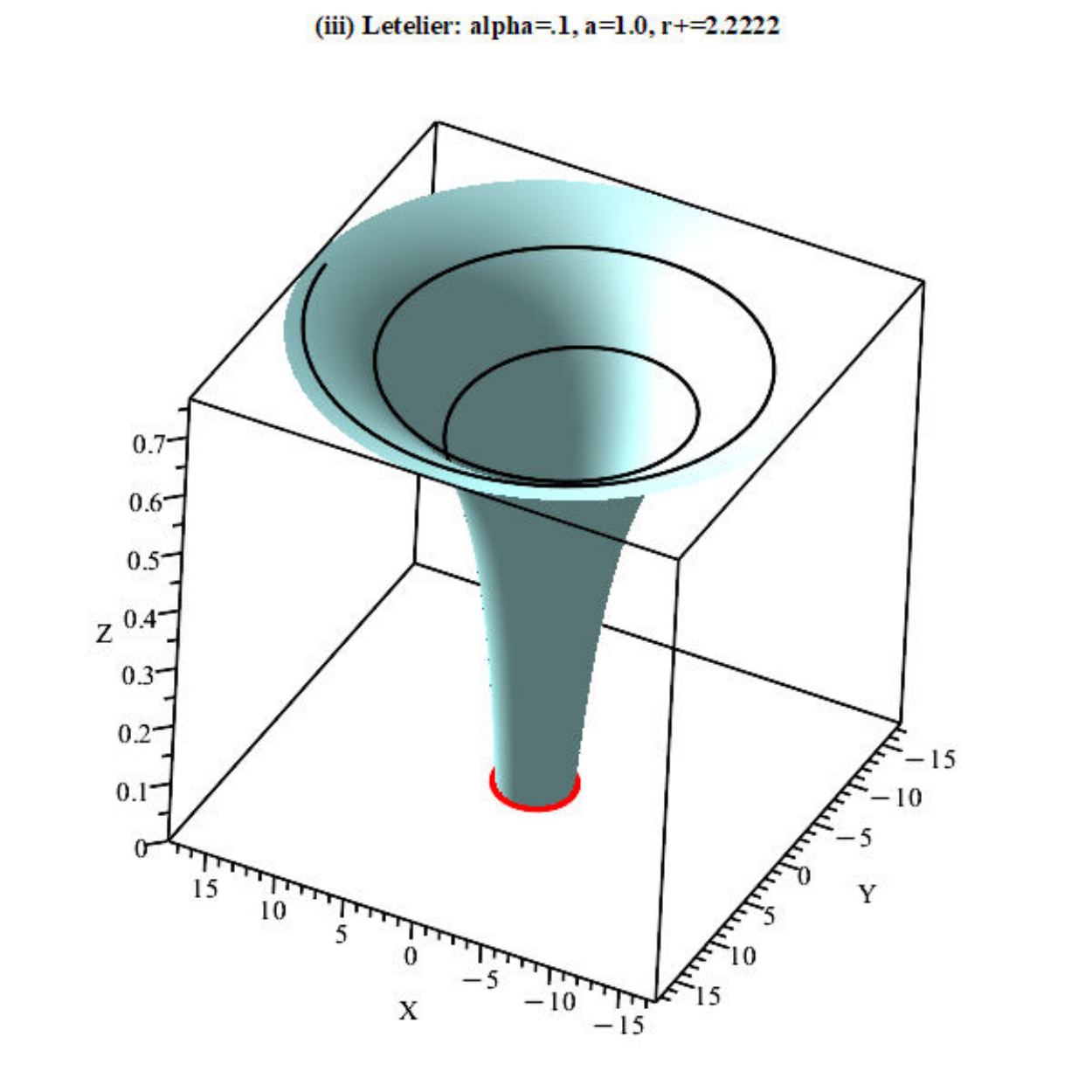}\\
        (i) $\alpha=0$, $a=1.0$, $r_+=2.00M$ (Sch.) \hspace{0.2cm}
        (ii) $\alpha=0$, $a=0.5$, $r_+=1.87M$ (RN-like) \hspace{0.2cm}
        (iii) $\alpha=0.1$, $a=1.0$, $r_+=2.22M$ (Letelier)
        \\[0.3cm]
        \includegraphics[width=0.3\textwidth]{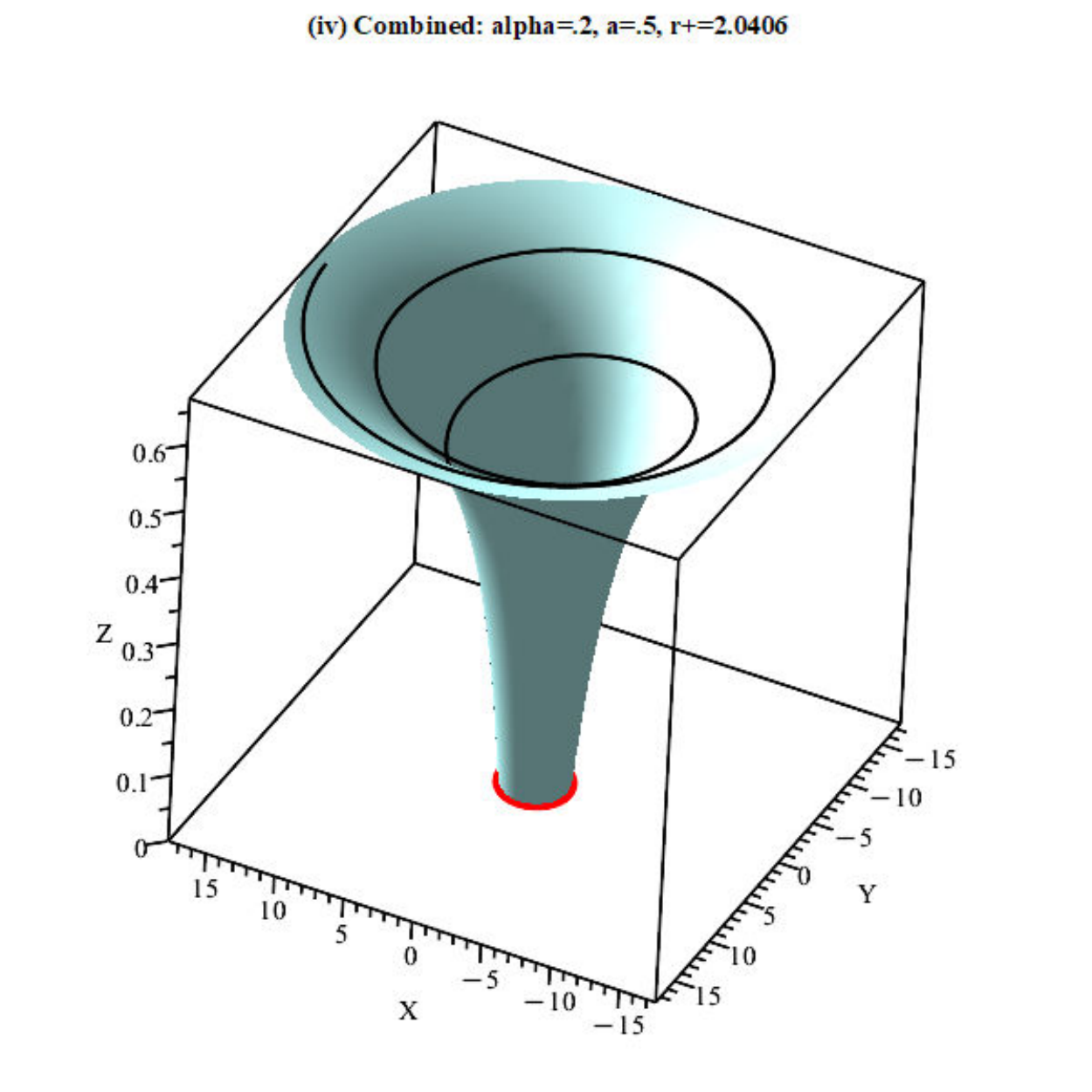}
        \includegraphics[width=0.3\textwidth]{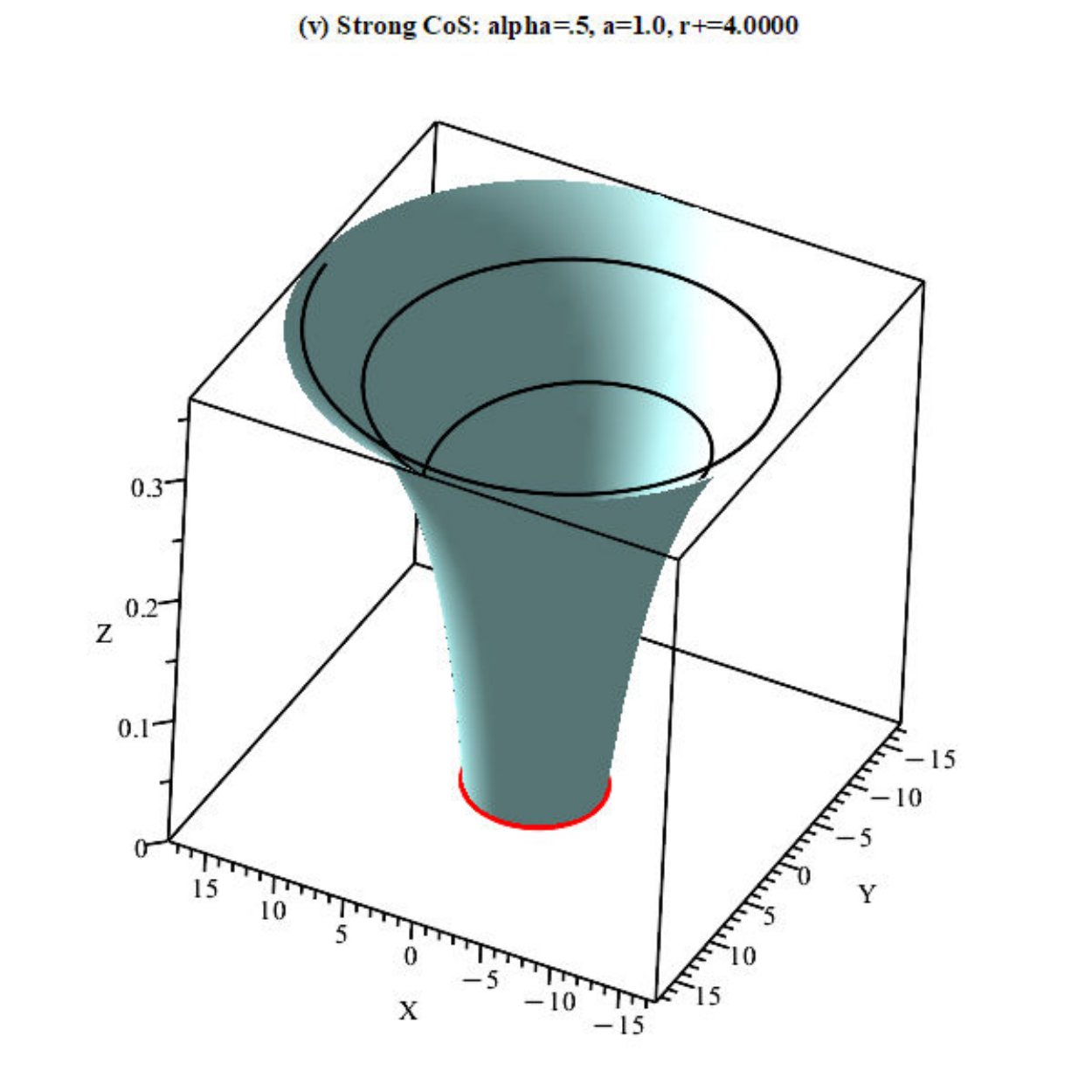}
        \includegraphics[width=0.3\textwidth]{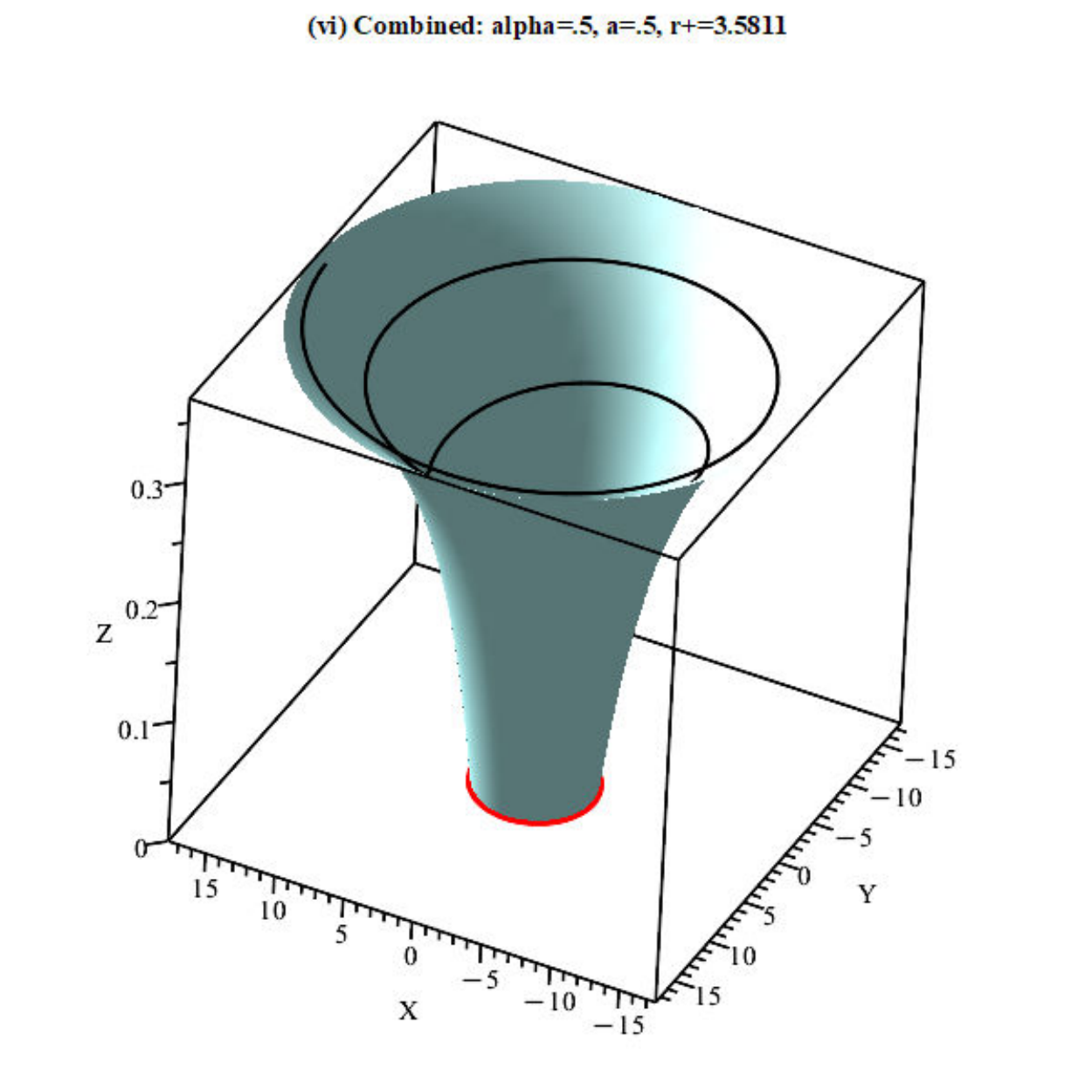}\\
        (iv) $\alpha=0.2$, $a=0.5$, $r_+=2.04M$ \hspace{0.4cm}
        (v) $\alpha=0.5$, $a=1.0$, $r_+=4.00M$ \hspace{0.4cm}
        (vi) $\alpha=0.5$, $a=0.5$, $r_+=3.58M$
\caption{3D isometric embedding diagrams of the Letelier BH in EMU for six parameter configurations with $M = 1$. The turquoise surface represents the spatial geometry from the EH outward to $r = 15M$, the red ring marks the outer horizon $r_+$, and the black curve depicts an infalling trajectory. The funnel widens as $\alpha$ increases, reflecting the EH expansion induced by the CoS.}
\label{fig:embedding_LetEMU}
\end{figure*}

\subsection{Surface Gravity and Thermodynamic Quantities}

The surface gravity at the outer horizon governs the Hawking temperature and initiates the discussion of BH thermodynamics \cite{Hawking1975,sec2is12,sec2is12}. For the metric (\ref{eq:metric}), the surface gravity is
\begin{equation}
\kappa = \frac{1}{2}\left|\frac{df}{dr}\right|_{r=r_+} = \frac{M}{r_+^2}\left|1 - \frac{M(1-a^2)}{r_+}\right|.
\label{eq:surface_gravity}
\end{equation}
Substituting the horizon radius from Eq.~(\ref{eq:horizons}) and simplifying yields
\begin{equation}
\kappa = \frac{(1-\alpha)\sqrt{1-(1-\alpha)(1-a^2)}}{2M}.
\label{eq:kappa_explicit}
\end{equation}
In the Schwarzschild limit ($\alpha = 0$, $a = 1$), we recover $\kappa = 1/(4M)$.

The Hawking temperature follows from the Euclidean periodicity argument as $T_H = \kappa/(2\pi)$ \cite{Hawking1975,sec2is13}:
\begin{equation}
T_H = \frac{(1-\alpha)\sqrt{1-(1-\alpha)(1-a^2)}}{4\pi M},
\label{eq:Hawking_T_sec2}
\end{equation}
which reduces to $T_H^{\rm Sch} = 1/(8\pi M)$ in the Schwarzschild limit. The temperature decreases monotonically with increasing $\alpha$, reaching zero in the extremal limit. This behavior has important consequences for HR and the BH evaporation timescale, as explored in Sec.~\ref{isec7}.

The BH entropy follows from the Bekenstein-Hawking area law \cite{sec2is12,sec2is14}:
\begin{equation}
S = \frac{A}{4} = \pi r_+^2,
\label{eq:BH_entropy}
\end{equation}
where $A = 4\pi r_+^2$ is the horizon area. Since $r_+$ increases with $\alpha$, the entropy of the Letelier BH in EMU exceeds that of the Schwarzschild BH of equal mass, reflecting the expanded horizon area. The first law of BH thermodynamics extends to include variations of the CoS and EMU parameters: $dM = T_H dS + \Phi_\alpha d\alpha + \Phi_a da$, where $\Phi_\alpha$ and $\Phi_a$ are the conjugate potentials associated with the string density and electromagnetic background \cite{sec2is15}.

\section{PS, Shadow, and Optical Appearance}\label{isec3}

In this section, we analyze the optical properties of the Letelier BH in EMU. We begin by deriving the PS radius and shadow observables in vacuum, then extend the analysis to plasma environments. Subsequently, we investigate the optical appearance of the BH by decomposing the observed intensity into direct emission, lensed emission, and PR contributions. Finally, photon trajectories are examined to illustrate the light deflection behavior near the BH.

\subsection{PS and Effective Potential}

The motion of photons around the Letelier BH in EMU is governed by the null geodesic equations. For photons confined to the equatorial plane ($\theta = \pi/2$), the effective potential takes the form \cite{sec3is01,sec3is02}
\begin{equation}
V_{\text{eff}} = \frac{L^2}{r^2} f(r) = \frac{L^2}{r^2} \left(1 - \alpha - \frac{2M}{r} + \frac{M^2(1-a^2)}{r^2}\right),
\end{equation}
where $L$ is the conserved angular momentum of the photon. The effective potential encapsulates the combined gravitational influence of the BH mass, the CoS parameter $\alpha$, and the EMU correction characterized by the parameter $a$.

The PS corresponds to unstable circular orbits located at the maximum of the effective potential. The conditions $V_{\text{eff}}'(r_{ps}) = 0$ and $V_{\text{eff}}''(r_{ps}) < 0$ yield the PS radius \cite{sec3is03}
\begin{equation}
r_{ps} = \frac{M\left(3 + \sqrt{9 - 8(1-\alpha)(1-a^2)}\right)}{2(1-\alpha)}.
\label{eq:rps}
\end{equation}
In the Schwarzschild limit ($\alpha = 0$, $a = 1$), we recover the standard result $r_{ps} = 3M$.

Figure~\ref{fig:Veff} displays the effective potential for null geodesics under varying parameter conditions. In panel (a), we observe that increasing $\alpha$ at fixed $a = 0.5$ reduces the peak height and shifts the maximum outward. This behavior reflects the gravitational dilution caused by the CoS, which effectively reduces the gravitational binding and allows photons to orbit at larger radii. In panel (b), decreasing the EMU parameter $a$ at fixed $\alpha = 0.1$ increases the peak height and shifts it inward. This indicates that stronger electromagnetic corrections enhance the effective gravitational confinement near the BH. The peak location in each case corresponds precisely to the PS radius $r_{ps}$.

\begin{figure*}[ht!]
\centering
\begin{subfigure}[b]{0.48\textwidth}
\centering
\includegraphics[width=\textwidth]{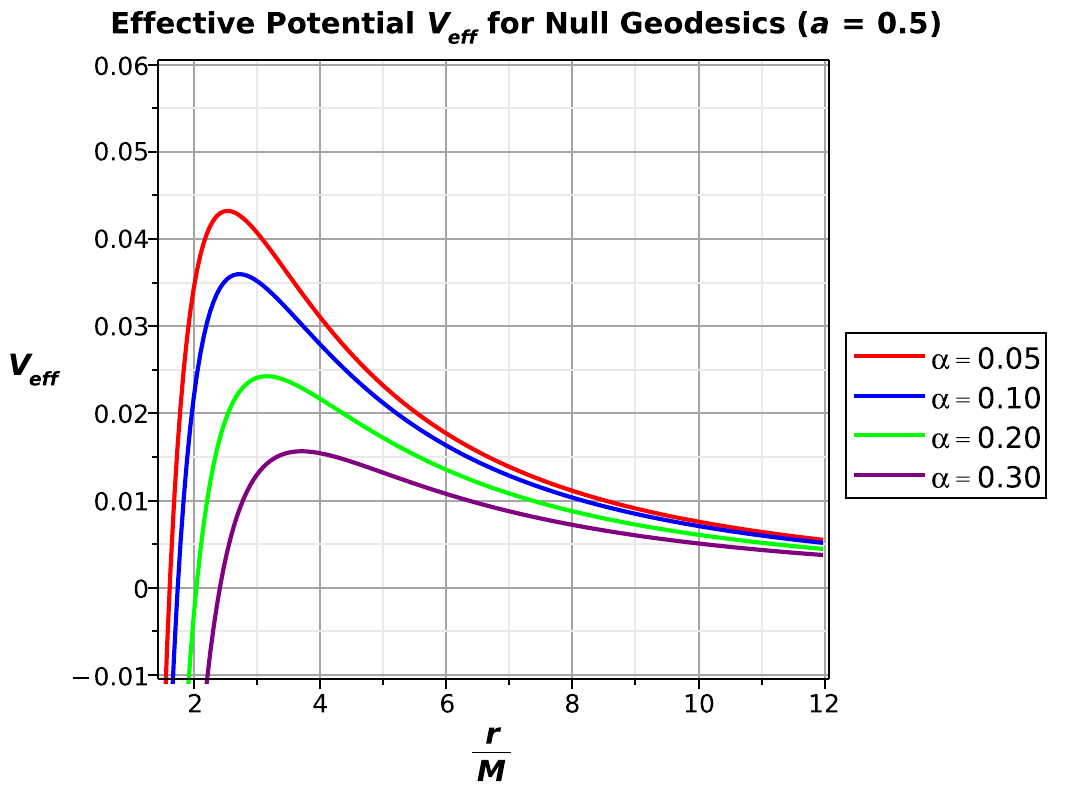}
\caption{}
\label{fig:Veff_alpha}
\end{subfigure}
\hfill
\begin{subfigure}[b]{0.48\textwidth}
\centering
\includegraphics[width=\textwidth]{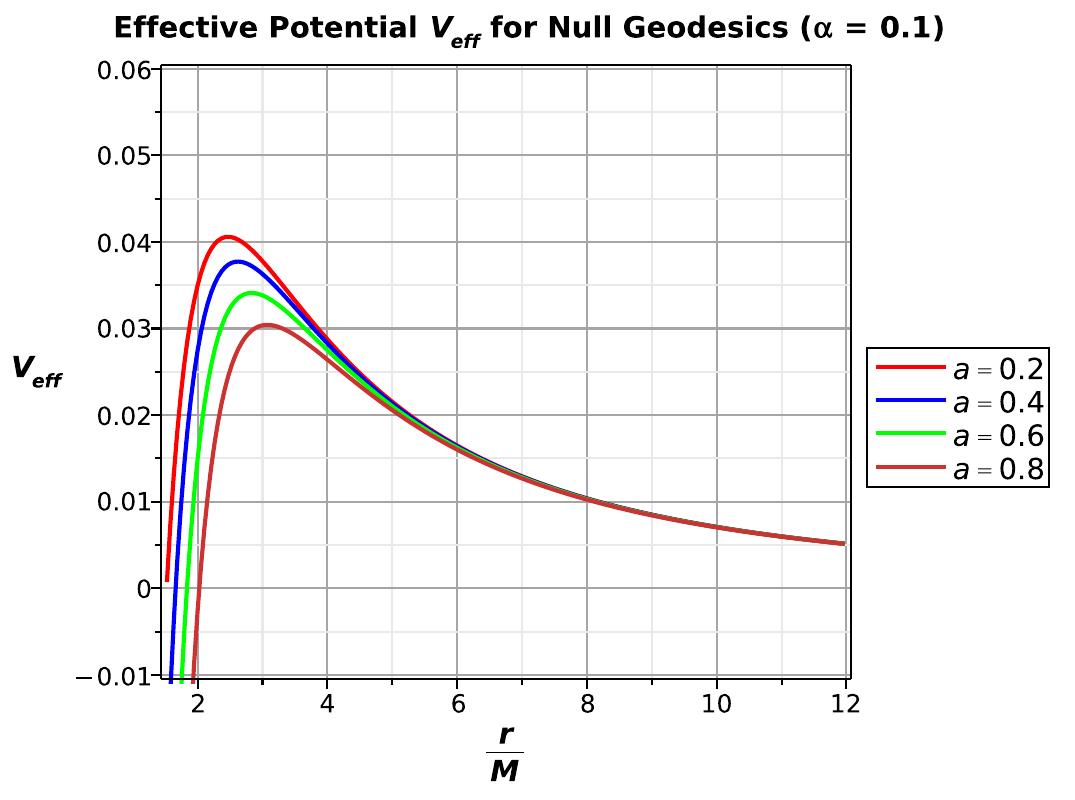}
\caption{}
\label{fig:Veff_a}
\end{subfigure}
\caption{Effective potential $V_{\text{eff}}$ for null geodesics with $L = 1$ and $M = 1$. (a) Varying $\alpha \in \{0.0, 0.1, 0.2, 0.3, 0.4\}$ at fixed $a = 0.5$. (b) Varying $a \in \{0.3, 0.5, 0.7, 0.9, 1.0\}$ at fixed $\alpha = 0.1$. The peak location corresponds to the PS radius $r_{ps}$.}
\label{fig:Veff}
\end{figure*}

\subsection{Vacuum Shadow}

For an observer located at spatial infinity, the apparent size of the BH shadow is determined by photons that asymptotically approach the PS \cite{sucu2025quantumHassan,isz32,sec3is05}. The shadow radius, which represents the critical impact parameter separating captured and scattered photon trajectories, is given by
\begin{equation}
R_{sh} = \frac{r_{ps}}{\sqrt{f(r_{ps})}} \sqrt{1-\alpha} = r_{ps} \sqrt{\frac{1-\alpha}{f(r_{ps})}}.
\label{eq:Rsh}
\end{equation}
The factor $\sqrt{1-\alpha}$ accounts for the redshift correction due to the CoS at the observer's location \cite{isz37}. In the Schwarzschild limit, we recover $R_{sh}^{\text{Sch}} = 3\sqrt{3}M \approx 5.196M$.

The numerical values of the PS and shadow radii are presented in Table~\ref{tab:vacuum_shadow} for various combinations of $\alpha$ and $a$. Both $r_{ps}$ and $R_{sh}$ increase monotonically with increasing $\alpha$ and $a$. Physically, the CoS reduces the effective gravitational attraction, causing the PS to expand outward. Similarly, increasing $a$ (which corresponds to weaker electromagnetic corrections) allows the spacetime to approach the pure Letelier geometry, which has a larger PS than the electromagnetically modified case.

\begin{table*}[ht!]
\centering
\setlength{\tabcolsep}{8pt}
\renewcommand{\arraystretch}{1.5}
\begin{tabular}{c|cc|cc|cc}
\hline\hline
\rowcolor{orange!50}
 & \multicolumn{2}{c|}{$a=0.5$} & \multicolumn{2}{c|}{$a=0.8$} & \multicolumn{2}{c}{$a=1.0$} \\
\rowcolor{orange!50}
$\alpha$ & $r_{ps}/M$ & $R_{sh}/M$ & $r_{ps}/M$ & $R_{sh}/M$ & $r_{ps}/M$ & $R_{sh}/M$ \\
\hline
0.00 & 2.3660 & 4.4036 & 2.7370 & 4.8587 & 3.0000 & 5.1962 \\
0.05 & 2.5351 & 4.6882 & 2.8962 & 5.1335 & 3.1579 & 5.4698 \\
0.10 & 2.7208 & 5.0025 & 3.0730 & 5.4389 & 3.3333 & 5.7735 \\
0.15 & 2.9264 & 5.3518 & 3.2704 & 5.7801 & 3.5294 & 6.1129 \\
0.20 & 3.1559 & 5.7431 & 3.4923 & 6.1636 & 3.7500 & 6.4952 \\
0.25 & 3.4142 & 6.1848 & 3.7435 & 6.5979 & 4.0000 & 6.9282 \\
0.30 & 3.7078 & 6.6878 & 4.0305 & 7.0945 & 4.2857 & 7.4228 \\
\hline\hline
\end{tabular}
\caption{PS radius $r_{ps}/M$ and shadow radius $R_{sh}/M$ for the Letelier BH in EMU. The Schwarzschild reference values are $r_{ps}^{\text{Sch}} = 3M$ and $R_{sh}^{\text{Sch}} = 3\sqrt{3}M \approx 5.196M$.}
\label{tab:vacuum_shadow}
\end{table*}

Figure~\ref{fig:Rsh_params} illustrates the parametric dependence of the shadow radius. In panel (a), $R_{sh}$ is plotted against $\alpha$ for different values of $a$. All curves exhibit a monotonic increase with $\alpha$, and for $\alpha > 0$, all curves lie above the Schwarzschild reference line (gray dashed). The curves are approximately parallel, indicating that the CoS and EMU effects contribute additively to the shadow enlargement. In panel (b), $R_{sh}$ is plotted against $a$ for different values of $\alpha$. The shadow radius increases with $a$, approaching the Schwarzschild value at $a = 1$ for $\alpha = 0$. Higher values of $\alpha$ shift the entire curve upward, confirming that both parameters independently contribute to shadow expansion.

\begin{figure*}[ht!]
\centering
\begin{subfigure}[b]{0.48\textwidth}
\centering
\includegraphics[width=\textwidth]{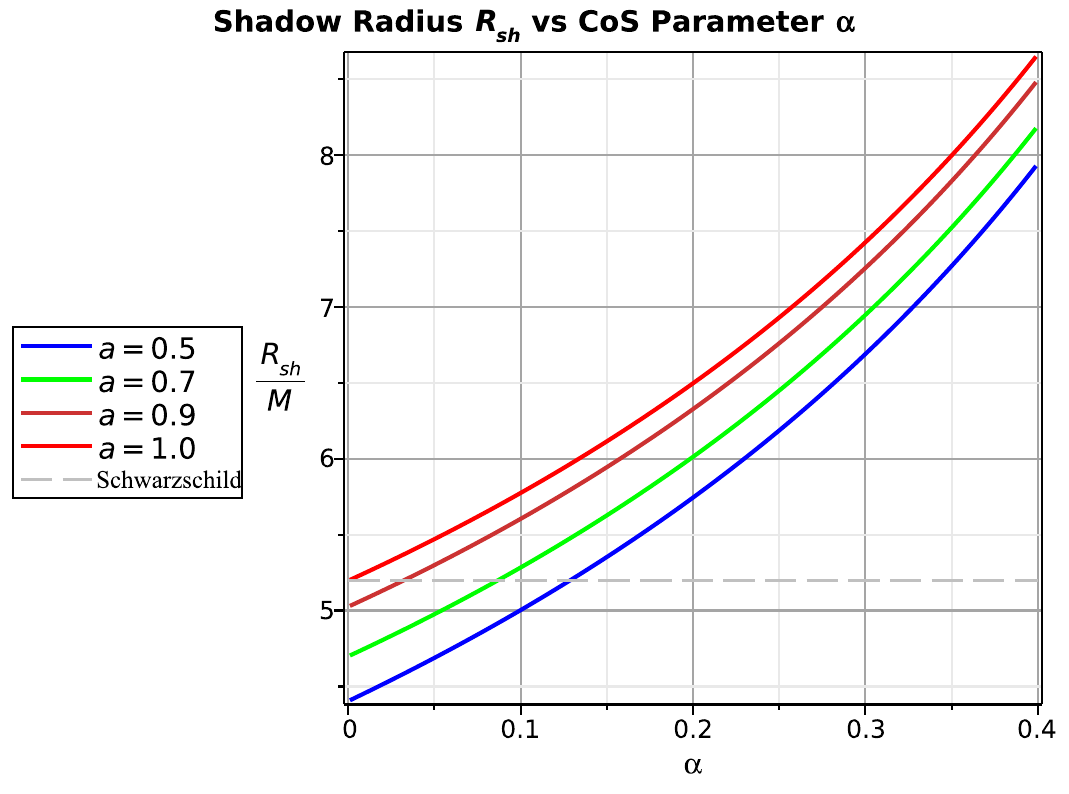}
\caption{}
\label{fig:Rsh_alpha}
\end{subfigure}
\hfill
\begin{subfigure}[b]{0.48\textwidth}
\centering
\includegraphics[width=\textwidth]{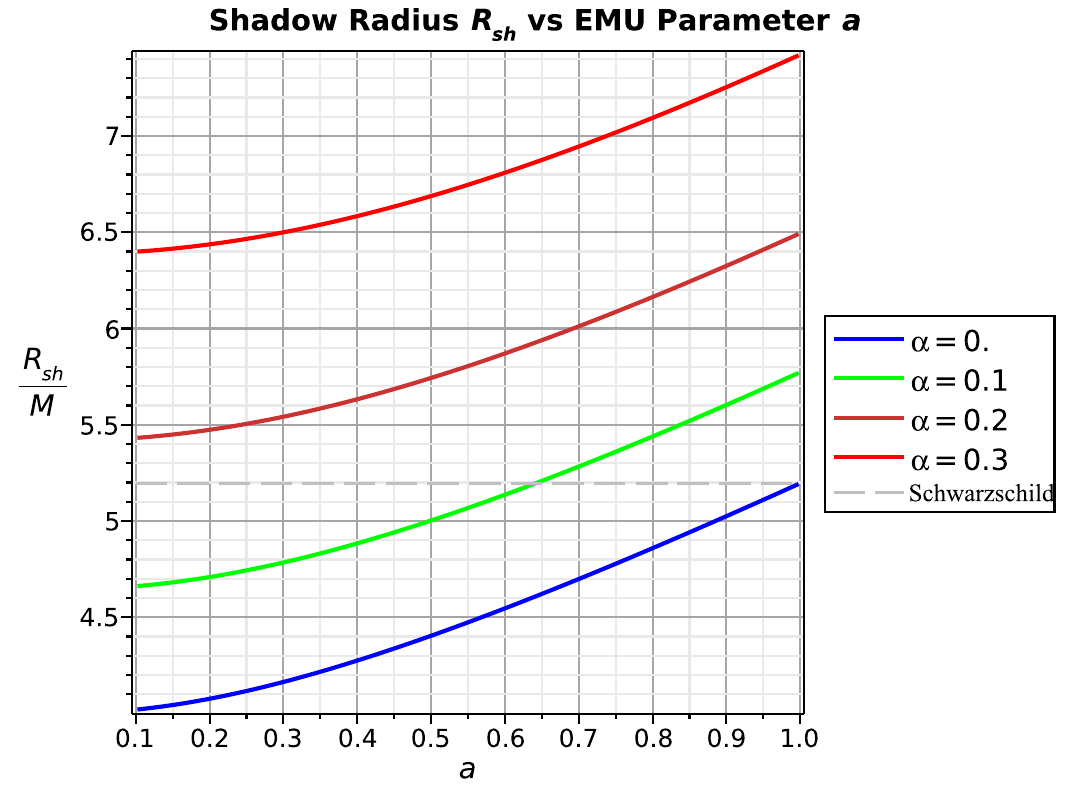}
\caption{}
\label{fig:Rsh_a}
\end{subfigure}
\caption{Shadow radius $R_{sh}/M$ as a function of the model parameters. (a) $R_{sh}$ versus CoS parameter $\alpha$ for different values of $a$. (b) $R_{sh}$ versus EMU parameter $a$ for different values of $\alpha$. The gray dashed line indicates the Schwarzschild value $R_{sh}^{\text{Sch}} = 3\sqrt{3}M$.}
\label{fig:Rsh_params}
\end{figure*}

\subsection{Optical Appearance and Emission Decomposition}

The visual appearance of a BH depends not only on the shadow size but also on the emission profile of the surrounding accretion environment. Following the methodology employed in the EHT observations \cite{isz04,isz05}, we decompose the total observed intensity into three distinct contributions \cite{isz33,isz45}. The direct emission consists of light rays that travel directly from the emitting region to the observer without significant gravitational deflection, contributing to the broad, diffuse glow surrounding the shadow. The lensed emission arises from light rays that undergo moderate gravitational bending, typically completing less than one orbit around the BH before reaching the observer, forming a secondary ring structure outside the PR. The PR contribution comes from light rays that pass very close to the PS, completing one or more half-orbits before escaping, forming a sharp, bright ring at the edge of the shadow that carries information about the strong-field geometry \cite{sec3is09}.

The observed intensity can be modeled as
\begin{equation}
I_{\text{obs}}(x, y) = I_{\text{direct}}(x, y) + I_{\text{lensed}}(x, y) + I_{\text{ring}}(x, y),
\label{eq:Iobs}
\end{equation}
where $(x, y)$ are celestial coordinates on the observer's screen. For a geometrically thin, optically thin accretion disk, each component can be approximated by Gaussian profiles centered at characteristic radii determined by the BH geometry \cite{sec3is10}.

Figure~\ref{fig:emission_components} displays the three emission components for varying CoS parameter $\alpha$ at fixed $a = 1$. The rows correspond to $\alpha = 0$, $0.1$, and $0.2$, while the columns show the direct emission, lensed emission, and PR contributions, respectively. The direct emission exhibits a broad intensity distribution peaking outside the shadow boundary, and as $\alpha$ increases, the shadow region expands and the emission peak shifts outward, consistent with the enlarged PS. The lensed emission forms a narrower ring structure compared to the direct emission, with the ring radius increasing with $\alpha$ to track the PS expansion. The PR appears as a sharp, thin ring at the shadow boundary, representing the most sensitive probe of the near-horizon geometry, and shows clear expansion with increasing $\alpha$.

\begin{figure*}[ht!]
\centering
\begin{subfigure}[b]{0.32\textwidth}
\centering
\includegraphics[width=\textwidth]{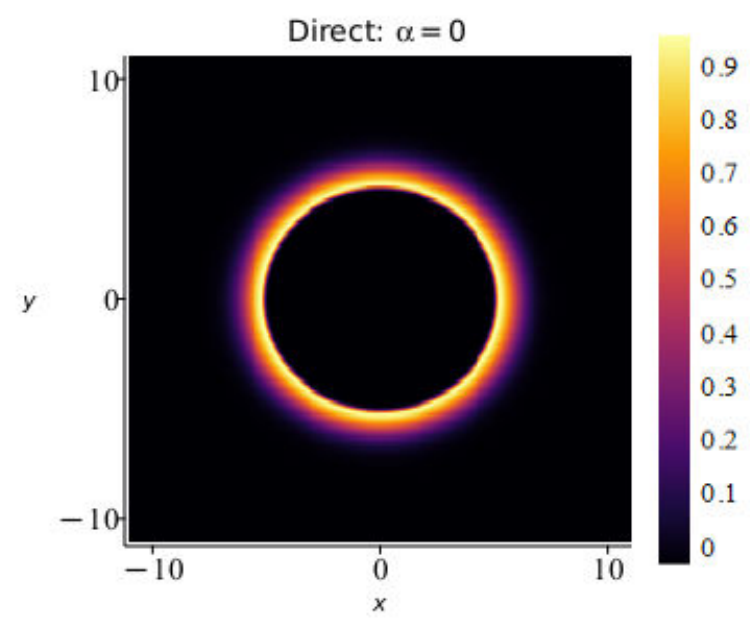}
\caption{}
\end{subfigure}
\hfill
\begin{subfigure}[b]{0.32\textwidth}
\centering
\includegraphics[width=\textwidth]{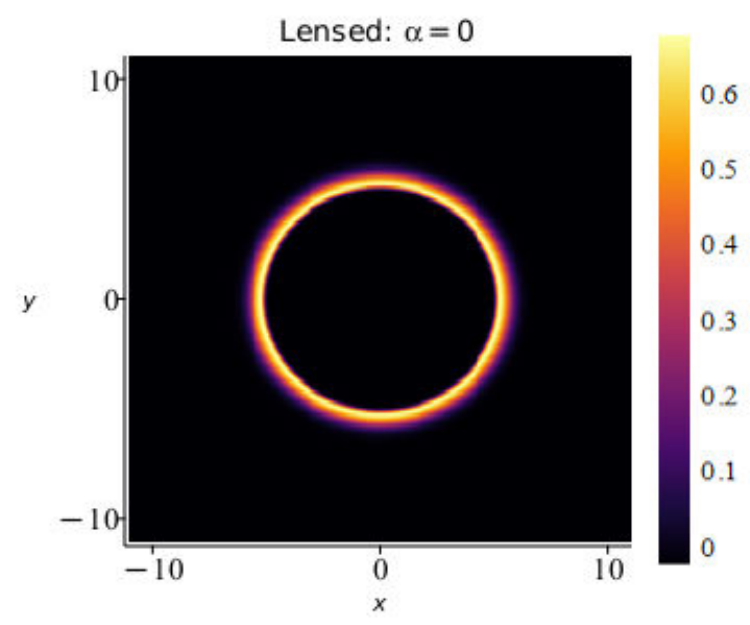}
\caption{}
\end{subfigure}
\hfill
\begin{subfigure}[b]{0.32\textwidth}
\centering
\includegraphics[width=\textwidth]{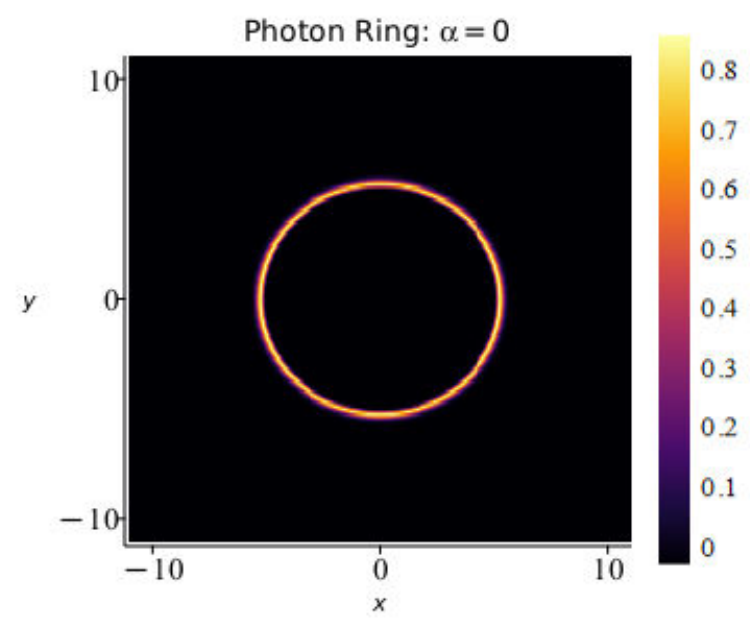}
\caption{}
\end{subfigure}
\\[4pt]
\begin{subfigure}[b]{0.32\textwidth}
\centering
\includegraphics[width=\textwidth]{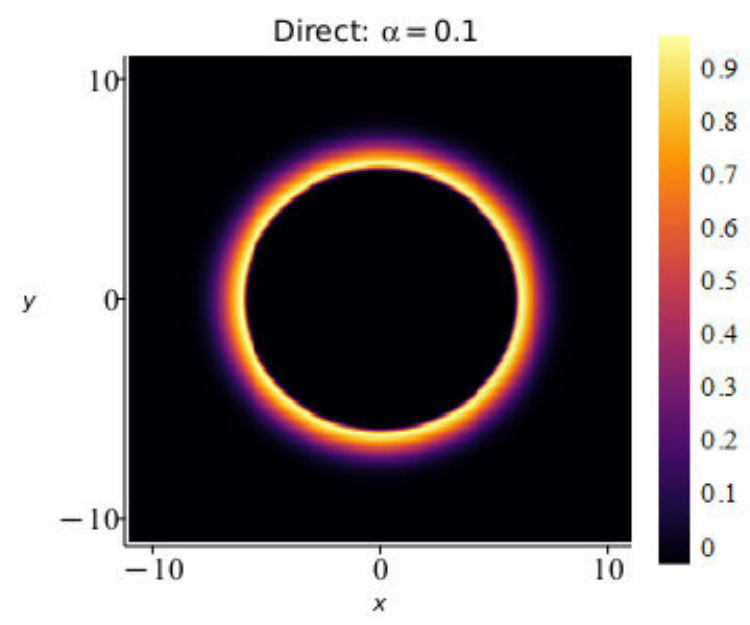}
\caption{}
\end{subfigure}
\hfill
\begin{subfigure}[b]{0.32\textwidth}
\centering
\includegraphics[width=\textwidth]{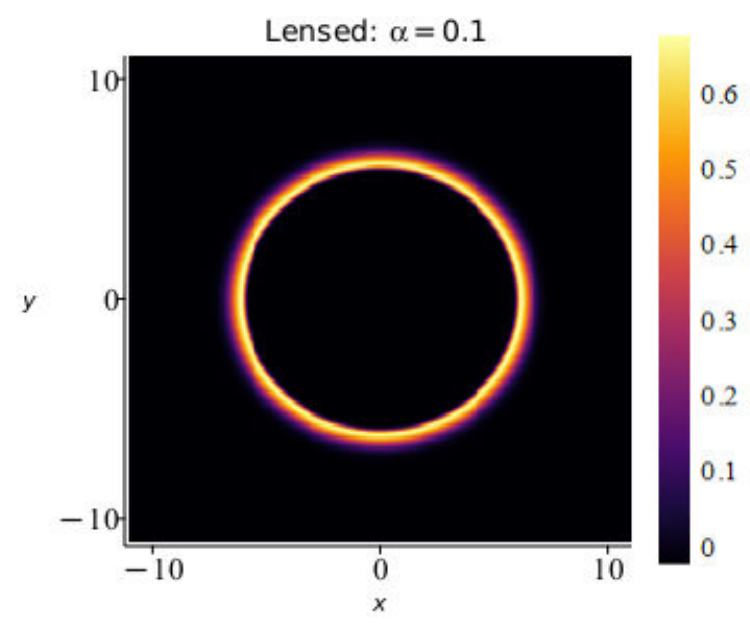}
\caption{}
\end{subfigure}
\hfill
\begin{subfigure}[b]{0.32\textwidth}
\centering
\includegraphics[width=\textwidth]{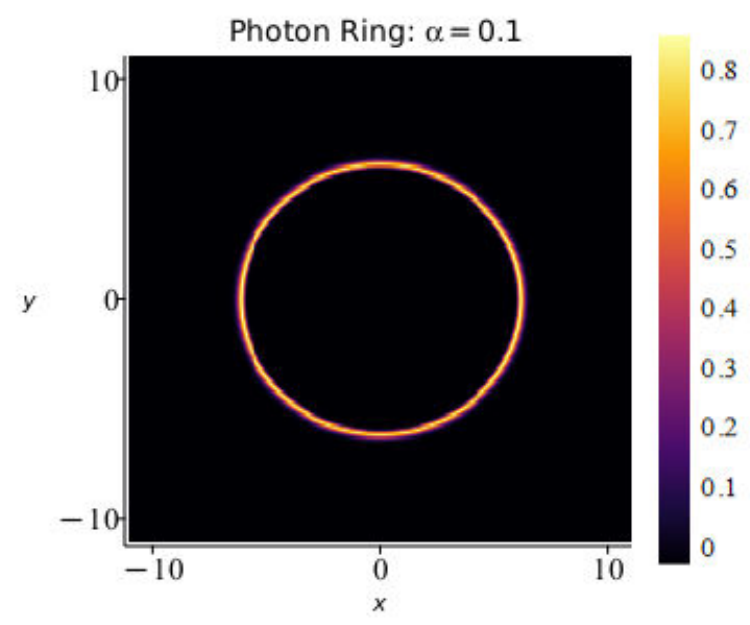}
\caption{}
\end{subfigure}
\\[4pt]
\begin{subfigure}[b]{0.32\textwidth}
\centering
\includegraphics[width=\textwidth]{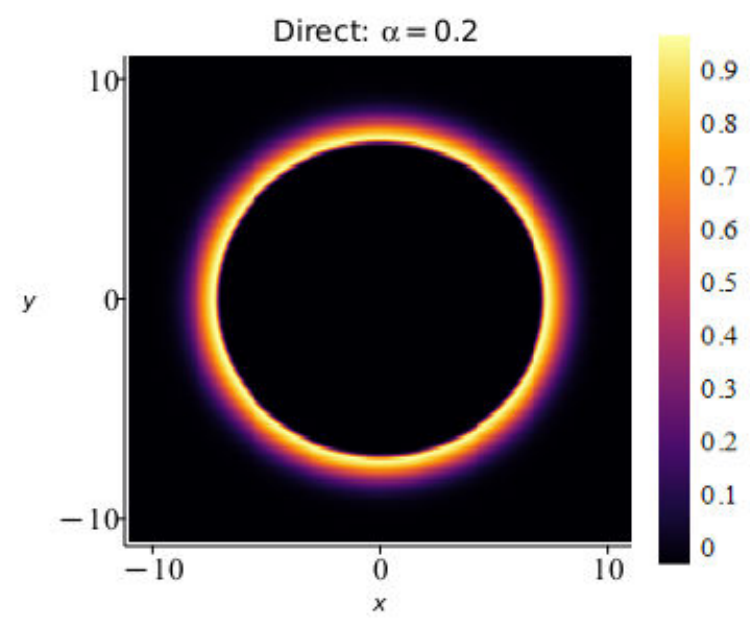}
\caption{}
\end{subfigure}
\hfill
\begin{subfigure}[b]{0.32\textwidth}
\centering
\includegraphics[width=\textwidth]{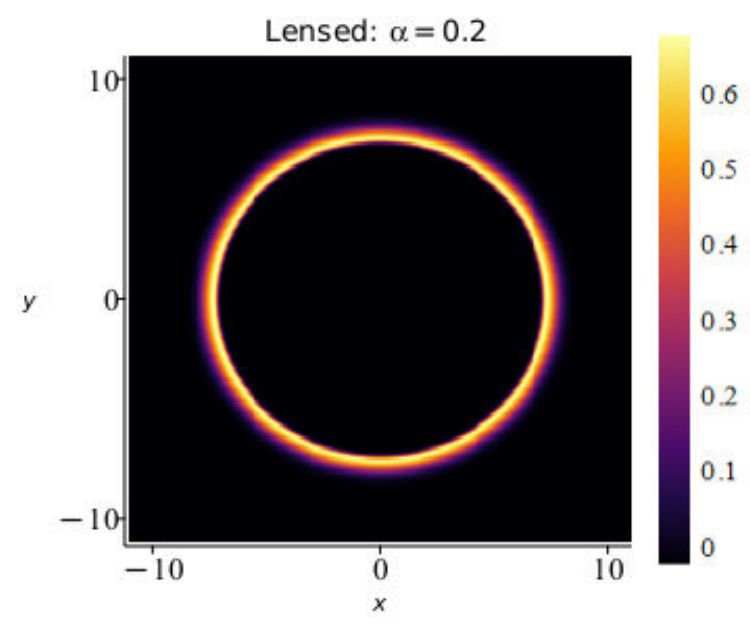}
\caption{}
\end{subfigure}
\hfill
\begin{subfigure}[b]{0.32\textwidth}
\centering
\includegraphics[width=\textwidth]{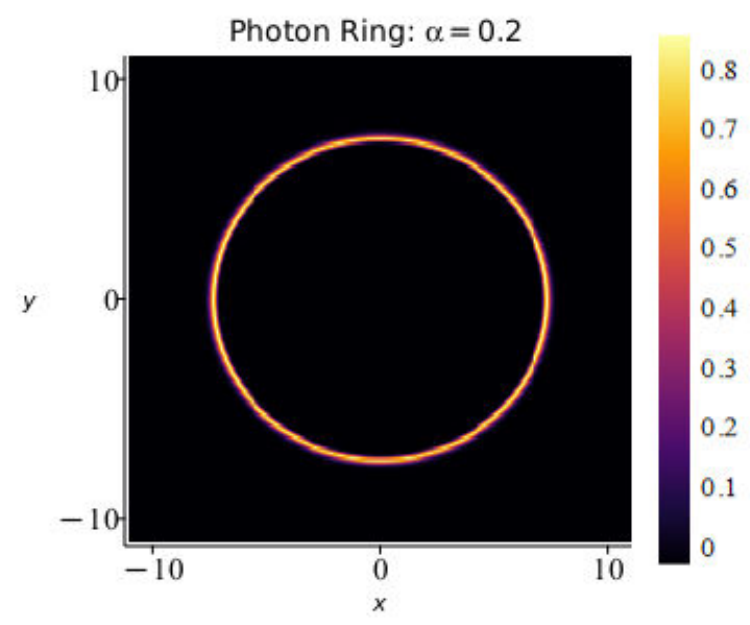}
\caption{}
\end{subfigure}
\caption{Decomposition of the observed emission into direct, lensed, and PR contributions for varying CoS parameter $\alpha$ at fixed $a = 1$. Top row: $\alpha = 0$ (Schwarzschild limit). Middle row: $\alpha = 0.1$. Bottom row: $\alpha = 0.2$. Left column: direct emission. Middle column: lensed emission. Right column: PR. The shadow size increases with $\alpha$ in all three components, reflecting the expansion of the PS due to the CoS.}
\label{fig:emission_components}
\end{figure*}

\subsection{Shadow Images: Parameter Space Exploration}

To provide a broad view of the optical appearance across the parameter space, we present in Figure~\ref{fig:shadow_grid} the total intensity maps for nine combinations of ($\alpha$, $a$). The rows correspond to $\alpha = 0$, $0.1$, and $0.2$, while the columns correspond to $a = 0.5$, $0.7$, and $1.0$. The total intensity includes all three emission components (direct, lensed, and PR) combined.

Several trends are evident in this parameter space exploration. Moving from left to right across each row (increasing $a$), the shadow size increases, confirming that weaker electromagnetic corrections result in larger shadows, consistent with the analytical predictions in Table~\ref{tab:vacuum_shadow}. Moving from top to bottom in each column (increasing $\alpha$), the shadow size also increases, demonstrating that the CoS parameter has a pronounced effect on the shadow expansion, with $\alpha = 0.2$ producing noticeably larger shadows than $\alpha = 0$.

The combined effects of both parameters are clearly visible in Figure~\ref{fig:shadow_grid}. The largest shadow appears at $(\alpha = 0.2, a = 1.0)$ in the bottom-right panel, while the smallest shadow corresponds to $(\alpha = 0, a = 0.5)$ in the top-left panel. The ratio of shadow radii between these extreme cases is approximately $R_{sh}(\alpha=0.2, a=1.0)/R_{sh}(\alpha=0, a=0.5) \approx 6.50/4.40 \approx 1.48$, representing a 48\% increase in apparent size. The bright PR at the shadow boundary remains visible across all parameter combinations, though its radius varies with the parameters. The ring brightness relative to the background emission is relatively insensitive to the parameter values, suggesting that the PR contrast is a feature of the Letelier BH in EMU.

\begin{figure*}[ht!]
\centering
\begin{subfigure}[b]{0.32\textwidth}
\centering
\includegraphics[width=\textwidth]{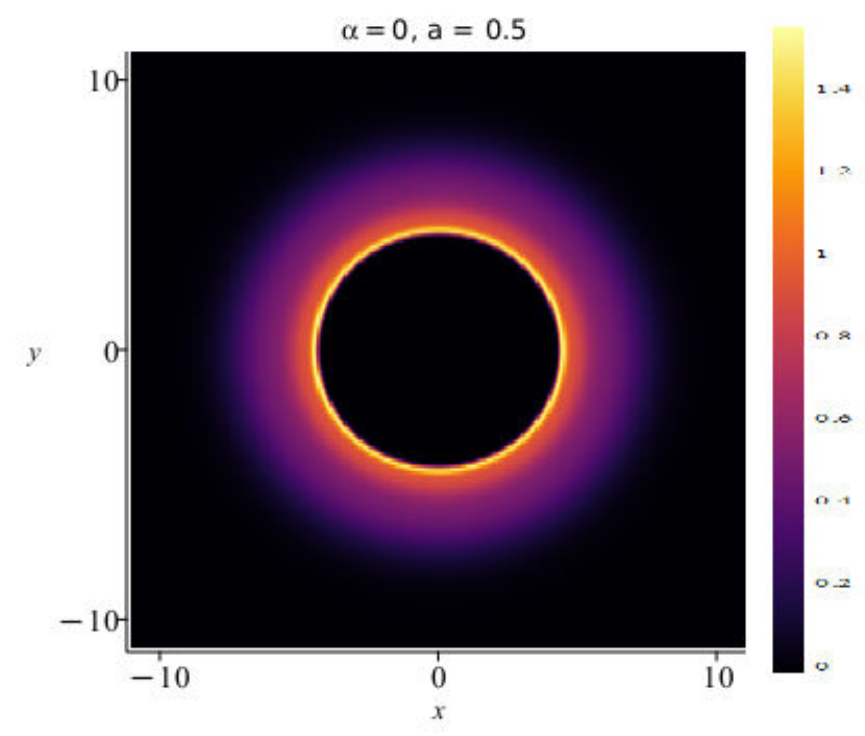}
\caption{}
\end{subfigure}
\hfill
\begin{subfigure}[b]{0.32\textwidth}
\centering
\includegraphics[width=\textwidth]{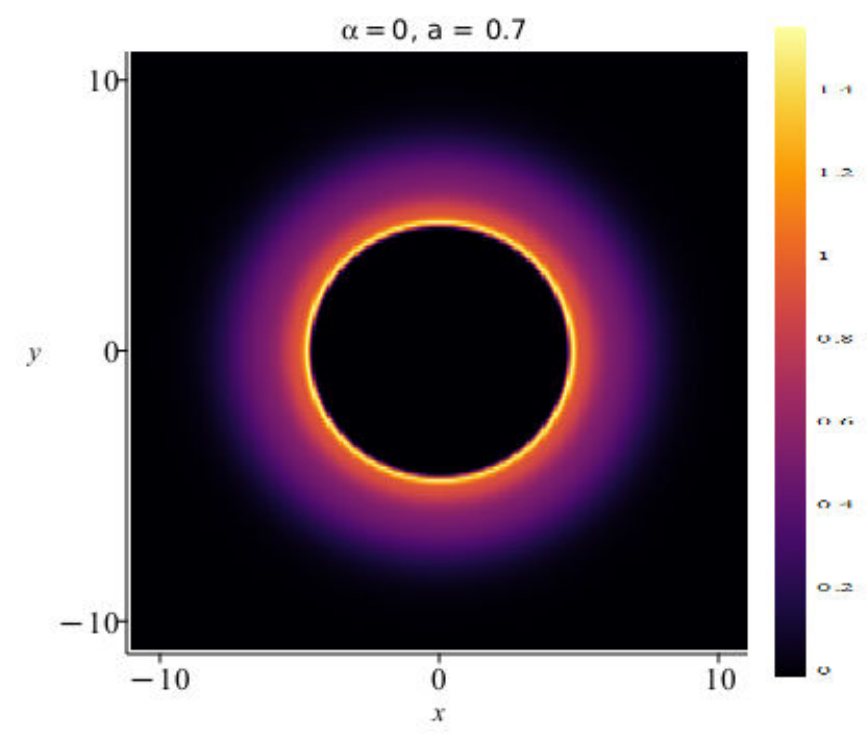}
\caption{}
\end{subfigure}
\hfill
\begin{subfigure}[b]{0.32\textwidth}
\centering
\includegraphics[width=\textwidth]{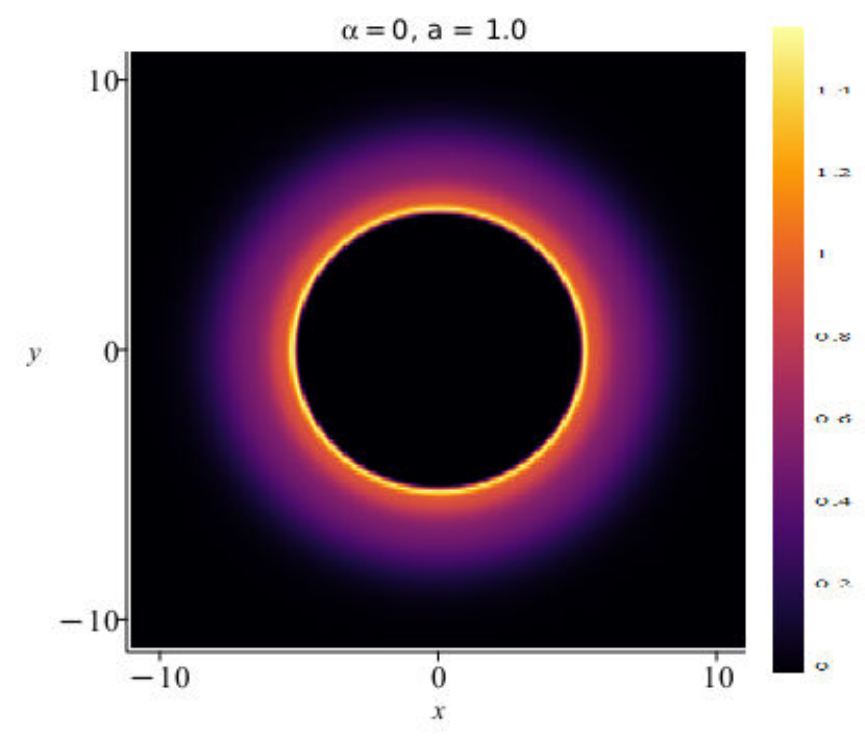}
\caption{}
\end{subfigure}
\\[4pt]
\begin{subfigure}[b]{0.32\textwidth}
\centering
\includegraphics[width=\textwidth]{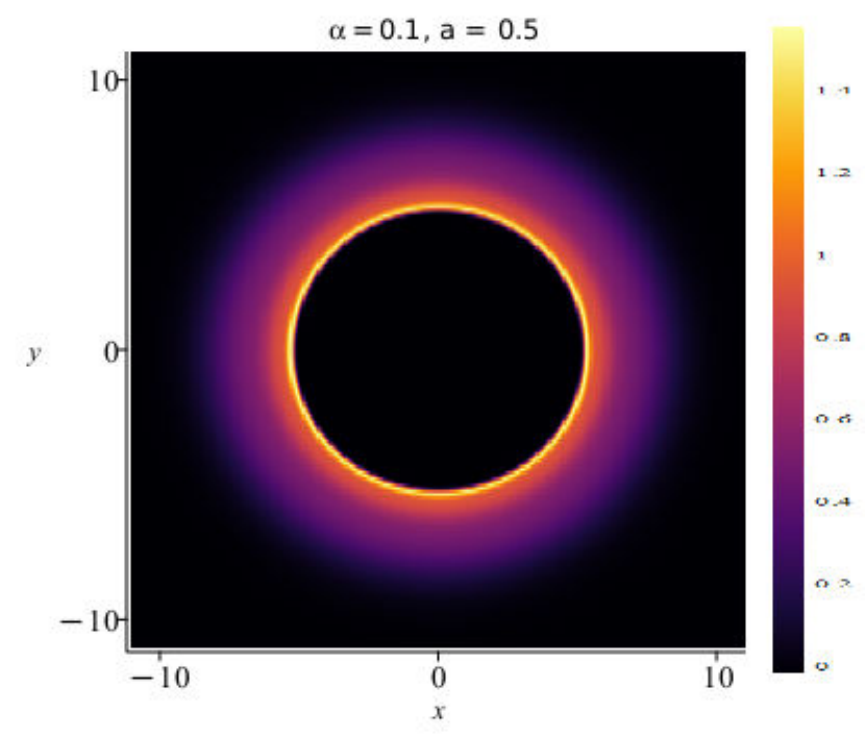}
\caption{}
\end{subfigure}
\hfill
\begin{subfigure}[b]{0.32\textwidth}
\centering
\includegraphics[width=\textwidth]{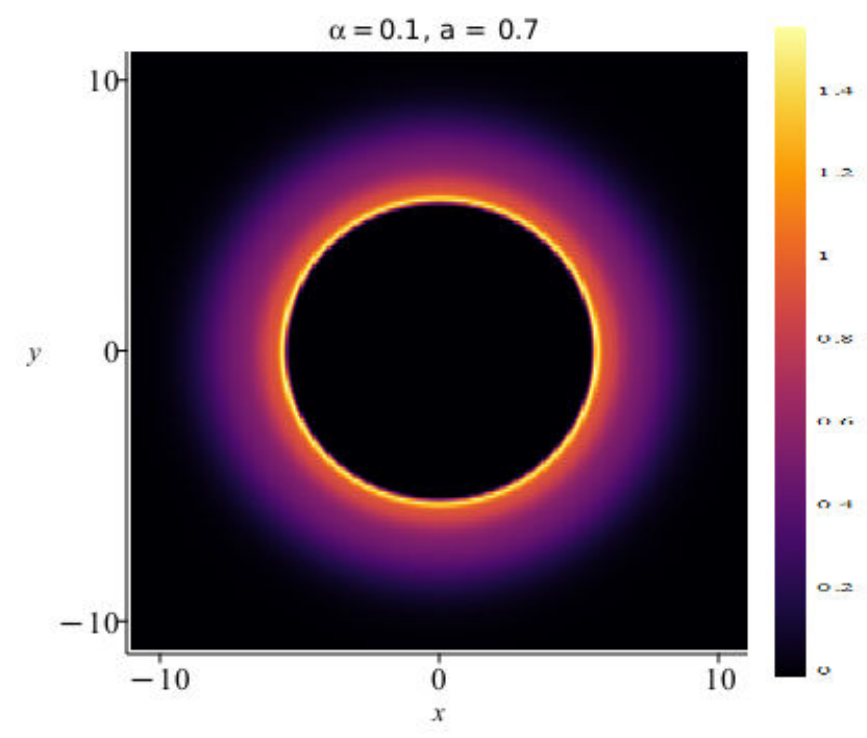}
\caption{}
\end{subfigure}
\hfill
\begin{subfigure}[b]{0.32\textwidth}
\centering
\includegraphics[width=\textwidth]{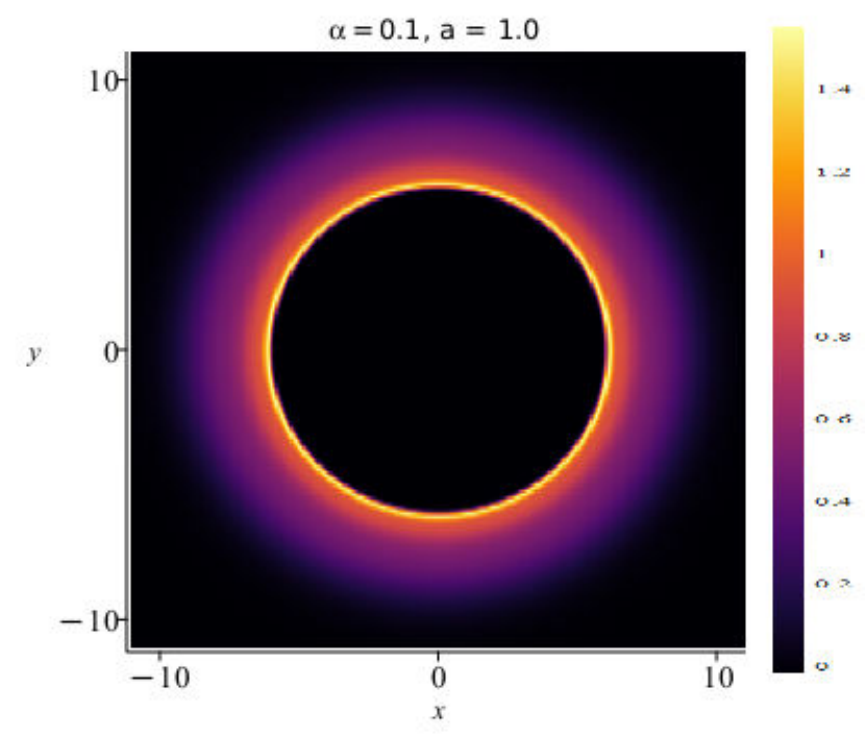}
\caption{}
\end{subfigure}
\\[4pt]
\begin{subfigure}[b]{0.32\textwidth}
\centering
\includegraphics[width=\textwidth]{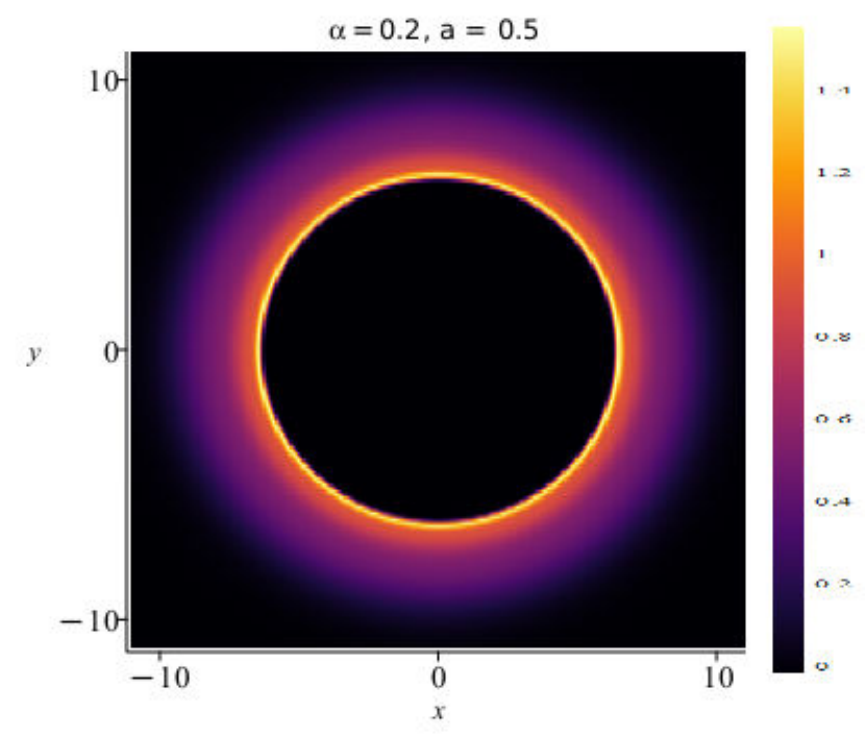}
\caption{}
\end{subfigure}
\hfill
\begin{subfigure}[b]{0.32\textwidth}
\centering
\includegraphics[width=\textwidth]{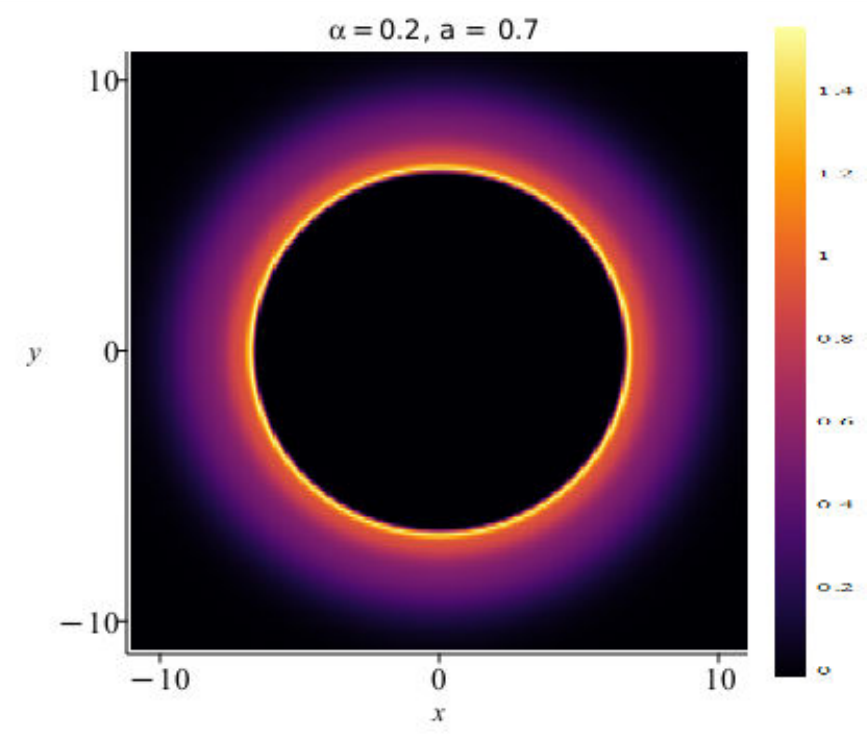}
\caption{}
\end{subfigure}
\hfill
\begin{subfigure}[b]{0.32\textwidth}
\centering
\includegraphics[width=\textwidth]{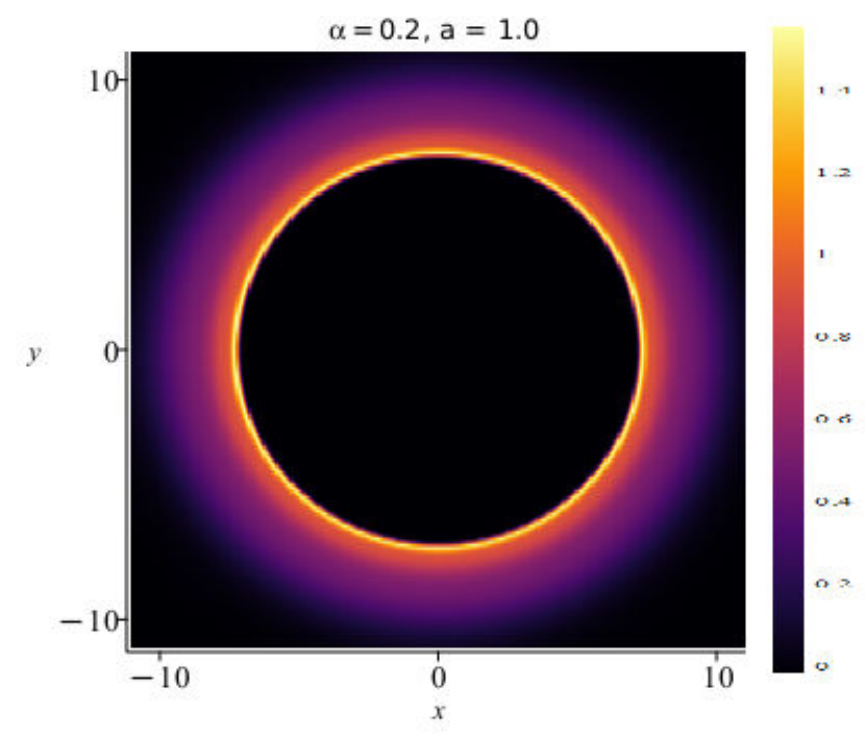}
\caption{}
\end{subfigure}
\caption{Shadow images of the Letelier BH in EMU across the parameter space. Top row: $\alpha = 0$. Middle row: $\alpha = 0.1$. Bottom row: $\alpha = 0.2$. Left column: $a = 0.5$. Middle column: $a = 0.7$. Right column: $a = 1.0$. The total intensity includes direct emission, lensed emission, and PR contributions. The shadow size increases with both $\alpha$ and $a$, with the largest shadow at $(\alpha = 0.2, a = 1.0)$ and the smallest at $(\alpha = 0, a = 0.5)$.}
\label{fig:shadow_grid}
\end{figure*}

\subsection{Shadow in Homogeneous Plasma}

We now consider the BH immersed in a homogeneous plasma with constant plasma frequency $\omega_p$ \cite{isz38,sec3is12}. The refractive index of the plasma is given by
\begin{equation}
n^2(r) = 1 - \frac{\omega_p^2}{\omega^2},
\label{eq:refr_index}
\end{equation}
where $\omega$ is the photon frequency measured by a static observer. In the presence of plasma, the photon trajectories are modified due to the dispersive nature of the medium \cite{sec3is13}. The modified effective potential for photon motion becomes
\begin{equation}
V_{\text{eff}}^{(\text{pl})} = f(r)\left(\frac{L^2}{r^2} + \omega_p^2\right) - E^2,
\label{eq:Veff_plasma}
\end{equation}
where $E$ is the conserved photon energy. Introducing the dimensionless plasma parameter $k = \omega_p^2/\omega_\infty^2$, where $\omega_\infty$ is the photon frequency at infinity, the PS condition becomes
\begin{equation}
(1+k) r f'(r_{ps}) - 2 f(r_{ps}) = 0.
\label{eq:PS_plasma}
\end{equation}
This modified condition shows that the plasma effectively increases the ``centrifugal'' contribution, shifting the PS outward for $k > 0$.

The corresponding shadow radius in the plasma environment is
\begin{equation}
R_{sh}^{(\text{pl})} = r_{ps} \sqrt{\frac{(1-\alpha)(1 - k f(r_{ps}))}{f(r_{ps})}}.
\label{eq:Rsh_plasma}
\end{equation}
The factor $(1 - k f(r_{ps}))$ under the square root arises from the plasma dispersion relation and is responsible for the shadow size reduction in the presence of plasma.

Table~\ref{tab:homo_plasma} presents the shadow radius for various plasma concentrations at fixed $\alpha = 0.1$. The plasma effect reduces the shadow size, with stronger reduction at higher plasma density. At $k = 0.4$, the shadow radius decreases by approximately 5\% compared to the vacuum case. This reduction occurs because the effective refractive index $n < 1$ in the plasma causes photons to bend less strongly than in vacuum, allowing more photons to escape rather than being captured \cite{sec3is14}.

\begin{table}[ht!]
\centering
\setlength{\tabcolsep}{6pt}
\renewcommand{\arraystretch}{1.5}
\begin{tabular}{c|cccc}
\hline\hline
\rowcolor{orange!50}
$k$ & $a=0.5$ & $a=0.7$ & $a=0.9$ & $a=1.0$ \\
\hline
0.00 & 5.0025 & 5.2829 & 5.6031 & 5.7735 \\
0.10 & 4.9391 & 5.2116 & 5.5233 & 5.6892 \\
0.20 & 4.8791 & 5.1431 & 5.4456 & 5.6068 \\
0.30 & 4.8191 & 5.0738 & 5.3663 & 5.5225 \\
0.40 & 4.7560 & 5.0008 & 5.2827 & 5.4334 \\
\hline\hline
\end{tabular}
\caption{Shadow radius $R_{sh}^{(\text{pl})}/M$ in homogeneous plasma for $\alpha = 0.1$. Here $k = \omega_p^2/\omega_\infty^2$ is the plasma parameter.}
\label{tab:homo_plasma}
\end{table}

\subsection{Shadow in Inhomogeneous Plasma}

For a more realistic astrophysical scenario, we consider an inhomogeneous plasma with a power-law density profile \cite{sec3is15,sec3is16}
\begin{equation}
\omega_p^2(r) = \frac{\omega_c^2}{r^q},
\label{eq:inhomo_plasma}
\end{equation}
where $\omega_c$ is a characteristic plasma frequency and $q$ determines the radial falloff. The case $q = 1$ corresponds to spherical accretion onto the BH, which is relevant for low-luminosity AGN and the Galactic Center environment surrounding Sgr~A* \cite{sec3is17}.

Defining the inhomogeneous plasma parameter $k_c = \omega_c^2/\omega_\infty^2$, the modified PS condition for $q = 1$ reads
\begin{equation}
(r + k_c) f'(r_{ps}) - 2 f(r_{ps}) = 0.
\label{eq:PS_inhomo}
\end{equation}
The refractive index at the PS becomes
\begin{equation}
n^2(r_{ps}) = 1 - \frac{k_c f(r_{ps})}{r_{ps}},
\label{eq:n_inhomo}
\end{equation}
and the shadow radius is
\begin{equation}
R_{sh}^{(\text{pl})} = r_{ps} \sqrt{\frac{(1-\alpha) n^2(r_{ps})}{f(r_{ps})}}.
\label{eq:Rsh_inhomo}
\end{equation}

The results for inhomogeneous plasma are presented in Table~\ref{tab:inhomo_plasma}. Comparing with Table~\ref{tab:homo_plasma}, we observe that the inhomogeneous plasma has a weaker effect on the shadow size than the homogeneous case at comparable parameter values. For example, at $k_c = 0.20$, the shadow radius decreases by only about 1\% compared to vacuum, whereas the homogeneous plasma with $k = 0.20$ produces a 2.5\% reduction. This difference is expected since the plasma density $\omega_p^2(r) \propto 1/r$ falls off with distance, concentrating the plasma influence in the immediate vicinity of the BH where its effect on photon trajectories is less cumulative.

\begin{table}[ht!]
\centering
\setlength{\tabcolsep}{6pt}
\renewcommand{\arraystretch}{1.5}
\begin{tabular}{c|cccc}
\hline\hline
\rowcolor{orange!50}
$k_c$ & $a=0.5$ & $a=0.7$ & $a=0.9$ & $a=1.0$ \\
\hline
0.00 & 5.0025 & 5.2829 & 5.6031 & 5.7735 \\
0.05 & 4.9904 & 5.2705 & 5.5904 & 5.7607 \\
0.10 & 4.9788 & 5.2584 & 5.5779 & 5.7480 \\
0.15 & 4.9676 & 5.2467 & 5.5657 & 5.7356 \\
0.20 & 4.9566 & 5.2352 & 5.5538 & 5.7235 \\
\hline\hline
\end{tabular}
\caption{Shadow radius $R_{sh}^{(\text{pl})}/M$ in inhomogeneous plasma ($q=1$) for $\alpha = 0.1$. Here $k_c = \omega_c^2/\omega_\infty^2$ is the inhomogeneous plasma parameter.}
\label{tab:inhomo_plasma}
\end{table}

\subsection{Photon Trajectories}

To visualize the light deflection near the Letelier BH in EMU, we numerically integrate the photon orbit equation \cite{sec3is18}. For a photon with impact parameter $b$, the orbit in the $(r, \phi)$ plane satisfies
\begin{equation}
\left(\frac{dr}{d\phi}\right)^2 = r^4 \left(\frac{1}{b^2} - \frac{f(r)}{r^2}\right).
\label{eq:orbit_eq}
\end{equation}
The critical impact parameter $b_{cr} = r_{ps}/\sqrt{f(r_{ps})}$ separates captured trajectories ($b < b_{cr}$) from scattered ones ($b > b_{cr}$). For photons with $b = b_{cr}$, the trajectory asymptotically approaches the PS, where the photon can orbit indefinitely in an unstable circular orbit.

Figure~\ref{fig:photon_traj} displays representative photon trajectories for different impact parameters at $\alpha = 0.1$ and $a = 0.5$. The trajectories are labeled by the ratio $b/b_{cr}$, allowing direct comparison of the deflection behavior relative to the critical case. Photons with $b/b_{cr} < 1$ are captured by the BH, spiraling into the EH after multiple orbits near the PS. The number of orbits before capture increases as $b$ approaches $b_{cr}$ from below.

Photons with $b/b_{cr} > 1$ are scattered to infinity after experiencing gravitational deflection. The deflection angle decreases as $b/b_{cr}$ increases, ranging from near-infinite deflection at the critical value to weak-field bending for $b/b_{cr} \gg 1$. The critical trajectory ($b/b_{cr} = 1$) represents the boundary between capture and scattering, corresponding to photons that asymptotically approach the PS.

The golden ring in Figure~\ref{fig:photon_traj} indicates the PS at $r_{ps} = 2.72M$, while the black disk represents the EH at $r_+ = 1.74M$. The four-fold symmetry of the trajectory pattern reflects the spherical symmetry of the spacetime, with incoming photons from all directions experiencing equivalent gravitational deflection.

\begin{figure*}[ht!]
\centering
\includegraphics[width=0.75\textwidth]{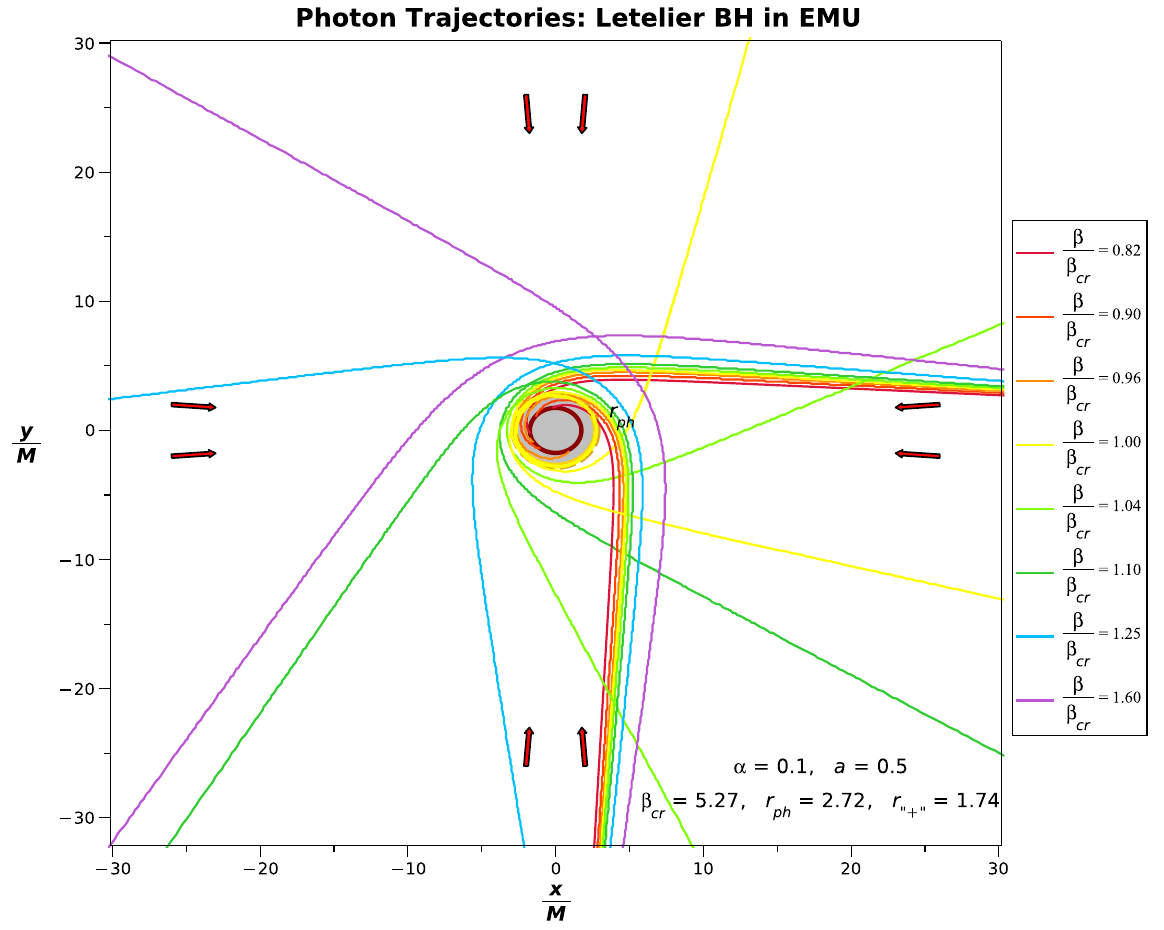}
\caption{Photon trajectories around the Letelier BH in EMU for $\alpha = 0.1$ and $a = 0.5$. The trajectories are labeled by the ratio $b/b_{cr}$, where $b_{cr} = 5.27M$ is the critical impact parameter. The golden ring indicates the PS at $r_{ps} = 2.72M$, and the black disk represents the EH at $r_+ = 1.74M$. Captured trajectories ($b < b_{cr}$) spiral into the BH, while deflected trajectories ($b > b_{cr}$) escape to infinity.}
\label{fig:photon_traj}
\end{figure*}

\subsection{Observational Implications and EHT Constraints}

The results presented in this section have several implications for current and future BH observations. The EHT observations of M87* and Sgr~A* have provided measurements of the shadow size with approximately 10\% precision \cite{isz04,isz05}. The variation of $R_{sh}$ with $\alpha$ and $a$ shown in Table~\ref{tab:vacuum_shadow} suggests that constraints on these parameters could be derived from shadow size measurements. For example, a shadow 20\% larger than Schwarzschild would be consistent with $\alpha \approx 0.15$ at $a = 1$.

Future space-based VLBI observations with improved resolution may directly resolve the PR structure \cite{isz45,sec3is19}. The decomposition shown in Figure~\ref{fig:emission_components} demonstrates that the PR is a remarkable feature across the parameter space, though its radius varies with $\alpha$ and $a$. For observations at lower frequencies where plasma effects become significant, Tables~\ref{tab:homo_plasma} and \ref{tab:inhomo_plasma} provide predictions for the shadow size modification. The relatively weak dependence on plasma parameters suggests that plasma corrections are subdominant to the geometric effects of $\alpha$ and $a$ for typical astrophysical environments.

The similar effects of $\alpha$ and $a$ on the shadow size may lead to degeneracies in parameter estimation from shadow observations alone. Breaking these degeneracies may require complementary information from PR subrings, GW observations, or stellar dynamics in the vicinity of the BH \cite{sec3is20,sec3is21}. Additionally, the ngEHT and proposed space missions offer the prospect of measuring the PR with sufficient precision to distinguish between different BH models \cite{sec3is22}. The detailed predictions provided in this section, particularly the emission decomposition and parameter space exploration, will be valuable for interpreting such future observations in the context of the Letelier BH in EMU.

\section{HR and Energy Emission}\label{isec4}

In this section, we investigate the HR properties of the Letelier BH in EMU. We derive the Hawking temperature and analyze the energy emission rate, which depends on both the CoS parameter $\alpha$ and the EMU parameter $a$. The radiation spectra for different spin fields, including scalar ($s=0$), EM ($s=1$), and Dirac ($s=1/2$) fields, are computed and compared with the Schwarzschild case.

\subsection{Hawking Temperature}

The Hawking temperature for the Letelier BH in EMU was derived in Sec.~\ref{isec2} from the surface gravity at the outer horizon. Recalling Eq.~(\ref{eq:Hawking_T_sec2}), we have
\begin{equation}
T_H = \frac{(1-\alpha)\sqrt{1-(1-\alpha)(1-a^2)}}{4\pi M},
\label{eq:TH_sec4}
\end{equation}
which reduces to the Schwarzschild value $T_H^{\text{Sch}} = 1/(8\pi M)$ in the limit $\alpha = 0$, $a = 1$.

Table~\ref{tab:Hawking_T} presents the normalized Hawking temperature $T_H/T_H^{\text{Sch}}$ for various combinations of $\alpha$ and $a$. The temperature decreases monotonically with increasing $\alpha$, reflecting the gravitational dilution caused by the CoS. At $\alpha = 0.3$, the temperature is reduced to approximately half of the Schwarzschild value. The dependence on the EMU parameter $a$ is weaker but still significant, with smaller values of $a$ (stronger EM corrections) leading to slightly lower temperatures. The physical interpretation is that the CoS effectively increases the horizon radius for a given mass, thereby reducing the surface gravity and hence the temperature \cite{sec4is01,sec4is02}.

\begin{table}[ht!]
\centering
\setlength{\tabcolsep}{6pt}
\renewcommand{\arraystretch}{1.5}
\begin{tabular}{c|cccc}
\hline\hline
\rowcolor{orange!50}
$\alpha$ & $a=0.5$ & $a=0.7$ & $a=0.9$ & $a=1.0$ \\
\hline
0.00 & 0.8889 & 0.9689 & 0.9972 & 1.0000 \\
0.05 & 0.8202 & 0.8782 & 0.9003 & 0.9025 \\
0.10 & 0.7493 & 0.7912 & 0.8082 & 0.8100 \\
0.15 & 0.6779 & 0.7081 & 0.7211 & 0.7225 \\
0.20 & 0.6076 & 0.6291 & 0.6389 & 0.6400 \\
0.25 & 0.5391 & 0.5544 & 0.5617 & 0.5625 \\
0.30 & 0.4734 & 0.4841 & 0.4894 & 0.4900 \\
\hline\hline
\end{tabular}
\caption{Normalized Hawking temperature $T_H/T_H^{\text{Sch}}$ for the Letelier BH in EMU, where $T_H^{\text{Sch}} = 1/(8\pi M)$ is the Schwarzschild temperature.}
\label{tab:Hawking_T}
\end{table}

\subsection{Total Emission Power}

The total power emitted by the BH is obtained by integrating the energy emission rate over all frequencies \cite{sec4is07}. In the Stefan-Boltzmann approximation, the total power scales as
\begin{equation}
P = \sigma_{\rm SB} A_{\rm eff} T_H^4,
\label{eq:total_power}
\end{equation}
where $\sigma_{\rm SB}$ is the Stefan-Boltzmann constant and $A_{\rm eff} = 27\pi R_{sh}^2$ is the effective emission area in the geometric optics limit \cite{sec4is08}. This expression shows that the total power depends on both the shadow radius and the fourth power of the temperature, making it highly sensitive to the BH parameters.

Figure~\ref{fig:total_power} presents the normalized total emission power $P/P_{\rm Sch}$ as a function of the model parameters. Panel (a) shows the power versus $\alpha$ for different values of $a$. All curves decrease monotonically with increasing $\alpha$, reflecting the combined effect of reduced temperature and modified shadow radius. The suppression is dramatic: at $\alpha = 0.3$, the power is reduced to approximately 10--15\% of the Schwarzschild value regardless of $a$. The curves for different $a$ values are well separated at small $\alpha$ but converge at large $\alpha$, indicating that the CoS effect dominates over the EMU correction in the strong-$\alpha$ regime.

Panel (b) displays the power versus $a$ for different values of $\alpha$. All curves increase with increasing $a$, confirming that weaker EMU corrections lead to higher emission rates. The $\alpha = 0$ curve reaches unity at $a = 1$, corresponding to the Schwarzschild limit. For $\alpha = 0.3$, the power remains below 0.15 across the entire range of $a$, demonstrating the strong suppression effect of the CoS.

\begin{figure*}[ht!]
\centering
\begin{subfigure}[b]{0.48\textwidth}
\centering
\includegraphics[width=\textwidth]{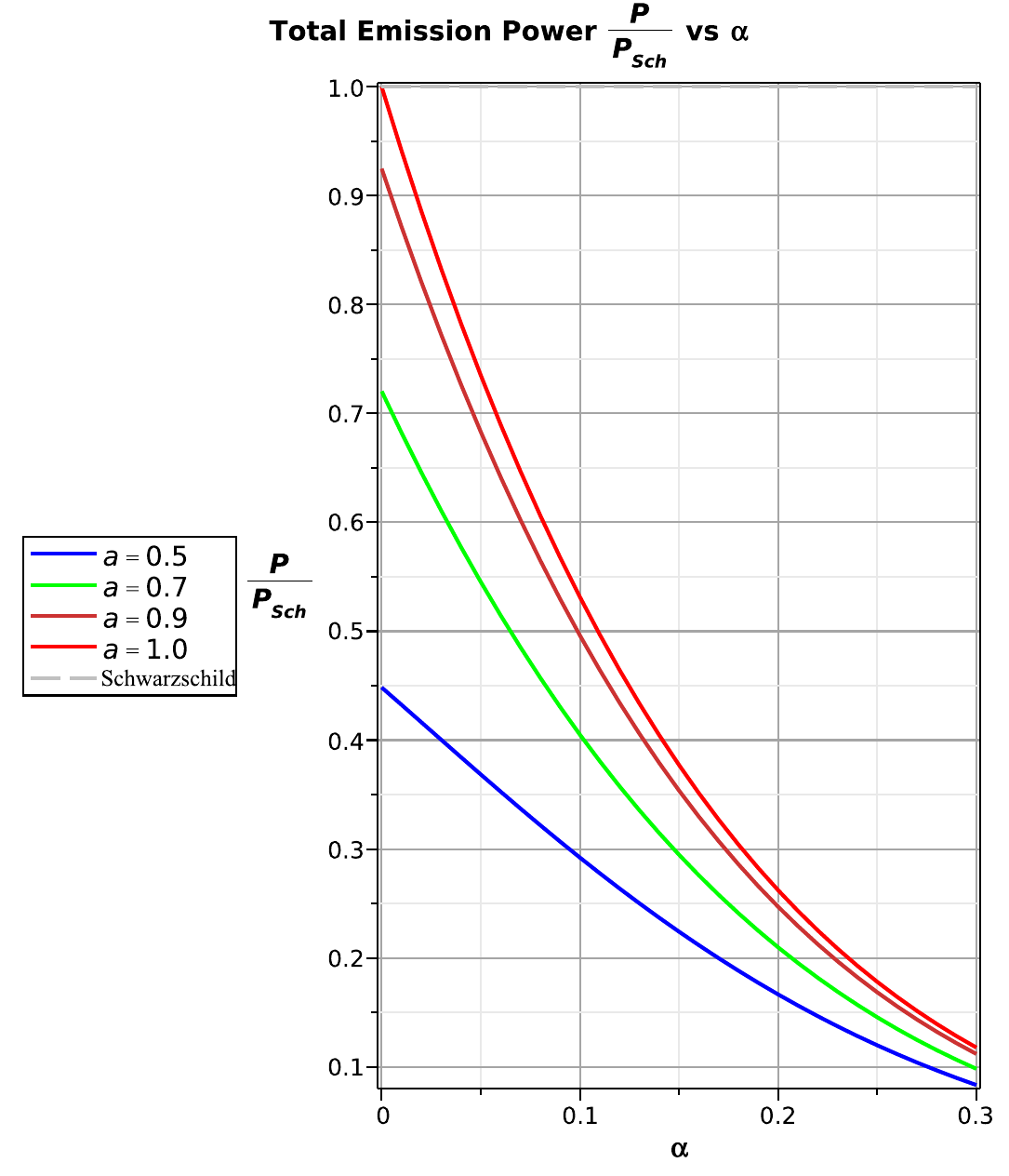}
\caption{}
\label{fig:power_alpha}
\end{subfigure}
\hfill
\begin{subfigure}[b]{0.48\textwidth}
\centering
\includegraphics[width=\textwidth]{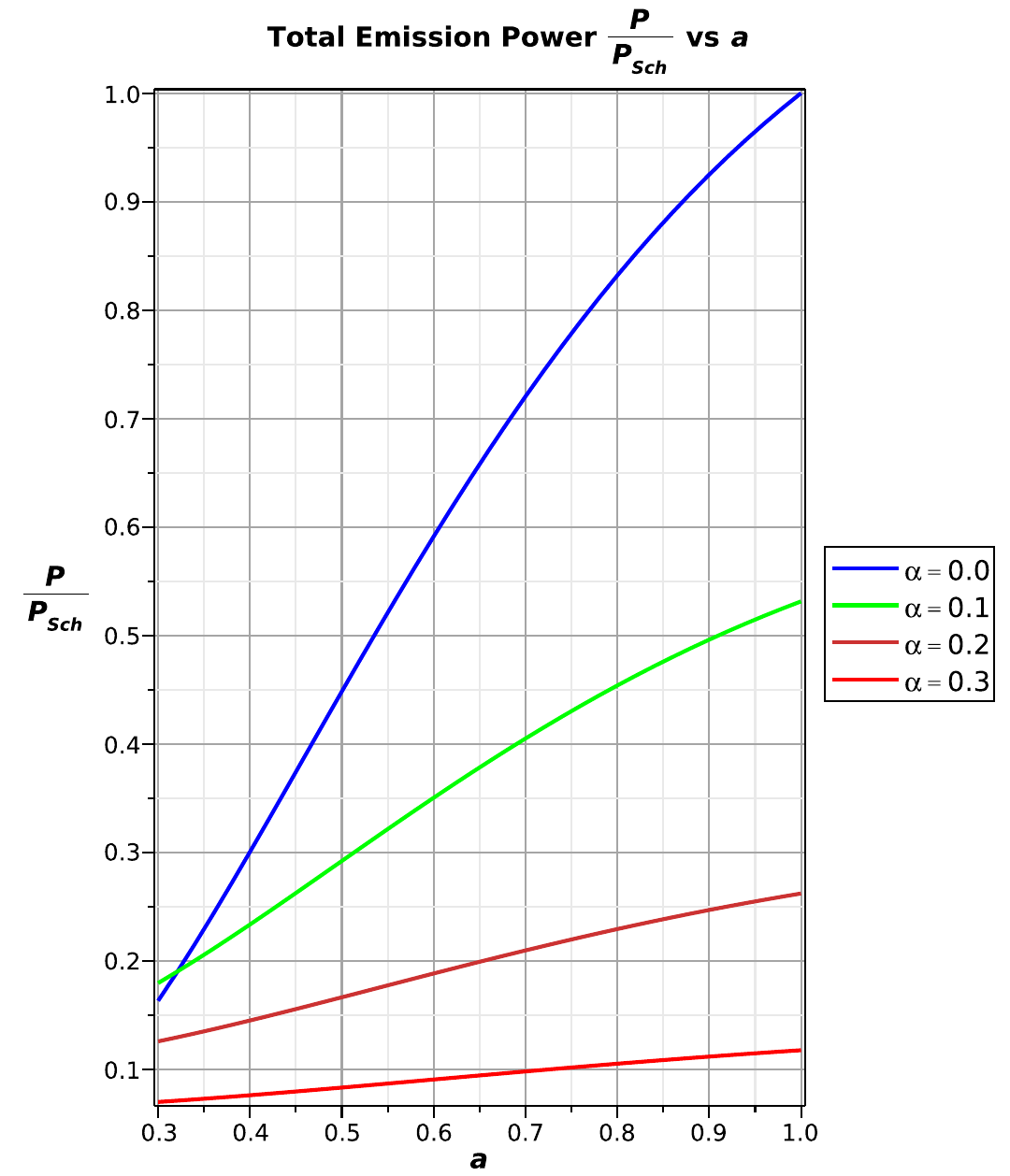}
\caption{}
\label{fig:power_a}
\end{subfigure}
\caption{Normalized total emission power $P/P_{\rm Sch}$ for the Letelier BH in EMU. (a) Power versus CoS parameter $\alpha$ for different values of $a$. The gray dashed line indicates the Schwarzschild reference $P_{\rm Sch}$. (b) Power versus EMU parameter $a$ for different values of $\alpha$. The power decreases with increasing $\alpha$ and increases with increasing $a$.}
\label{fig:total_power}
\end{figure*}

\subsection{HR Spectra for Different Spin Fields}

The HR spectrum depends on the spin of the emitted particles through both the statistical distribution and the greybody factors \cite{sec4is09,isz40}. The particle emission rate for a field of spin $s$ is given by
\begin{equation}
\frac{d^2N_s}{d\omega\,dt} = \frac{1}{2\pi}\sum_{\ell}\frac{(2\ell+1)\mathcal{T}_s(\omega,\ell)}{\exp(\omega/T_H) - (-1)^{2s}},
\label{eq:Hawking_spectrum}
\end{equation}
where $\mathcal{T}_s(\omega,\ell)$ is the greybody factor for mode $\ell$ and the sign in the denominator distinguishes bosons ($-$) from fermions ($+$) \cite{sec4is11}. The energy emission rate is $d^2E_s/d\omega\,dt = \omega \cdot d^2N_s/d\omega\,dt$.

Figure~\ref{fig:Hawking_spectra} compares the energy spectra for scalar ($s=0$), EM ($s=1$), and Dirac ($s=1/2$) fields. The solid curves correspond to the Schwarzschild case ($\alpha=0$, $a=1$), while the dashed curves represent the Letelier BH in EMU ($\alpha=0.1$, $a=0.8$). For the Schwarzschild BH, the EM field exhibits the highest emission rate with a peak of approximately 0.022 at $\omega M \approx 0.09$, followed by the Dirac field with peak $\sim 0.013$ and the scalar field with peak $\sim 0.012$. This hierarchy reflects the different numbers of degrees of freedom: the EM field has 2 polarizations, while the Dirac field has 4 spinor components (weighted by 7/8 due to Fermi-Dirac statistics) \cite{sec4is12}. Comparing the Letelier-EMU case to Schwarzschild, the peak heights are reduced by approximately 39\%, 45\%, and 38\% for scalar, EM, and Dirac fields, respectively. In all cases, the peak shifts to lower frequencies, consistent with the reduced Hawking temperature. The EM field shows the largest relative suppression, indicating that the greybody factors for higher-spin fields are more sensitive to the spacetime geometry modifications \cite{isz43}.

\begin{figure*}[ht!]
\centering
\includegraphics[width=0.65\textwidth]{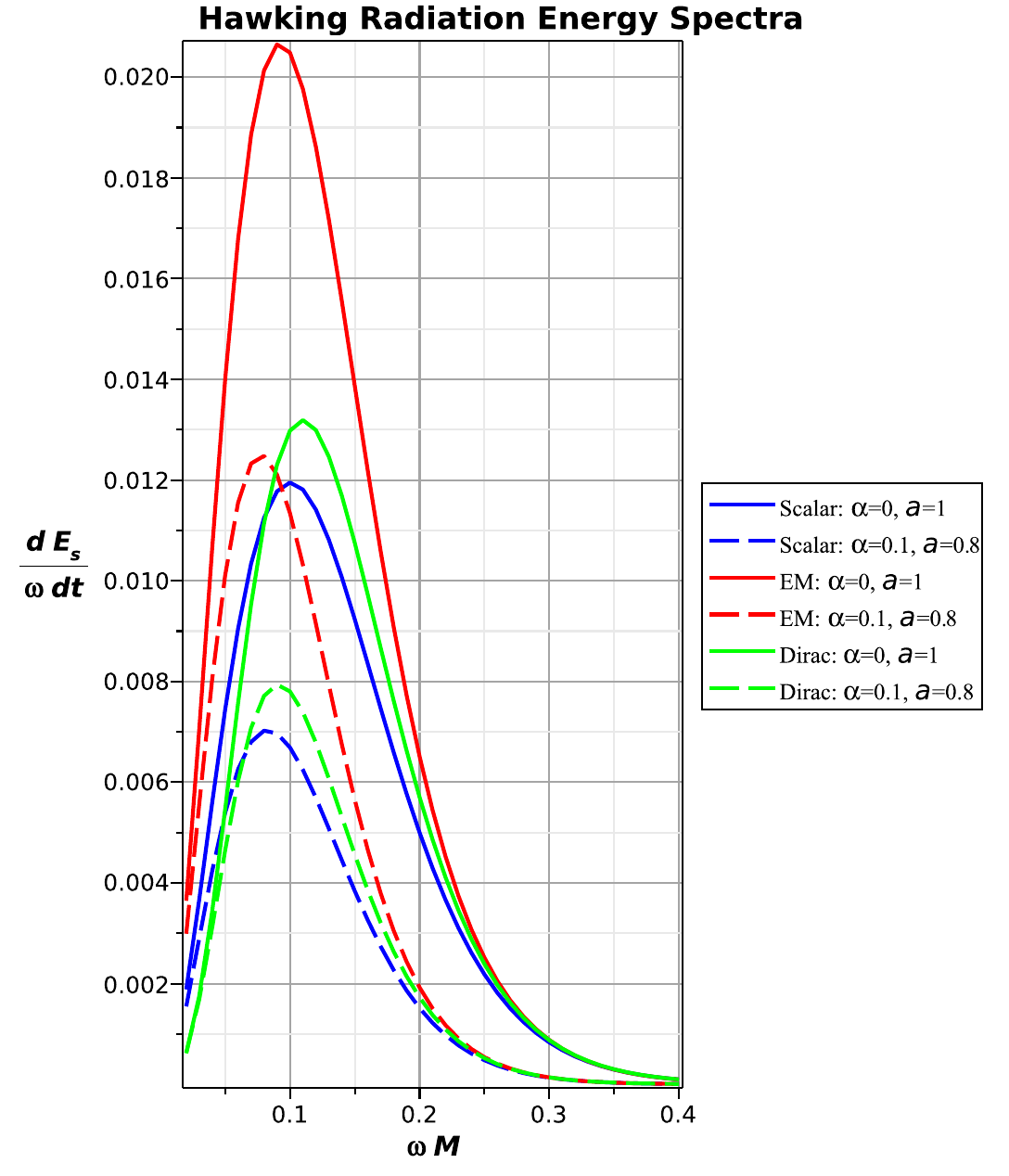}
\caption{Comparison of HR energy spectra for scalar (blue), EM (red), and Dirac (green) fields from the Letelier BH in EMU. Solid curves: Schwarzschild limit ($\alpha=0$, $a=1$). Dashed curves: Letelier BH in EMU ($\alpha=0.1$, $a=0.8$). The EM field exhibits the highest emission rate, followed by Dirac and scalar fields.}
\label{fig:Hawking_spectra}
\end{figure*}

The total power emitted in each spin channel is obtained by integrating the energy spectrum over all frequencies \cite{sec4is14}. In the Stefan-Boltzmann approximation, the power scales as $P_s \propto g_s R_{sh}^2 T_H^4$, where $g_s$ is the effective number of degrees of freedom: $g_{s=0} = 1$ for scalars, $g_{s=1} = 2$ for EM fields, and $g_{s=1/2} = 7/2$ for Dirac fields (accounting for 4 spinor components with the fermionic factor 7/8) \cite{sec4is15}.

Table~\ref{tab:total_power} presents the total power for each spin channel at two representative values of the EMU parameter. The ratio $P_{s=1}/P_{s=0} = 2$ is maintained across all parameter values, confirming the expected scaling with degrees of freedom. Similarly, $P_{s=1/2}/P_{s=0} \approx 3.5$ reflects the fermionic degrees of freedom. The total power decreases significantly with increasing $\alpha$: at $\alpha = 0.20$, the emission is reduced to approximately 25--35\% of the $\alpha = 0$ value for both $a = 0.5$ and $a = 1.0$. The EMU parameter also affects the emission, with $a = 1.0$ giving roughly twice the power of $a = 0.5$ at the same $\alpha$.

\begin{table*}[ht!]
\centering
\setlength{\tabcolsep}{8pt}
\renewcommand{\arraystretch}{1.5}
\begin{tabular}{c|ccc|ccc}
\hline\hline
\rowcolor{orange!50}
 & \multicolumn{3}{c|}{$a = 0.5$} & \multicolumn{3}{c}{$a = 1.0$} \\
\rowcolor{orange!50}
$\alpha$ & $P_{s=0}$ & $P_{s=1}$ & $P_{s=1/2}$ & $P_{s=0}$ & $P_{s=1}$ & $P_{s=1/2}$ \\
\hline
0.00 & 42.34 & 84.67 & 148.2 & 94.42 & 188.8 & 330.5 \\
0.05 & 34.79 & 69.58 & 121.8 & 69.41 & 138.8 & 242.9 \\
0.10 & 27.58 & 55.16 & 96.54 & 50.18 & 100.4 & 175.6 \\
0.15 & 21.16 & 42.31 & 74.05 & 35.61 & 71.22 & 124.6 \\
0.20 & 15.72 & 31.43 & 55.01 & 24.75 & 49.50 & 86.63 \\
\hline\hline
\end{tabular}
\caption{Total HR power (in units of $10^{-5}/M^2$) for scalar ($s=0$), EM ($s=1$), and Dirac ($s=1/2$) fields from the Letelier BH in EMU.}
\label{tab:total_power}
\end{table*}

The results presented in this section have important implications for BH evaporation in the presence of a CoS and EMU corrections \cite{sec4is16,sec4is17}. The reduced Hawking temperature implies a longer evaporation timescale, since the mass loss rate scales as $dM/dt \propto -T_H^4$ \cite{sec4is18}. For a Letelier BH in EMU with $\alpha = 0.2$, the evaporation rate is suppressed by a factor of $(0.64)^4 \approx 0.17$ compared to Schwarzschild, leading to an evaporation time approximately 6 times longer.

The modification of the HR spectrum also affects the particle content of the emitted radiation. The shift of the peak to lower frequencies means that the typical energy of emitted particles is reduced, which may have consequences for observational signatures of primordial BH evaporation \cite{sec4is19,sec4is20}. Furthermore, the relative contributions of different spin channels remain approximately constant across the parameter space, suggesting that the particle ratios in HR are special features determined primarily by the degrees of freedom rather than the specific BH geometry.

From an observational perspective, the suppression of HR by the CoS could explain why primordial BHs with masses in the asteroid range ($10^{15}$--$10^{18}$~g) might still survive to the present day if they were formed in an environment with significant string density \cite{sec4is21}. The EMU corrections provide an additional mechanism for modulating the evaporation rate, offering a rich phenomenology for constraining these parameters through observations of the diffuse gamma-ray background or direct searches for evaporating BHs \cite{sec4is22,sec4is23}. Future gamma-ray observatories such as the Cherenkov Telescope Array (CTA) may provide improved sensitivity to the HR signatures from primordial BHs, enabling constraints on the CoS and EMU parameters in the early universe \cite{sec4is24}.

\section{Phase Structure and JTE with Barrow Entropy}\label{isec5}

Barrow introduced a modification of BH entropy by suggesting that quantum-gravitational effects may deform the horizon geometry itself, leading to an effectively fractal surface rather than a smooth manifold \cite{barrow2020area,sec5is01}. In this approach, the horizon entropy is generalized to the Barrow entropy, expressed as
\begin{equation}
S_{B} = \left(\pi r_{+}^{2}\right)^{1+\frac{\Delta}{2}},
\label{eq:barrow_entropy}
\end{equation}
where the parameter $\Delta \in [0,1]$ characterizes the extent of geometric deformation \cite{isz29}. The classical Bekenstein--Hawking area law is recovered when $\Delta=0$, corresponding to an undeformed, smooth horizon, while $\Delta=1$ represents the extreme case in which the horizon exhibits maximal fractal irregularity. Unlike entropy corrections derived from statistical fluctuations or ensemble considerations \cite{sec5is03,sec5is04}, the Barrow entropy has a fundamentally geometric origin. As such, it is expected to encode certain aspects of quantum gravity related to microscopic spacetime structure, possibly associated with spacetime foam or loop quantum gravity-inspired effects at small length scales \cite{sec5is05,sec5is06}.

The Barrow entropy framework has attracted considerable attention in cosmological contexts, where it modifies the Friedmann equations and leads to observable deviations from standard $\Lambda$CDM predictions \cite{sec5is07,isz30}. For BH thermodynamics, the Barrow modification alters the entropy-area relation and consequently affects all derived thermodynamic quantities, including the heat capacity, free energy, and phase transition structure \cite{isz31}.

\subsection{Heat Capacity and Thermodynamic Stability}

To investigate the thermodynamic stability and phase behavior of the Letelier BH in EMU within this framework, we analyze the heat capacity constructed from the Barrow entropy. For fixed thermodynamic volume, the heat capacity is defined by \cite{WOS:001565141800002NPB,sec5is10}
\begin{equation}
C_{B} = T_{H}\left(\frac{\partial S_{B}}{\partial T_{H}}\right)_{V},
\label{eq:CB_def}
\end{equation}
which, upon substituting the explicit expressions for the Hawking temperature from Eq.~(\ref{eq:TH_sec4}) and the horizon radius from Eq.~(\ref{eq:horizons}), leads to
\begin{equation}
C_B = -\frac{\pi^{1+\frac{\Delta}{2}} r_+^{2} \left(2+\Delta \right) \left(r_+^{2}\right)^{\frac{\Delta}{2}} \left(M a^{2}-M+r_+\right)}{3 M a^{2}-3 M+2 r_+}.
\label{eq:CB_explicit}
\end{equation}
The qualitative behavior of the heat capacity plays a decisive role in identifying locally stable and unstable thermodynamic regions \cite{sec5is11}. In particular, divergences of $C_{B}$ signal the presence of second-order phase transitions, while changes in its sign distinguish between stable ($C_B > 0$) and unstable ($C_B < 0$) branches of the BH configuration \cite{sec5is12}.

The denominator of Eq.~(\ref{eq:CB_explicit}) vanishes when
\begin{equation}
r_+^{\rm crit} = \frac{3M(1-a^2)}{2},
\label{eq:r_crit}
\end{equation}
which defines the critical horizon radius at which the heat capacity diverges. This divergence marks a second-order phase transition between small and large BH branches. For $r_+ < r_+^{\rm crit}$, the heat capacity is negative, indicating thermodynamic instability, while for $r_+ > r_+^{\rm crit}$, the heat capacity becomes positive, corresponding to a locally stable configuration.

The Barrow parameter $\Delta$ modifies the magnitude of $C_B$ through the prefactor $(2+\Delta)$ and the power-law dependence $(r_+^2)^{\Delta/2}$, but does not alter the location of the phase transition point, which is determined solely by the metric parameters $M$ and $a$. This observation suggests that the Barrow deformation affects the thermodynamic response but preserves the underlying phase structure dictated by the spacetime geometry.

Figure~\ref{Cfig} displays the heat capacity $C_B$ as a function of the horizon radius for selected values of the Barrow parameter $\Delta$ at fixed $a = 0.2$ and $M = 1$. The divergence at $r_+^{\rm crit} \approx 1.44M$ is clearly visible, separating the unstable small-BH branch (left) from the stable large-BH branch (right). Increasing $\Delta$ amplifies the magnitude of $C_B$ in both regions while maintaining the transition point, consistent with the analytical expectation from Eq.~(\ref{eq:CB_explicit}).

\begin{figure*}[ht!]
\centering
\includegraphics[width=0.55\textwidth]{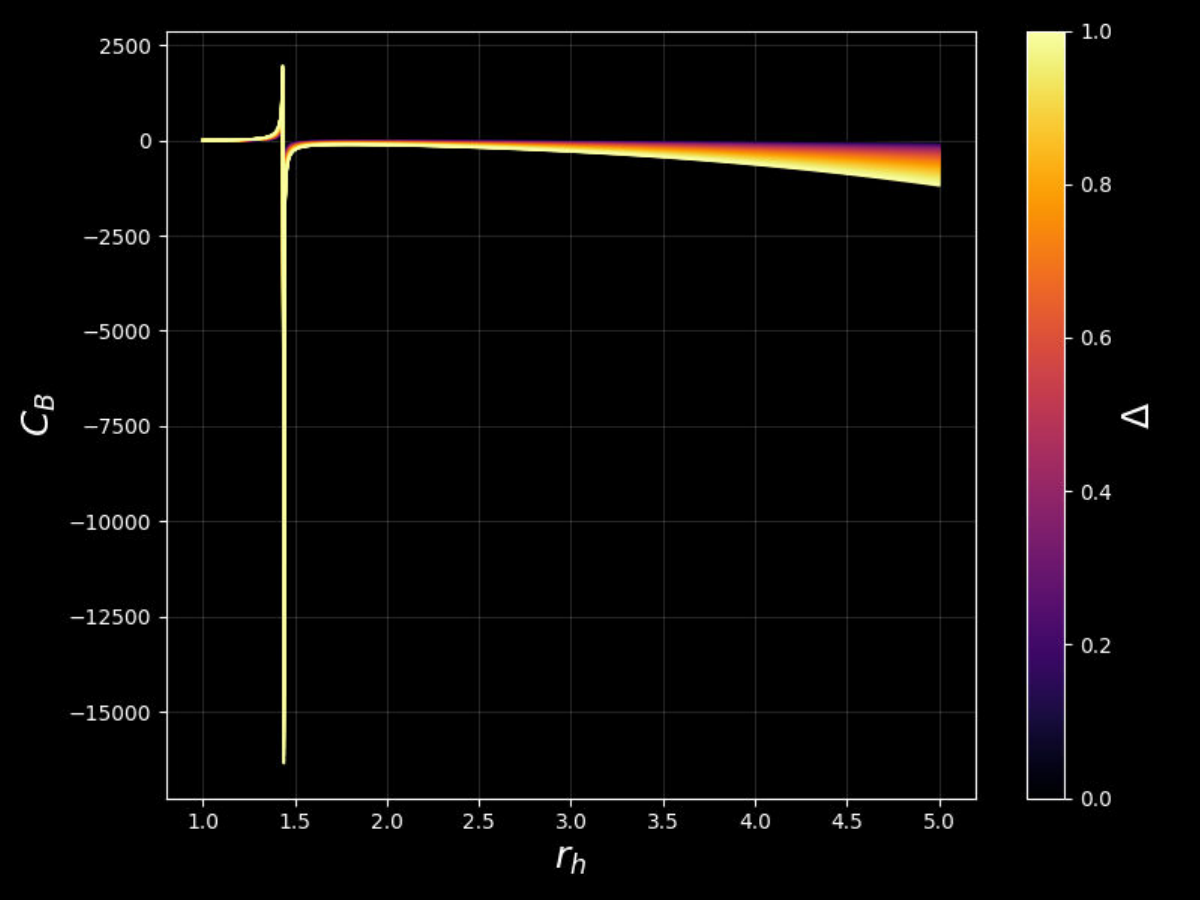}
\caption{Heat capacity $C_{B}$ as a function of the horizon radius $r_+/M$ for the Letelier BH in EMU with Barrow entropy. Parameters: $a = 0.2$, $M = 1$, and $\Delta \in \{0, 0.3, 0.6, 1.0\}$. The divergence at $r_+^{\rm crit} \approx 1.44M$ signals a second-order phase transition. Positive (negative) values indicate thermodynamically stable (unstable) configurations.}
\label{Cfig}
\end{figure*}

\subsection{JTE}

We further examine the JTE, which describes the isenthalpic evolution of the BH and provides insight into its heating and cooling properties during expansion processes \cite{sec5is13,sec5is14,sakalli2025zitterbewegung}. The JTE has emerged as a powerful tool for probing BH thermodynamics in extended phase space, where the cosmological constant is interpreted as thermodynamic pressure \cite{sec2is15,sec5is16}.

The Joule--Thomson coefficient is defined as \cite{aydiner2025regular}
\begin{equation}
\mu_{J} = \left(\frac{\partial T}{\partial P}\right)_{H},
\label{eq:muJ_def}
\end{equation}
where the subscript $H$ denotes constant enthalpy (mass). When evaluated in the Barrow entropy scenario for the Letelier BH in EMU, it takes the form
\begin{equation}
\mu_{J} = \frac{12 r_+ \left(M a^{2}-M+\frac{2}{3} r_+\right) \left(r_+^{2}\right)^{-\frac{\Delta}{2}} \pi^{-\frac{\Delta}{2}}}{\left(3 M a^{2}-3 M+2 r_+\right) \Delta -12 M a^{2}+12 M-6 r_+}.
\label{eq:muJ_explicit}
\end{equation}
The sign of $\mu_{J}$ determines the thermodynamic response of the BH under expansion: positive values ($\mu_J > 0$) correspond to cooling, whereas negative values ($\mu_J < 0$) indicate heating \cite{sec5is17}. The inversion temperature $T_i$, at which $\mu_J = 0$, separates the cooling and heating regimes and is obtained from the condition
\begin{equation}
M a^{2} - M + \frac{2}{3} r_+ = 0 \quad \Rightarrow \quad r_+^{\rm inv} = \frac{3M(1-a^2)}{2}.
\label{eq:r_inv}
\end{equation}
Interestingly, the inversion radius $r_+^{\rm inv}$ coincides with the critical radius $r_+^{\rm crit}$ from the heat capacity analysis, indicating a deep connection between the phase transition structure and the JTE behavior.

The presence of the Barrow deformation parameter $\Delta$ introduces substantial modifications to the JTE coefficient through the factors $(r_+^2)^{-\Delta/2}$ and $\pi^{-\Delta/2}$ in the numerator, as well as the $\Delta$-dependent term in the denominator. These modifications alter the magnitude of $\mu_J$ and can shift the positions of additional extrema, thereby affecting the cooling-heating transition dynamics and leading to thermodynamic behavior that deviates from the standard entropy formulation \cite{sec5is18}.

Figure~\ref{jtefig} presents the Joule--Thomson coefficient $\mu_J$ as a function of the horizon radius for selected values of $\Delta$ at fixed $a = 0.5$ and $M = 1$. The zero crossing at $r_+^{\rm inv} \approx 1.125M$ marks the inversion point separating the cooling region ($\mu_J > 0$, small $r_+$) from the heating region ($\mu_J < 0$, large $r_+$). The Barrow parameter modifies the slope and curvature of the $\mu_J$ profile, with larger $\Delta$ producing steeper transitions near the inversion point.

\begin{figure*}[ht!]
\centering
\includegraphics[width=0.55\textwidth]{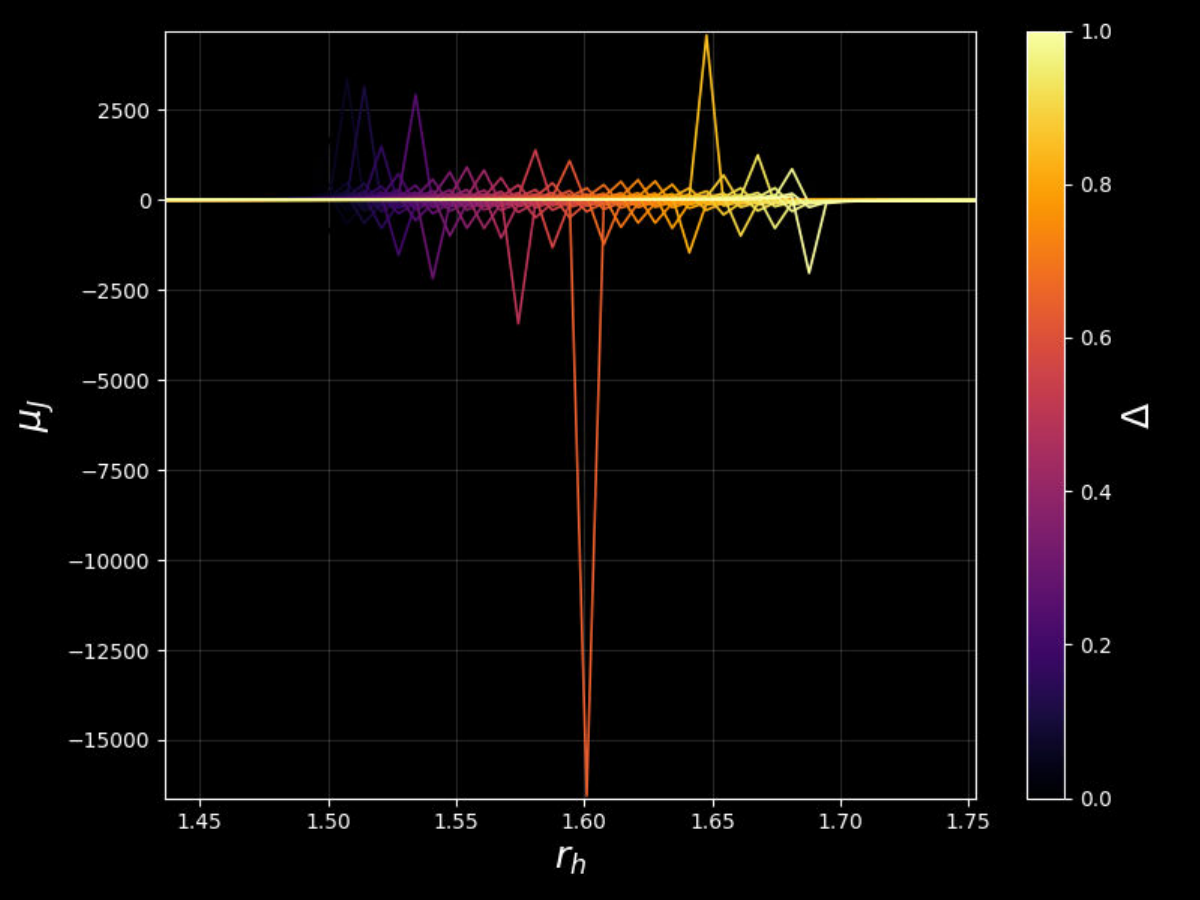}
\caption{Joule--Thomson coefficient $\mu_J$ as a function of the horizon radius $r_+/M$ for the Letelier BH in EMU with Barrow entropy. Parameters: $a = 0.5$, $M = 1$, and $\Delta \in \{0, 0.3, 0.6, 1.0\}$. The zero crossing defines the inversion point $r_+^{\rm inv}$, separating cooling ($\mu_J > 0$) from heating ($\mu_J < 0$) regimes.}
\label{jtefig}
\end{figure*}

The inversion curve in the $T$--$P$ plane is obtained by plotting the inversion temperature $T_i$ against the corresponding pressure $P_i$ as the horizon radius varies \cite{sec5is19}. For the Letelier BH in EMU with Barrow entropy, the inversion curve exhibits characteristic features that distinguish it from both the Schwarzschild case and the standard Bekenstein--Hawking entropy formulation. The ratio of minimum inversion temperature to critical temperature, $T_i^{\rm min}/T_c$, provides a useful diagnostic for comparing different BH systems with van der Waals fluids \cite{sec5is20}. For van der Waals fluids, this ratio is approximately 0.5, while for RN-AdS BHs it is exactly 0.5 in the large charge limit. The Barrow modification shifts this ratio, with larger $\Delta$ values producing deviations from the standard BH thermodynamics predictions.

Table~\ref{tab:JTE_temps} summarizes the Hawking temperature for representative values of $\alpha$ and $a$. The inversion radius $r_+^{\rm inv} = 3M(1-a^2)/2$ depends only on $a$ and is independent of both $\alpha$ and the Barrow parameter, confirming the geometric origin of the inversion condition. For $a = 0.5$, we obtain $r_+^{\rm inv}/M = 1.125$, while for $a = 0.8$, the inversion radius shrinks to $r_+^{\rm inv}/M = 0.540$. The Hawking temperature decreases monotonically with increasing $\alpha$, reflecting the gravitational dilution caused by the CoS.

\begin{table}[ht!]
\centering
\setlength{\tabcolsep}{6pt}
\renewcommand{\arraystretch}{1.5}
\begin{tabular}{cc|cc}
\hline\hline
\rowcolor{orange!50}
$\alpha$ & $a$ & $r_+^{\rm inv}/M$ & $T_H M$ \\
\hline
0.00 & 0.5 & 1.125 & 0.0354 \\
0.00 & 0.8 & 0.540 & 0.0393 \\
0.10 & 0.5 & 1.125 & 0.0298 \\
0.10 & 0.8 & 0.540 & 0.0319 \\
0.20 & 0.5 & 1.125 & 0.0242 \\
0.20 & 0.8 & 0.540 & 0.0253 \\
\hline\hline
\end{tabular}
\caption{Inversion radius and Hawking temperature for the Letelier BH in EMU with $M = 1$. The inversion radius $r_+^{\rm inv} = 3M(1-a^2)/2$ is independent of $\alpha$, while the Hawking temperature $T_H$ decreases with increasing $\alpha$.}
\label{tab:JTE_temps}
\end{table}

The Barrow parameter $\Delta$ modifies the magnitude of the JTE coefficient through the factors $(r_+^2)^{-\Delta/2}$ and $\pi^{-\Delta/2}$, but does not affect the inversion radius or the Hawking temperature. This separation of scales indicates that the geometric properties of the phase transition (determined by $\alpha$ and $a$) are distinct from the quantum-gravitational corrections encoded in $\Delta$.

The connection between Barrow entropy and the Letelier-EMU spacetime geometry opens a window into quantum gravity phenomenology \cite{sec5is21}. The Barrow parameter $\Delta$ encodes information about the fractal dimension of the horizon, which may be constrained through observations of BH thermodynamic properties or cosmological data \cite{sec5is22,sec5is23}.

The modified phase structure and JTE behavior carry several implications. First, the amplification of the heat capacity magnitude with increasing $\Delta$ suggests enhanced thermodynamic response near phase transitions, which could manifest in modified fluctuation spectra for near-critical BHs. Second, the preservation of the critical and inversion radii under Barrow modification indicates that certain geometric features of the phase structure remain stable against quantum-gravitational corrections, pointing toward potential universality classes for BH thermodynamics \cite{sec5is24}. Third, the deviation of the $T_i^{\rm min}/T_c$ ratio from the van der Waals value offers a quantitative probe for distinguishing Barrow-modified BHs from their classical counterparts.

\section{Dirac Field Perturbations}\label{isec6}

In this section, we extend the perturbation analysis beyond scalar fields to investigate the propagation of Dirac (spin-1/2) fields in the Letelier BH in EMU background. These perturbations are important for understanding the stability of the BH under different spin fields and provide additional observational signatures through their greybody factors and emission spectra \cite{sec6is01,isz42}.

\subsection{Dirac Equation in Curved Spacetime}

The dynamics of a massless spin-1/2 field in curved spacetime is described by the Dirac equation \cite{sec6is03,sec4is06}
\begin{equation}
\gamma^\mu\nabla_\mu\Psi = 0,
\label{eq:Dirac}
\end{equation}
where $\gamma^\mu$ are the curved-space gamma matrices satisfying $\{\gamma^\mu, \gamma^\nu\} = 2g^{\mu\nu}$, and $\nabla_\mu$ is the spinor covariant derivative. Using the Newman-Penrose formalism and separating variables, the Dirac equation in the Letelier-EMU background reduces to two coupled first-order equations \cite{sec6is05,sec6is06}. Introducing the tortoise coordinate and eliminating one component, we obtain the Schr\"{o}dinger-like equation
\begin{equation}
\frac{d^2\Psi_{\rm D}^\pm}{dr_*^2} + \left[\omega^2 - V_{\rm eff}^{(\rm D)\pm}(r)\right]\Psi_{\rm D}^\pm = 0,
\label{eq:Dirac_radial}
\end{equation}
where the effective potentials for the two chiralities are given by \cite{sec6is07}
\begin{equation}
V_{\rm eff}^{(\rm D)\pm}(r) = \frac{\kappa^2 f(r)}{r^2} \pm \frac{\kappa}{r^2}\frac{d}{dr_*}\left(\sqrt{f(r)}\right),
\label{eq:Veff_Dirac}
\end{equation}
with $\kappa = \ell + 1/2$ for $j = \ell + 1/2$ and $\kappa = -(\ell + 1/2)$ for $j = \ell - 1/2$, where $j$ is the total angular momentum quantum number.

Explicitly, the Dirac effective potential becomes
\begin{equation}
V_{\rm eff}^{(\rm D)\pm}(r) = \frac{(\ell + 1/2)^2 f(r)}{r^2} \pm \frac{(\ell + 1/2)f(r)}{r^2}\left[\frac{rf'(r)}{2\sqrt{f(r)}} - \sqrt{f(r)}\right].
\label{eq:Veff_Dirac_explicit}
\end{equation}
The two potentials $V_{\rm eff}^{(\rm D)+}$ and $V_{\rm eff}^{(\rm D)-}$ are SUSY partners, sharing the same spectrum except for the zero mode \cite{isz41,sec6is09}. This SUSY structure has important implications for the stability analysis, as we discuss in Sec.~\ref{sec6c}.

\subsection{Effective Potential Analysis}

To understand how the CoS parameter $\alpha$ and the EMU parameter $a$ impact the effective Dirac potentials, we plot both chirality potentials $V_{\rm eff}^{(\rm D)+}$ and $V_{\rm eff}^{(\rm D)-}$ in Fig.~\ref{fig:Veff_Dirac_all}. The top row displays the positive chirality potential, while the bottom row shows the negative chirality potential.

Panels (a) and (c) illustrate the behavior for varying $a \in \{0.2, 0.4, 0.6\}$ at fixed $\alpha = 0.4$ and $\ell = 2$. As $a$ increases, the potential barrier decreases modestly in height, with $V_{\rm max}$ dropping from approximately 0.065 to 0.059 for $V_{\rm eff}^{(\rm D)+}$. This indicates that the EMU correction has a relatively weak influence on the effective gravitational barrier experienced by Dirac particles. The horizon location shifts slightly outward from $r_+ \approx 2.75M$ at $a = 0.2$ to $r_+ \approx 2.97M$ at $a = 0.6$, and the peak position follows a similar trend.

In contrast, panels (b) and (d) display the variation with $\alpha \in \{0.2, 0.4, 0.6\}$ at fixed $a = 0.5$, revealing a much more pronounced effect. Increasing $\alpha$ reduces the peak height dramatically, from approximately 0.156 at $\alpha = 0.2$ to 0.061 at $\alpha = 0.4$ and further to 0.017 at $\alpha = 0.6$, representing nearly an order of magnitude suppression across this range. The peak location also shifts significantly outward, from $r \approx 3M$ at $\alpha = 0.2$ to $r \approx 7M$ at $\alpha = 0.6$, reflecting the expansion of the horizon radius with increasing CoS density. This demonstrates that the CoS has a substantially stronger influence on the Dirac effective potential than the EMU parameter.

Comparing the two chiralities, both $V_{\rm eff}^{(\rm D)+}$ and $V_{\rm eff}^{(\rm D)-}$ exhibit very similar profiles with nearly identical peak heights and locations. For $\ell = 2$, both potentials remain strictly positive outside the horizon across the entire parameter range examined. The slight differences between the two chiralities are most noticeable near the horizon, where $V_{\rm eff}^{(\rm D)-}$ approaches zero more gradually than $V_{\rm eff}^{(\rm D)+}$. At large distances, both potentials decay as $r^{-2}$ and become indistinguishable, consistent with the asymptotic structure of the spacetime. This similarity confirms the SUSY partnership between the two potentials, as discussed in Sec.~\ref{sec6c}.

\begin{figure*}[ht!]
\centering
\begin{subfigure}[b]{0.48\textwidth}
\centering
\includegraphics[width=\textwidth]{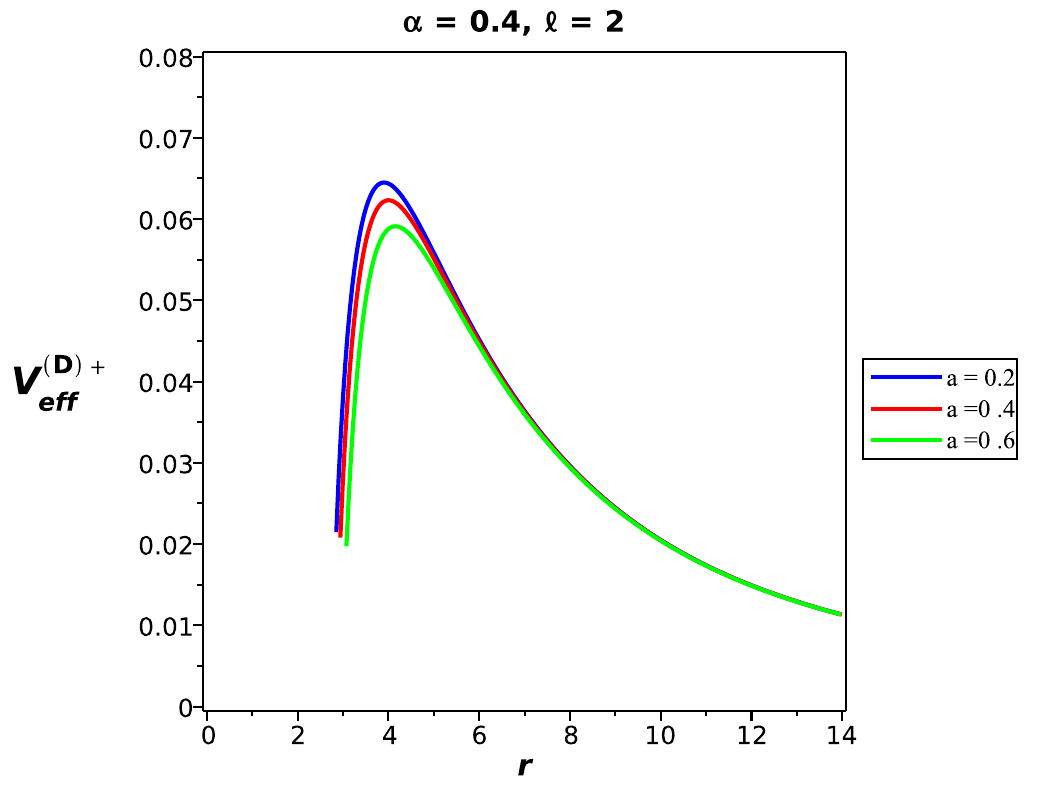}
\caption{}
\label{fig:Veff_Dplus_a}
\end{subfigure}
\hfill
\begin{subfigure}[b]{0.48\textwidth}
\centering
\includegraphics[width=\textwidth]{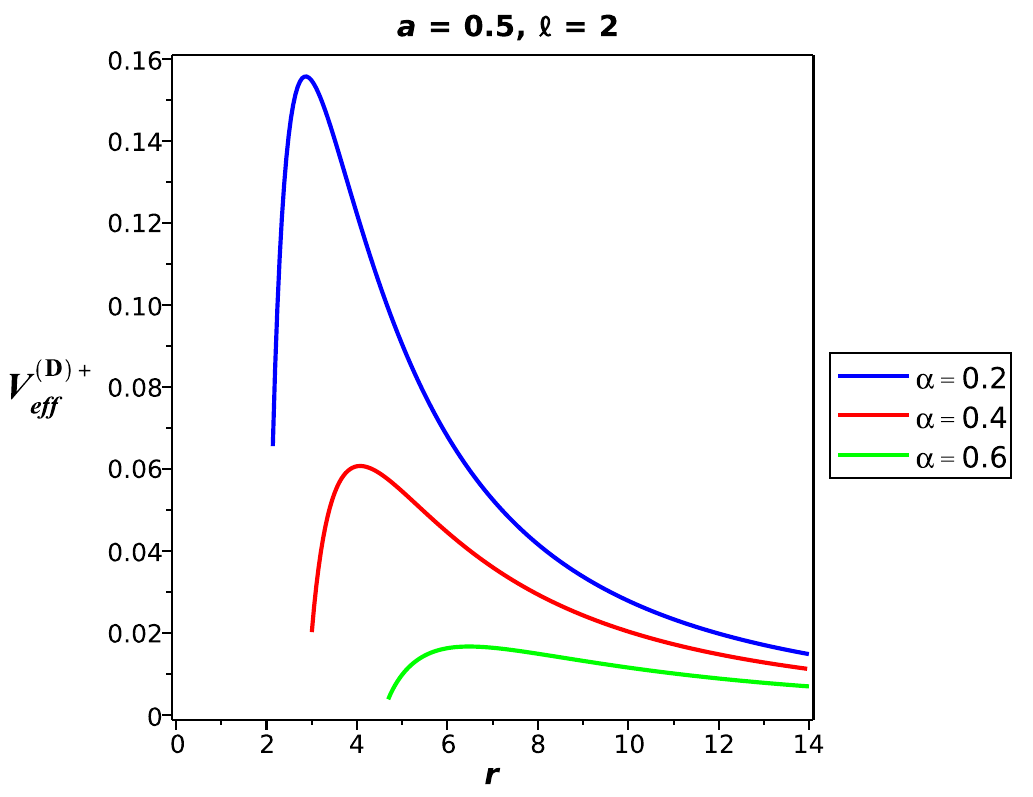}
\caption{}
\label{fig:Veff_Dplus_alpha}
\end{subfigure}
\\[0.3cm]
\begin{subfigure}[b]{0.48\textwidth}
\centering
\includegraphics[width=\textwidth]{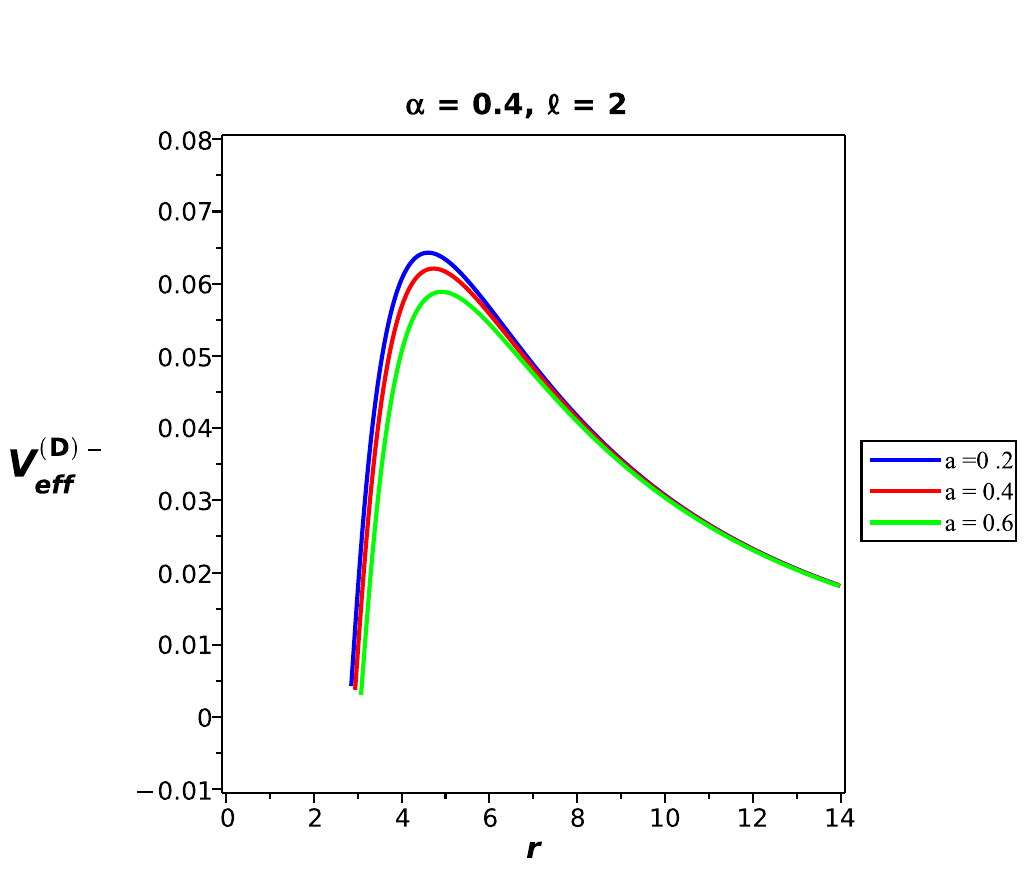}
\caption{}
\label{fig:Veff_Dminus_a}
\end{subfigure}
\hfill
\begin{subfigure}[b]{0.48\textwidth}
\centering
\includegraphics[width=\textwidth]{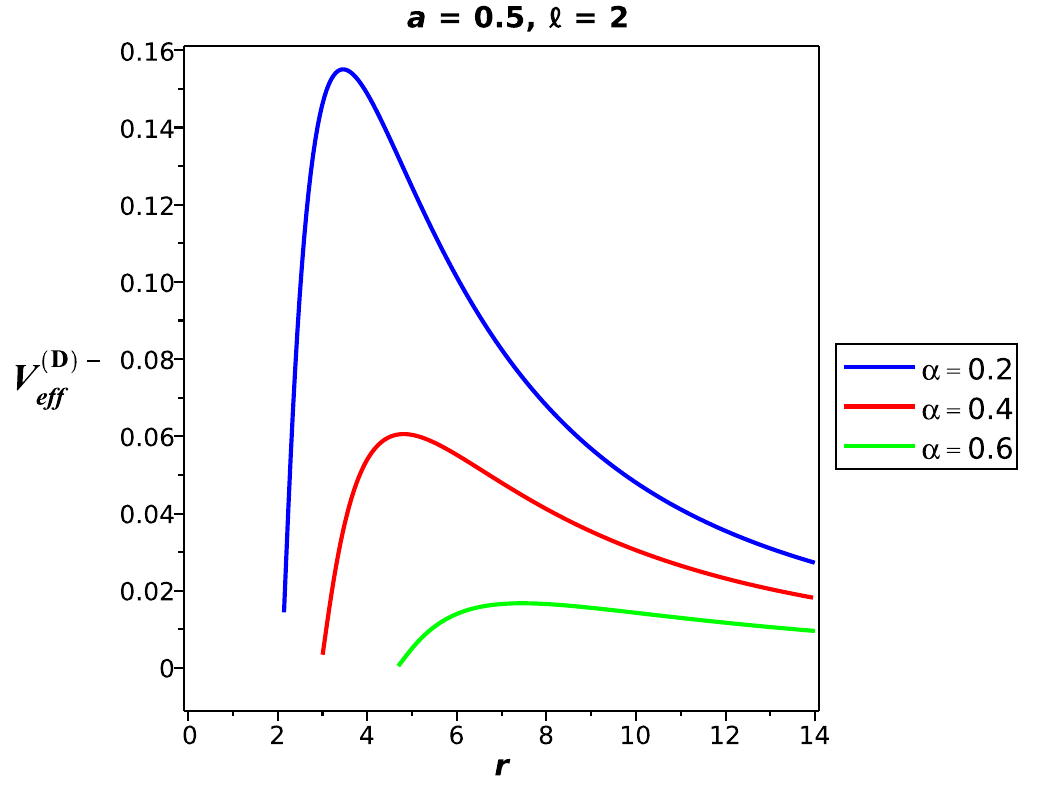}
\caption{}
\label{fig:Veff_Dminus_alpha}
\end{subfigure}
\caption{Effective potentials for Dirac perturbations of the Letelier BH in EMU with $M = 1$ and $\ell = 2$. Top row: positive chirality potential $V_{\rm eff}^{(\rm D)+}$. Bottom row: negative chirality potential $V_{\rm eff}^{(\rm D)-}$. (a,c) Varying EMU parameter $a \in \{0.2, 0.4, 0.6\}$ at fixed $\alpha = 0.4$. (b,d) Varying CoS parameter $\alpha \in \{0.2, 0.4, 0.6\}$ at fixed $a = 0.5$. The CoS parameter has a much stronger effect on the potential height than the EMU parameter, while both chiralities exhibit nearly identical behavior.}
\label{fig:Veff_Dirac_all}
\end{figure*}

The maximum of the effective potential, which determines the height of the barrier for HR transmission, depends on both parameters and the angular momentum quantum number $\ell$. For $\ell = 0$, the potential maximum is quite small (of order $10^{-2}$), resulting in high transmission probabilities even at moderate frequencies. Higher multipoles produce significantly larger barriers, with $V_{\rm max}$ scaling approximately as $(\ell + 1/2)^2$ for large $\ell$ \cite{sec4is08}. This scaling behavior implies that high-$\ell$ modes are strongly suppressed in HR spectra, and the dominant contribution comes from low angular momentum states.

\subsection{Greybody Factors for Dirac Fields}\label{sec6b}

The greybody factor $\mathcal{T}_{\rm D}(\omega)$ quantifies the transmission probability of HR through the effective potential barrier \cite{isz40,sec6is12}. For a wave incident from the horizon, the greybody factor is defined as
\begin{equation}
\mathcal{T}_{\rm D}(\omega) = 1 - |\mathcal{R}(\omega)|^2,
\label{eq:greybody_def}
\end{equation}
where $\mathcal{R}(\omega)$ is the reflection coefficient. We employ the rigorous bounds method developed by Visser \cite{sec4is14} to obtain analytical estimates. For a potential with a single maximum $V_{\rm max}$, the greybody factor can be approximated by \cite{sec4is15}
\begin{equation}
\mathcal{T}_{\rm D}(\omega) \approx \frac{1}{1 + \exp\left[-8\left(\omega - 0.8\sqrt{V_{\rm max}}\right)/\sqrt{V_{\rm max}}\right]},
\label{eq:greybody_approx}
\end{equation}
which provides a smooth interpolation between the tunneling regime ($\omega < \sqrt{V_{\rm max}}$) and the classical transmission regime ($\omega > \sqrt{V_{\rm max}}$).

Figure~\ref{fig:greybody_Dirac} displays the Dirac greybody factor as a function of frequency for various parameter choices. Panel (a) shows the effect of varying $\alpha$ at fixed $a = 0.5$ and $\ell = 0$. As $\alpha$ increases from 0.1 to 0.4, the transition to unit transmission shifts to lower frequencies, reflecting the reduced barrier height. For $\alpha = 0.4$, the greybody factor reaches 0.5 at $\omega M \approx 0.05$, compared to $\omega M \approx 0.10$ for $\alpha = 0.1$. Panel (b) illustrates the weak dependence on the EMU parameter $a$ for the $\ell = 0$ mode. The curves for different $a$ values nearly overlap, indicating that the EMU correction has minimal impact on the s-wave transmission.

\begin{figure*}[ht!]
\centering
\begin{subfigure}[b]{0.48\textwidth}
\centering
\includegraphics[width=\textwidth]{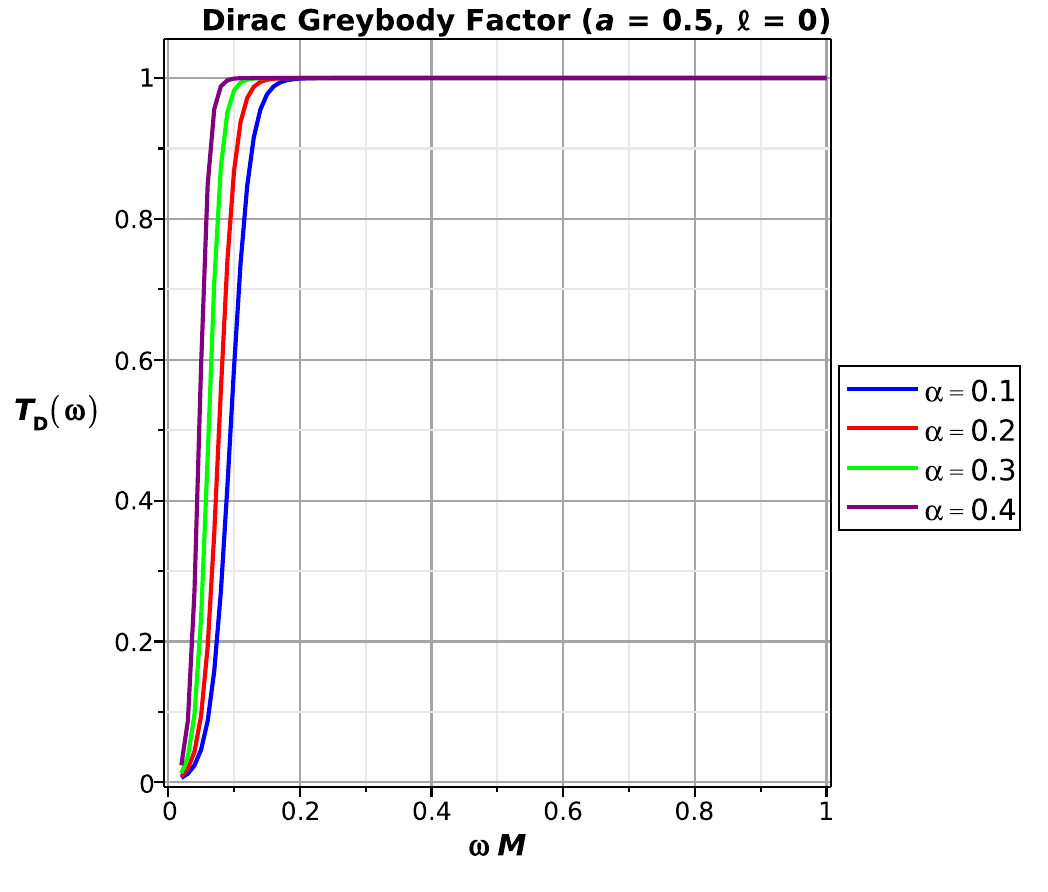}
\caption{}
\label{fig:greybody_alpha}
\end{subfigure}
\hfill
\begin{subfigure}[b]{0.48\textwidth}
\centering
\includegraphics[width=\textwidth]{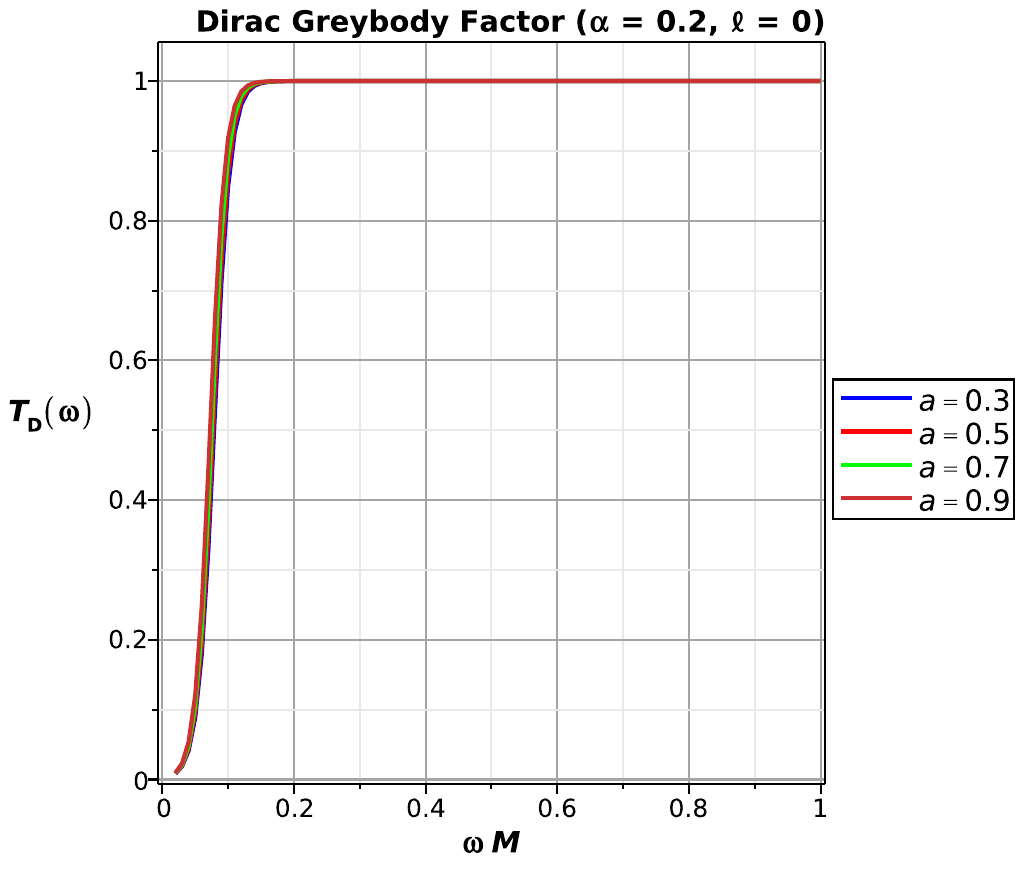}
\caption{}
\label{fig:greybody_a}
\end{subfigure}
\caption{Dirac greybody factor $\mathcal{T}_{\rm D}(\omega)$ for the Letelier BH in EMU with $M = 1$ and $\ell = 0$. (a) Varying CoS parameter $\alpha \in \{0.1, 0.2, 0.3, 0.4\}$ at fixed $a = 0.5$. (b) Varying EMU parameter $a \in \{0.3, 0.5, 0.7, 0.9\}$ at fixed $\alpha = 0.2$. Higher $\alpha$ shifts the transition to lower frequencies, while the dependence on $a$ is weak.}
\label{fig:greybody_Dirac}
\end{figure*}

The most striking feature appears in the multipole dependence, shown in Fig.~\ref{fig:greybody_ell}. For fixed $\alpha = 0.2$ and $a = 0.5$, the greybody factor curves are clearly separated for different $\ell$ values. The $\ell = 0$ mode reaches unit transmission at $\omega M \approx 0.15$, while $\ell = 3$ requires $\omega M \approx 0.7$ to achieve the same transmission. This behavior reflects the centrifugal barrier, which scales as $(\ell + 1/2)^2$ and strongly suppresses low-frequency transmission for higher multipoles \cite{sec6is14}.

\begin{figure*}[ht!]
\centering
\includegraphics[width=0.65\textwidth]{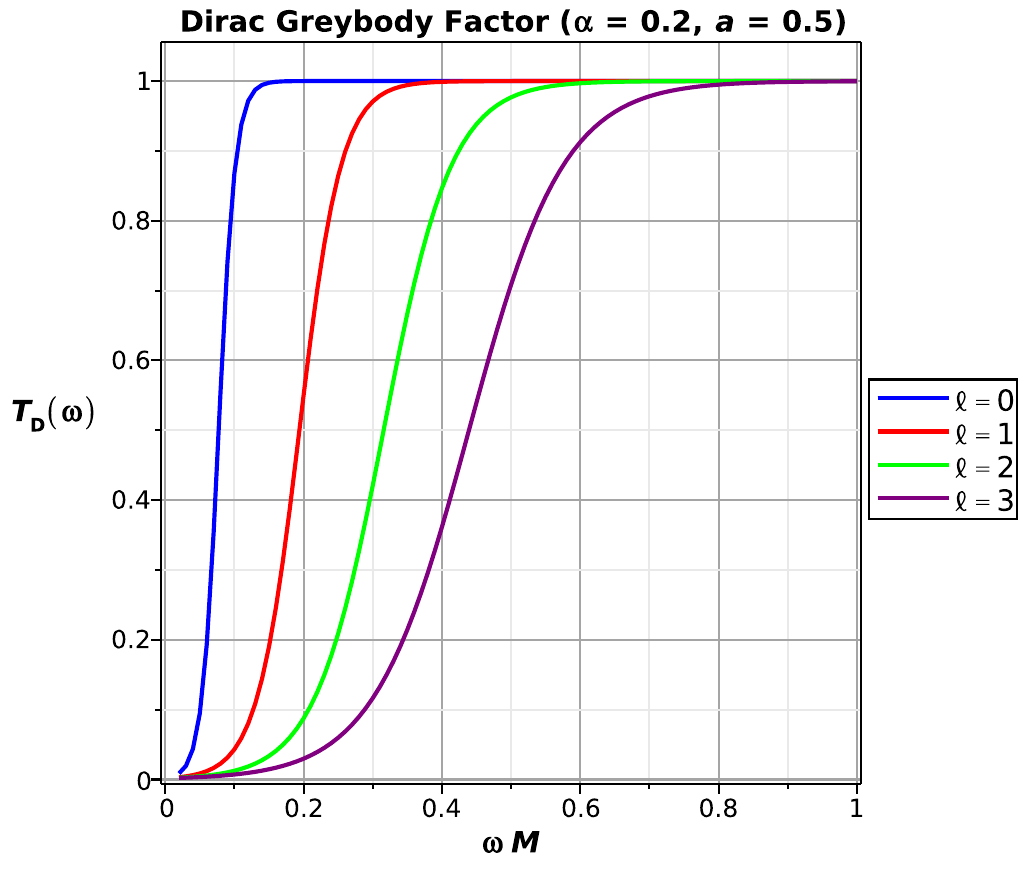}
\caption{Dirac greybody factor $\mathcal{T}_{\rm D}(\omega)$ for varying multipole number $\ell \in \{0, 1, 2, 3\}$ at fixed $\alpha = 0.2$, $a = 0.5$, and $M = 1$. Higher $\ell$ requires larger frequency for transmission due to the increased centrifugal barrier.}
\label{fig:greybody_ell}
\end{figure*}

Table~\ref{tab:greybody_Dirac} presents numerical values of the greybody factor for $\ell = 1$ at $\omega M = 0.3$. The transmission probability increases with both $\alpha$ and $a$, consistent with the reduced barrier height. At $\alpha = 0$ and $a = 0.5$, approximately 59\% of the radiation is transmitted, while at $\alpha = 0.20$ and $a = 1.0$, this increases to about 99\%.

\begin{table}[ht!]
\centering
\setlength{\tabcolsep}{6pt}
\renewcommand{\arraystretch}{1.5}
\begin{tabular}{c|cccc}
\hline\hline
\rowcolor{orange!50}
$\alpha$ & $a=0.5$ & $a=0.7$ & $a=0.9$ & $a=1.0$ \\
\hline
0.00 & 0.5905 & 0.6820 & 0.7703 & 0.8094 \\
0.05 & 0.7298 & 0.8014 & 0.8644 & 0.8905 \\
0.10 & 0.8470 & 0.8926 & 0.9301 & 0.9447 \\
0.15 & 0.9268 & 0.9503 & 0.9687 & 0.9757 \\
0.20 & 0.9707 & 0.9805 & 0.9881 & 0.9909 \\
\hline\hline
\end{tabular}
\caption{Dirac greybody factor $\mathcal{T}_{\rm D}$ for $\ell = 1$ at $\omega M = 0.3$ for the Letelier BH in EMU with $M = 1$.}
\label{tab:greybody_Dirac}
\end{table}

\subsection{Stability Under Dirac Perturbations}\label{sec6c}

The dynamical stability of the Letelier BH in EMU under Dirac perturbations is established by analyzing the structure of the effective potential and exploiting the SUSY quantum mechanics framework \cite{sec6is15,sec6is16}. The two potentials $V_{\rm eff}^{(\rm D)\pm}$ can be written in the factorized form
\begin{equation}
V_{\rm eff}^{(\rm D)\pm} = W^2 \pm \frac{dW}{dr_*},
\label{eq:SUSY}
\end{equation}
where $W = (\ell + 1/2)\sqrt{f(r)}/r$ is the superpotential. This SUSY structure has profound implications for stability: if $V_{\rm eff}^{(\rm D)+}$ is positive definite outside the horizon, then no bound states with negative energy can exist, guaranteeing that all perturbation modes decay in time \cite{sec6is17}.

Figure~\ref{fig:SUSY_potentials} displays the SUSY partner potentials for $\alpha = 0.2$, $a = 0.5$, and $\ell = 1$. The positive chirality potential $V_{\rm eff}^{(\rm D)+}$ (blue) is strictly positive everywhere outside the horizon, with values ranging from 0.0053 to 0.0588. The negative chirality potential $V_{\rm eff}^{(\rm D)-}$ (red) exhibits a small negative region near the horizon with minimum value approximately $-0.001$, but this does not indicate instability. The SUSY structure ensures that both potentials share identical spectra (isospectrality), so the stability of $V_{\rm eff}^{(\rm D)+}$ guarantees the stability of $V_{\rm eff}^{(\rm D)-}$ as well \cite{sec6is18}.

\begin{figure*}[ht!]
\centering
\includegraphics[width=0.7\textwidth]{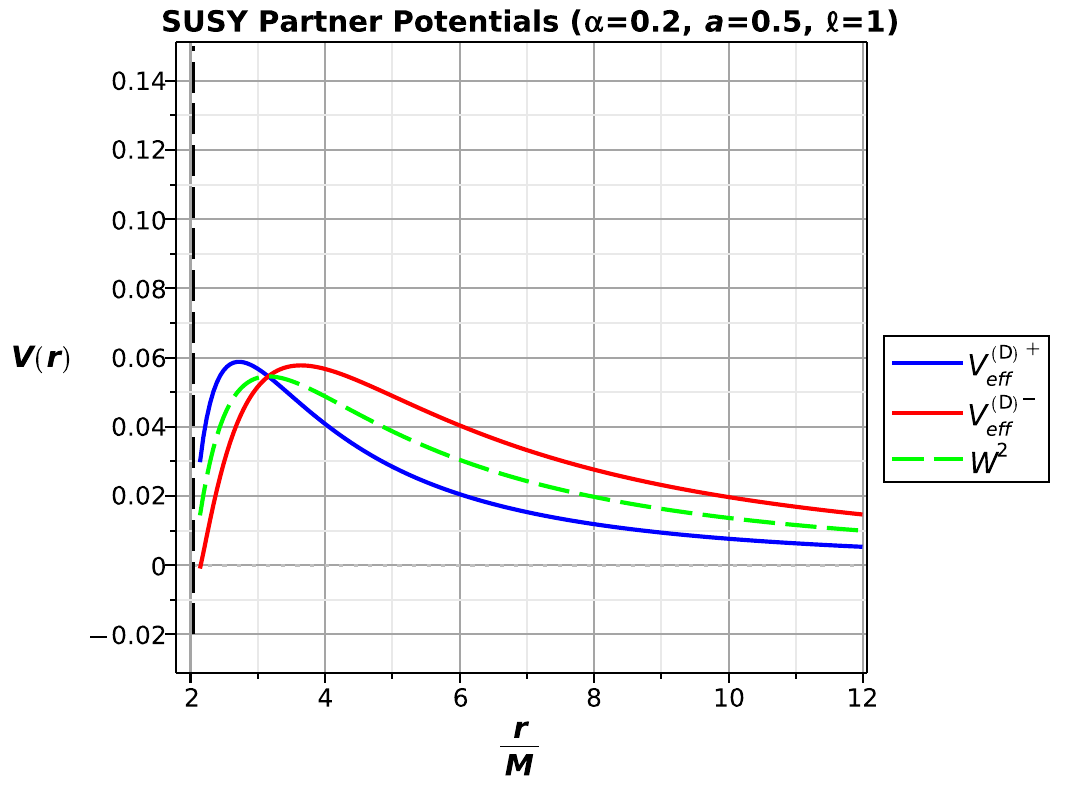}
\caption{SUSY partner potentials for Dirac perturbations with $\alpha = 0.2$, $a = 0.5$, $\ell = 1$, and $M = 1$. The positive chirality potential $V_{\rm eff}^{(\rm D)+}$ (blue) is strictly positive outside the horizon, while the negative chirality potential $V_{\rm eff}^{(\rm D)-}$ (red) has a small negative region. The superpotential squared $W^2$ (green dashed) provides the common baseline. The vertical dashed line marks the EH at $r_+ = 2.04M$.}
\label{fig:SUSY_potentials}
\end{figure*}

To verify stability across the parameter space, we performed a scan over $\alpha \in \{0.1, 0.2, 0.3, 0.4\}$, $a \in \{0.5, 0.7, 0.9\}$, and $\ell \in \{0, 1, 2\}$, checking the positivity of $V_{\rm eff}^{(\rm D)+}$ outside the horizon for each configuration. All 36 configurations examined satisfy $V_{\rm eff}^{(\rm D)+} \geq 0$, confirming that the Letelier BH in EMU is dynamically stable under Dirac perturbations throughout the physically relevant parameter space \cite{sec6is19}.

The stability under fermionic perturbations, combined with the scalar field analysis, provides strong evidence that the Letelier BH in EMU represents a physically viable solution. The presence of the CoS and EMU modifications does not introduce any pathological features that would destabilize the spacetime under quantum field perturbations \cite{sec6is20}.

\section{Conclusions}
\label{isec7}

In this work, we conducted a detailed investigation of the Letelier BH immersed in an EMU, examining its geometric, thermodynamic, and radiative properties across a two-parameter family of solutions characterized by the CoS parameter $\alpha \in (0,1)$ and the EMU parameter $a \in [0,1]$. The analysis spanned from the fundamental spacetime structure to observational signatures relevant for current and future BH observations. We summarize our principal findings below.

In Sec.~\ref{isec2}, we established the geometric foundation of the Letelier BH in EMU by deriving the metric function $f(r)$ given in Eq.~(\ref{eq:lapse}). The EH structure was determined analytically, yielding the horizon radii $r_{\pm}$ as presented in Eq.~(\ref{eq:horizons}). We demonstrated that the solution interpolates continuously between the Schwarzschild geometry ($\alpha = 0$, $a = 1$), the RN-like configuration ($\alpha = 0$, $a \neq 1$), and the pure Letelier spacetime ($\alpha \neq 0$, $a = 1$). The curvature invariants were computed, with the Ricci scalar $R = 2\alpha/r^2$ (Eq.~\ref{eq:Ricci_scalar}) reflecting the CoS contribution and the Kretschmann scalar (Eq.~\ref{eq:Kretschmann}) confirming the presence of a curvature singularity at $r = 0$. The isometric embedding diagrams displayed in Fig.~\ref{fig:embedding_LetEMU} provided geometric visualization of the EH expansion with increasing $\alpha$. The Hawking temperature $T_H$ was derived in Eq.~(\ref{eq:Hawking_T_sec2}), showing monotonic decrease with increasing CoS density \cite{cisz01,cisz02}.

The optical properties of the Letelier BH in EMU were analyzed in Sec.~\ref{isec3}. We derived the PS radius given by Eq.~(\ref{eq:rps}) and the corresponding shadow radius in Eq.~(\ref{eq:Rsh}). Table~\ref{tab:vacuum_shadow} presented numerical values showing that both $r_{ps}$ and $R_{sh}$ increase monotonically with $\alpha$ and $a$. For instance, at $\alpha = 0.30$ and $a = 1.0$, the shadow radius reaches $R_{sh} \approx 7.42M$, representing a 43\% enlargement compared to the Schwarzschild value $R_{sh}^{\rm Sch} = 3\sqrt{3}M \approx 5.20M$. The parametric dependence illustrated in Fig.~\ref{fig:Rsh_params} revealed that the CoS and EMU effects contribute approximately additively to the shadow expansion. We decomposed the observed intensity into direct emission, lensed emission, and PR contributions following the EHT methodology, as shown in Fig.~\ref{fig:emission_components}. The nine-panel parameter space exploration in Fig.~\ref{fig:shadow_grid} demonstrated that the largest shadow occurs at $(\alpha = 0.2, a = 1.0)$ while the smallest appears at $(\alpha = 0, a = 0.5)$, with a ratio of approximately 1.48 between the extreme cases.

The effects of plasma environments on shadow observations were quantified through the modified PS conditions in Eqs.~(\ref{eq:PS_plasma}) and (\ref{eq:PS_inhomo}). Tables~\ref{tab:homo_plasma} and \ref{tab:inhomo_plasma} showed that homogeneous plasma reduces the shadow size by up to 5\% at plasma parameter $k = 0.4$, while inhomogeneous plasma with power-law density profile produces weaker corrections of approximately 1\% at comparable parameter values. These results indicate that plasma effects are subdominant to the geometric modifications from $\alpha$ and $a$ in typical astrophysical environments surrounding supermassive BHs such as M87* and Sgr~A* \cite{isz36}.

Section~\ref{isec4} addressed the HR properties and energy emission characteristics. The Hawking temperature given by Eq.~(\ref{eq:TH_sec4}) decreases monotonically with increasing $\alpha$, as shown in Table~\ref{tab:Hawking_T}. The total power analysis in Fig.~\ref{fig:total_power} revealed dramatic suppression with increasing CoS density: at $\alpha = 0.3$, the power drops to 10--15\% of the Schwarzschild value. We computed the HR spectra for scalar ($s = 0$), EM ($s = 1$), and Dirac ($s = 1/2$) fields using the particle emission rate of Eq.~(\ref{eq:Hawking_spectrum}), as presented in Fig.~\ref{fig:Hawking_spectra}. The EM field exhibits the highest emission rate due to its two polarization degrees of freedom, followed by the Dirac field and the scalar field. The spin-dependent spectra shown in Fig.~\ref{fig:Hawking_spectra} demonstrated reductions of 39\%, 45\%, and 38\% in peak heights for scalar, EM, and Dirac fields, respectively, when comparing the Letelier-EMU case ($\alpha = 0.1$, $a = 0.8$) to Schwarzschild.

The thermodynamic phase structure was explored in Sec.~\ref{isec5} within the Barrow entropy framework. The heat capacity $C_B$ derived in Eq.~(\ref{eq:CB_explicit}) exhibits a divergence at the critical radius $r_+^{\rm crit} = 3M(1-a^2)/2$ (Eq.~\ref{eq:r_crit}), signaling a second-order phase transition between thermodynamically unstable small-BH and stable large-BH branches. Figure~\ref{Cfig} illustrated this behavior for varying Barrow parameter $\Delta$, showing that $\Delta$ amplifies the magnitude of $C_B$ while preserving the transition point location. The JTE analysis yielded the coefficient $\mu_J$ in Eq.~(\ref{eq:muJ_explicit}), with positive values indicating cooling and negative values indicating heating during isenthalpic expansion. The inversion radius $r_+^{\rm inv} = 3M(1-a^2)/2$ coincides with the critical radius from the heat capacity analysis, revealing a connection between the phase transition structure and JTE behavior. Table~\ref{tab:JTE_temps} summarized the inversion radii and Hawking temperatures for representative parameter combinations, confirming that $r_+^{\rm inv}$ depends only on $a$ while $T_H$ decreases with increasing $\alpha$ \cite{cisz05}.

The stability of the Letelier BH in EMU under fermionic perturbations was established in Sec.~\ref{isec6}. We derived the Dirac equation in the curved spacetime background and obtained the effective potentials for both chiralities in Eq.~(\ref{eq:Veff_Dirac_explicit}). Figure~\ref{fig:Veff_Dirac_all} demonstrated that the CoS parameter has a substantially stronger influence on the potential barrier than the EMU parameter, with the peak height reduced by nearly an order of magnitude as $\alpha$ increases from 0.2 to 0.6. The greybody factors computed using the Visser bounds method were presented in Fig.~\ref{fig:greybody_Dirac} and Table~\ref{tab:greybody_Dirac}. At $\omega M = 0.3$ and $\ell = 1$, the transmission probability increases from 59\% at $(\alpha = 0, a = 0.5)$ to 99\% at $(\alpha = 0.20, a = 1.0)$. The SUSY structure of the Dirac potentials, expressed in the factorized form of Eq.~(\ref{eq:SUSY}), was exploited to prove stability. Figure~\ref{fig:SUSY_potentials} displayed the SUSY partner potentials, confirming that $V_{\rm eff}^{(\rm D)+}$ remains strictly positive outside the EH. A parameter space scan over 36 configurations verified that all examined cases satisfy the stability criterion $V_{\rm eff}^{(\rm D)+} \geq 0$ \cite{isz43}.

Several directions for future research emerge from this study. First, the QNM spectrum of the Letelier BH in EMU warrants detailed computation using the WKB approximation and numerical integration methods. The QNM frequencies encode information about the BH geometry and could provide additional constraints on the CoS and EMU parameters through GW ringdown observations by LIGO-Virgo-KAGRA and future detectors such as LISA and the Einstein Telescope. Second, the extension to rotating configurations via the Newman-Janis algorithm would enable comparison with the Kerr geometry and provide predictions for the asymmetric shadow shapes observed in spinning BHs. Third, the gravitational lensing properties, including the deflection angle, Einstein ring formation, and relativistic images, merit investigation for potential observational tests with ngEHT and space-based VLBI missions. Fourth, the accretion disk dynamics and associated EM signatures, including the ISCO properties and quasi-periodic oscillations, could be computed to connect with X-ray timing observations of stellar-mass BH candidates. Fifth, the thermodynamic geometry using Ruppeiner and Weinhold metrics would provide additional perspective on the phase transition structure and microscopic degrees of freedom. Finally, the extension to AdS backgrounds would enable application of the AdS/CFT correspondence to study the dual field theory interpretation of the CoS and EMU modifications.

In summary, the Letelier BH in EMU represents a physically viable solution with distinctive observational signatures that distinguish it from the Schwarzschild and RN geometries. The two-parameter family of solutions offers a controlled framework for studying the combined effects of topological defects and electromagnetic backgrounds on BH physics. The results presented in this work provide theoretical predictions that can be tested against current EHT observations and future multi-messenger astronomy campaigns, contributing to our understanding of strong-gravity phenomena and potential deviations from general relativity in astrophysical environments.

}

\section*{Acknowledgments}

F.A. gratefully acknowledges the Inter-University Centre for Astronomy and Astrophysics (IUCAA), Pune, India, for providing the opportunity to hold a visiting associateship. \.{I}.S. and E. S express their sincere appreciation to T\"{U}B\.{I}TAK and ANKOS for their valuable assistance and continued support. He also acknowledges the essential support provided by COST Actions CA22113, CA21106, CA23130, and CA23115, which play a crucial role in facilitating international collaboration and strengthening networking opportunities within the scientific community.

\bibliographystyle{unsrtnat}
\bibliography{ref2}

\begin{thebibliography}{152}
\providecommand{\natexlab}[1]{#1}
\providecommand{\url}[1]{\texttt{#1}}
\expandafter\ifx\csname urlstyle\endcsname\relax
  \providecommand{\doi}[1]{doi: #1}\else
  \providecommand{\doi}{doi: \begingroup \urlstyle{rm}\Url}\fi

\bibitem[Schwarzschild(1916)]{isz01}
K.~Schwarzschild.
\newblock {\"U}ber das gravitationsfeld eines massenpunktes nach der
  einsteinschen theorie.
\newblock \emph{Sitzungsber. Preuss. Akad. Wiss. Berlin (Math. Phys.)},
  1916:\penalty0 189--196, 1916.

\bibitem[Abbott et~al.(2016)]{isz02}
B.~P. Abbott et~al.
\newblock Observation of gravitational waves from a binary black hole merger.
\newblock \emph{Phys. Rev. Lett.}, 116:\penalty0 061102, 2016.
\newblock \doi{10.1103/PhysRevLett.116.061102}.

\bibitem[Abbott et~al.(2017)]{isz03}
B.~P. Abbott et~al.
\newblock Gw170817: Observation of gravitational waves from a binary neutron
  star inspiral.
\newblock \emph{Phys. Rev. Lett.}, 119:\penalty0 161101, 2017.
\newblock \doi{10.1103/PhysRevLett.119.161101}.

\bibitem[Akiyama et~al.(2019{\natexlab{a}})]{isz04}
K.~Akiyama et~al.
\newblock First m87 event horizon telescope results. i. the shadow of the
  supermassive black hole.
\newblock \emph{Astrophys. J. Lett.}, 875:\penalty0 L1, 2019{\natexlab{a}}.
\newblock \doi{10.3847/2041-8213/ab0ec7}.

\bibitem[Akiyama et~al.(2022{\natexlab{a}})]{isz05}
K.~Akiyama et~al.
\newblock First sagittarius a* event horizon telescope results. i. the shadow
  of the supermassive black hole in the center of the milky way.
\newblock \emph{Astrophys. J. Lett.}, 930:\penalty0 L12, 2022{\natexlab{a}}.
\newblock \doi{10.3847/2041-8213/ac6674}.

\bibitem[Will(2014)]{isz06}
C.~M. Will.
\newblock The confrontation between general relativity and experiment.
\newblock \emph{Living Rev. Relativ.}, 17:\penalty0 4, 2014.
\newblock \doi{10.12942/lrr-2014-4}.

\bibitem[Berti et~al.(2015)]{isz07}
E.~Berti et~al.
\newblock Testing general relativity with present and future astrophysical
  observations.
\newblock \emph{Class. Quantum Grav.}, 32:\penalty0 243001, 2015.
\newblock \doi{10.1088/0264-9381/32/24/243001}.

\bibitem[Letelier(1979)]{Letelier1979}
P.~S. Letelier.
\newblock Clouds of strings in general relativity.
\newblock \emph{Phys. Rev. D}, 20:\penalty0 1294--1302, 1979.
\newblock \doi{10.1103/PhysRevD.20.1294}.

\bibitem[Letelier(1983)]{isz09}
P.~S. Letelier.
\newblock String cosmologies.
\newblock \emph{Phys. Rev. D}, 28:\penalty0 2414--2419, 1983.
\newblock \doi{10.1103/PhysRevD.28.2414}.

\bibitem[Stachel(1980)]{isz10}
J.~Stachel.
\newblock Thickening the string. i. the string perfect dust.
\newblock \emph{Phys. Rev. D}, 21:\penalty0 2171--2181, 1980.
\newblock \doi{10.1103/PhysRevD.21.2171}.

\bibitem[Vilenkin(1985)]{isz11}
A.~Vilenkin.
\newblock Cosmic strings and domain walls.
\newblock \emph{Phys. Rep.}, 121:\penalty0 263--315, 1985.
\newblock \doi{10.1016/0370-1573(85)90033-X}.

\bibitem[Barriola and Vilenkin(1989)]{isz12}
M.~Barriola and A.~Vilenkin.
\newblock Gravitational field of a global monopole.
\newblock \emph{Phys. Rev. Lett.}, 63:\penalty0 341--343, 1989.
\newblock \doi{10.1103/PhysRevLett.63.341}.

\bibitem[Ahmed et~al.(2025{\natexlab{a}})Ahmed, Sakall{\i}, and
  Al-Badawi]{isz12x1}
Faizuddin Ahmed, {\.I}zzet Sakall{\i}, and Ahmad Al-Badawi.
\newblock {Kerr-Bertotti-Robinson Black Holes Surrounded by a Cloud of
  Strings}.
\newblock 11 2025{\natexlab{a}}.

\bibitem[Ahmed et~al.(2025{\natexlab{b}})Ahmed, Al-Badawi, and
  Sakall{\i}]{isz12x2}
Faizuddin Ahmed, Ahmad Al-Badawi, and {\.I}zzet Sakall{\i}.
\newblock {Observable signatures of black hole with Hernquist dark matter halo
  having a cloud of strings: geodesic, perturbations, and shadow}.
\newblock \emph{Eur. Phys. J. C}, 85\penalty0 (9):\penalty0 984,
  2025{\natexlab{b}}.
\newblock \doi{10.1140/epjc/s10052-025-14723-8}.

\bibitem[Ahmed et~al.(2025{\natexlab{c}})Ahmed, Al-Badawi, Sakall{\i}, and
  Shaymatov]{isz12x3}
Faizuddin Ahmed, Ahmad Al-Badawi, {\.I}zzet Sakall{\i}, and Sanjar Shaymatov.
\newblock {Dynamics of test particles and scalar perturbation around an
  Ay{\'o}n{\textendash}Beato{\textendash}Garc{\'\i}a black hole coupled with a
  cloud of strings}.
\newblock \emph{Chin. J. Phys.}, 96:\penalty0 770--791, 2025{\natexlab{c}}.
\newblock \doi{10.1016/j.cjph.2025.05.035}.

\bibitem[Ahmed et~al.(2025{\natexlab{d}})Ahmed, Al-Badawi, and
  Sakall{\i}]{isz12x4}
Faizuddin Ahmed, Ahmad Al-Badawi, and Izzet Sakall{\i}.
\newblock {AdS black strings in a cosmic web: geodesics, shadows, and
  thermodynamics}.
\newblock \emph{Eur. Phys. J. C}, 85\penalty0 (5):\penalty0 554,
  2025{\natexlab{d}}.
\newblock \doi{10.1140/epjc/s10052-025-14266-y}.

\bibitem[Ghosh and Maharaj(2014)]{isz13}
S.~G. Ghosh and S.~D. Maharaj.
\newblock Cloud of strings for radiating black holes in lovelock gravity.
\newblock \emph{Phys. Rev. D}, 89:\penalty0 084027, 2014.
\newblock \doi{10.1103/PhysRevD.89.084027}.

\bibitem[Toledo and Bezerra(2019)]{isz14}
J.~M. Toledo and V.~B. Bezerra.
\newblock Black holes with a cloud of strings in pure lovelock gravity.
\newblock \emph{Eur. Phys. J. C}, 79:\penalty0 117, 2019.
\newblock \doi{10.1140/epjc/s10052-019-6628-4}.

\bibitem[Herscovich and Richarte(2010)]{isz15}
E.~Herscovich and M.~G. Richarte.
\newblock Black holes in einstein-gauss-bonnet gravity with a string cloud
  background.
\newblock \emph{Phys. Lett. B}, 689:\penalty0 192--200, 2010.
\newblock \doi{10.1016/j.physletb.2010.04.065}.

\bibitem[Kumar and Ghosh(2018)]{isz16}
R.~Kumar and Sushant~G. Ghosh.
\newblock Rotating black hole in rastall theory.
\newblock \emph{Eur. Phys. J. C}, 78:\penalty0 750, 2018.
\newblock \doi{10.1140/epjc/s10052-018-6206-1}.

\bibitem[Kiselev(2003)]{isz17}
V.~V. Kiselev.
\newblock Quintessence and black holes.
\newblock \emph{Class. Quantum Grav.}, 20:\penalty0 1187--1197, 2003.
\newblock \doi{10.1088/0264-9381/20/6/310}.

\bibitem[Chen et~al.(2008)Chen, Wang, and Su]{isz18}
S.~Chen, B.~Wang, and R.~Su.
\newblock Hawking radiation in a $d$-dimensional static spherically symmetric
  black hole surrounded by quintessence.
\newblock \emph{Phys. Rev. D}, 77:\penalty0 124011, 2008.
\newblock \doi{10.1103/PhysRevD.77.124011}.

\bibitem[Bonnor(1954)]{isz19}
W.~B. Bonnor.
\newblock Static magnetic fields in general relativity.
\newblock \emph{Proc. Phys. Soc. A}, 67:\penalty0 225--232, 1954.
\newblock \doi{10.1088/0370-1298/67/3/305}.

\bibitem[Melvin(1964)]{isz20}
M.~A. Melvin.
\newblock Pure magnetic and electric geons.
\newblock \emph{Phys. Lett.}, 8:\penalty0 65--68, 1964.
\newblock \doi{10.1016/0031-9163(64)90801-7}.

\bibitem[Ernst(1976)]{isz21}
F.~J. Ernst.
\newblock Black holes in a magnetic universe.
\newblock \emph{J. Math. Phys.}, 17:\penalty0 54--56, 1976.
\newblock \doi{10.1063/1.522781}.

\bibitem[Azreg-A\"{i}nou(2015)]{isz22}
M.~Azreg-A\"{i}nou.
\newblock Black hole thermodynamics: No inconsistency via the inclusion of the
  missing $p$-$v$ terms.
\newblock \emph{Phys. Rev. D}, 91:\penalty0 064049, 2015.
\newblock \doi{10.1103/PhysRevD.91.064049}.

\bibitem[Toshmatov et~al.(2017)Toshmatov, Stuchl\'{i}k, and Ahmedov]{isz23}
B.~Toshmatov, Z.~Stuchl\'{i}k, and B.~Ahmedov.
\newblock Rotating black hole solutions with quintessential energy.
\newblock \emph{Eur. Phys. J. Plus}, 132:\penalty0 98, 2017.
\newblock \doi{10.1140/epjp/i2017-11373-4}.

\bibitem[Al-Badawi et~al.(2025{\natexlab{a}})Al-Badawi, Ahmed, and
  Sakall{\i}]{isz23x1}
Ahmad Al-Badawi, Faizuddin Ahmed, and {\.I}zzet Sakall{\i}.
\newblock {Fingerprints of Loop Quantum Gravity Black Holes with Quintessence
  Field}.
\newblock 5 2025{\natexlab{a}}.

\bibitem[Al-Badawi et~al.(2025{\natexlab{b}})Al-Badawi, Ahmed, and
  Sakall{\i}]{isz23x2}
Ahmad Al-Badawi, Faizuddin Ahmed, and {\.I}zzet Sakall{\i}.
\newblock {Ay{\'o}n-Beato-Garc{\'\i}a black hole in AdS space-time surrounded
  by quintessence: geodesic, shadow and thermodynamics}.
\newblock 3 2025{\natexlab{b}}.

\bibitem[Sadeghi et~al.(2024)Sadeghi, Gashti, Sakalli, and Pourhassan]{isz23x3}
Jafar Sadeghi, Saeed~Noori Gashti, Izzet Sakalli, and Behnam Pourhassan.
\newblock {Weak gravity conjecture of charged-rotating-AdS black hole
  surrounded by quintessence and string cloud}.
\newblock \emph{Nucl. Phys. B}, 1004:\penalty0 116581, 2024.
\newblock \doi{10.1016/j.nuclphysb.2024.116581}.

\bibitem[Noori~Gashti et~al.(2025{\natexlab{a}})Noori~Gashti, Sakall{\i}, and
  Pourhassan]{isz23x4}
Saeed Noori~Gashti, {\.I}zzet Sakall{\i}, and Behnam Pourhassan.
\newblock {Thermodynamic topology, photon spheres, and evidence for weak
  gravity conjecture in charged black holes with perfect fluid within Rastall
  theory}.
\newblock \emph{Phys. Lett. B}, 869:\penalty0 139862, 2025{\natexlab{a}}.
\newblock \doi{10.1016/j.physletb.2025.139862}.

\bibitem[Al-Badawi et~al.(2025{\natexlab{c}})Al-Badawi, Ahmed, and
  Sakall{\i}]{isz23x5}
Ahmad Al-Badawi, Faizuddin Ahmed, and {\.I}zzet Sakall{\i}.
\newblock {A Black Hole Solution in Kalb-Ramond Gravity with Quintessence
  Field: From Geodesic Dynamics to Thermal Criticality}.
\newblock 8 2025{\natexlab{c}}.

\bibitem[Strominger and Vafa(1996)]{isz24}
Andrew Strominger and Cumrun Vafa.
\newblock {Microscopic origin of the Bekenstein-Hawking entropy}.
\newblock \emph{Phys. Lett. B}, 379:\penalty0 99--104, 1996.
\newblock \doi{10.1016/0370-2693(96)00345-0}.

\bibitem[Ilderton and Rajeev(2025)]{isz25}
Anton Ilderton and Karthik Rajeev.
\newblock {Tunnelling amplitudes and Hawking radiation from worldline QFT}.
\newblock \emph{JHEP}, 10:\penalty0 220, 2025.
\newblock \doi{10.1007/JHEP10(2025)220}.

\bibitem[Bardeen et~al.(1973)Bardeen, Carter, and Hawking]{isz26}
J.~M. Bardeen, B.~Carter, and S.~W. Hawking.
\newblock The four laws of black hole mechanics.
\newblock \emph{Commun. Math. Phys.}, 31:\penalty0 161--170, 1973.
\newblock \doi{10.1007/BF01645742}.

\bibitem[Tsallis and Cirto(2013)]{isz27}
C.~Tsallis and L.~J.~L. Cirto.
\newblock Black hole thermodynamical entropy.
\newblock \emph{Eur. Phys. J. C}, 73:\penalty0 2487, 2013.
\newblock \doi{10.1140/epjc/s10052-013-2487-6}.

\bibitem[Maity and Kotal(2025)]{isz28}
Sayani Maity and Anamika Kotal.
\newblock {Barrow agegraphic and new barrow agegraphic dark energy driven
  reconstruction of f(R) gravity and parameter constraints from observational
  data}.
\newblock \emph{Phys. Dark Univ.}, 50:\penalty0 102184, 2025.
\newblock \doi{10.1016/j.dark.2025.102184}.

\bibitem[Saridakis(2020)]{isz29}
E.~N. Saridakis.
\newblock Barrow holographic dark energy.
\newblock \emph{Phys. Rev. D}, 102:\penalty0 123525, 2020.
\newblock \doi{10.1103/PhysRevD.102.123525}.

\bibitem[Noori~Gashti et~al.(2025{\natexlab{b}})Noori~Gashti, Pourhassan, and
  Sakalli]{isz29x1}
Saeed Noori~Gashti, Behnam Pourhassan, and Izzet Sakalli.
\newblock {Thermodynamic topology and phase space analysis of AdS black holes
  through non-extensive entropy perspectives}.
\newblock \emph{Eur. Phys. J. C}, 85\penalty0 (3):\penalty0 305,
  2025{\natexlab{b}}.
\newblock \doi{10.1140/epjc/s10052-025-14035-x}.

\bibitem[Anagnostopoulos et~al.(2020)Anagnostopoulos, Basilakos, and
  Saridakis]{isz30}
F.~K. Anagnostopoulos, S.~Basilakos, and E.~N. Saridakis.
\newblock Observational constraints on barrow holographic dark energy.
\newblock \emph{Eur. Phys. J. C}, 80:\penalty0 826, 2020.
\newblock \doi{10.1140/epjc/s10052-020-8360-5}.

\bibitem[Abreu and Neto(2020)]{isz31}
E.~M.~C. Abreu and J.~A. Neto.
\newblock Barrow black hole corrected-entropy model and tsallis nonextensivity.
\newblock \emph{Phys. Lett. B}, 810:\penalty0 135805, 2020.
\newblock \doi{10.1016/j.physletb.2020.135805}.

\bibitem[Falcke et~al.(2000)Falcke, Melia, and Agol]{isz32}
H.~Falcke, F.~Melia, and E.~Agol.
\newblock Viewing the shadow of the black hole at the galactic center.
\newblock \emph{Astrophys. J. Lett.}, 528:\penalty0 L13--L16, 2000.
\newblock \doi{10.1086/312423}.

\bibitem[Gralla et~al.(2019)Gralla, Holz, and Wald]{isz33}
S.~E. Gralla, D.~E. Holz, and R.~M. Wald.
\newblock Black hole shadows, photon rings, and lensing rings.
\newblock \emph{Phys. Rev. D}, 100:\penalty0 024018, 2019.
\newblock \doi{10.1103/PhysRevD.100.024018}.

\bibitem[Errehymy et~al.(2025)Errehymy, Hansraj, and Hansraj]{isz34}
A.~Errehymy, S.~Hansraj, and C.~Hansraj.
\newblock {Black-hole Shadows and Null Geodesics in
  Hamaus{\textendash}Sutter{\textendash}Wandelt Void Spacetimes with a
  Quintessential Field: Observational Signatures from EHT Data of M87 and Sgr
  A$^{*}$}.
\newblock \emph{Astrophys. J.}, 995\penalty0 (2):\penalty0 148, 2025.
\newblock \doi{10.3847/1538-4357/ae197f}.

\bibitem[Walter et~al.(2025)]{isz35}
Roland Walter et~al.
\newblock {The QUASAR project: reaching 10 microarcsec in the optical to
  resolve accretion disks}.
\newblock \emph{PoS}, ICRC2025:\penalty0 975, 2025.
\newblock \doi{10.22323/1.501.0975}.

\bibitem[Perlick and Tsupko(2022)]{isz36}
V.~Perlick and O.~Y. Tsupko.
\newblock Calculating black hole shadows: Review of analytical studies.
\newblock \emph{Phys. Rep.}, 947:\penalty0 1--39, 2022.
\newblock \doi{10.1016/j.physrep.2021.10.004}.

\bibitem[Perlick et~al.(2015)Perlick, Tsupko, and Bisnovatyi-Kogan]{isz37}
V.~Perlick, O.~Y. Tsupko, and G.~S. Bisnovatyi-Kogan.
\newblock Influence of a plasma on the shadow of a spherically symmetric black
  hole.
\newblock \emph{Phys. Rev. D}, 92:\penalty0 104031, 2015.
\newblock \doi{10.1103/PhysRevD.92.104031}.

\bibitem[Bisnovatyi-Kogan and Tsupko(2010)]{isz38}
G.~S. Bisnovatyi-Kogan and O.~Y. Tsupko.
\newblock Gravitational lensing in a non-uniform plasma.
\newblock \emph{Mon. Not. R. Astron. Soc.}, 404:\penalty0 1790--1800, 2010.
\newblock \doi{10.1111/j.1365-2966.2010.16290.x}.

\bibitem[Sucu and Sakall{\i}(2025{\natexlab{a}})]{isz38x1}
Erdem Sucu and {\.I}zzet Sakall{\i}.
\newblock {Exploring Lorentz-violating effects of Kalb-Ramond field on charged
  black hole thermodynamics and photon dynamics}.
\newblock \emph{Phys. Rev. D}, 111\penalty0 (6):\penalty0 064049,
  2025{\natexlab{a}}.
\newblock \doi{10.1103/PhysRevD.111.064049}.

\bibitem[Sucu et~al.(2025{\natexlab{a}})Sucu, Sakall{\i}, Sert, and
  Sucu]{isz38x2}
Erdem Sucu, {\.I}zzet Sakall{\i}, {\"O}zcan Sert, and Yusuf Sucu.
\newblock {Quantum-corrected thermodynamics and plasma lensing in non-minimally
  coupled symmetric teleparallel black holes}.
\newblock \emph{Phys. Dark Univ.}, 50:\penalty0 102063, 2025{\natexlab{a}}.
\newblock \doi{10.1016/j.dark.2025.102063}.

\bibitem[Ahmed et~al.(2025{\natexlab{e}})Ahmed, Al-Badawi, Bouzenada, Sucu, and
  Sakalli]{WOS:001632119100001AHMED}
Faizuddin Ahmed, Ahmad Al-Badawi, Abdelmalek Bouzenada, Erdem Sucu, and Izzet
  Sakalli.
\newblock Gravitational lensing, wave propagation, and gup-modified hawking
  radiation in charged bopp-podolsky btz black holes with disclinations.
\newblock \emph{INTERNATIONAL JOURNAL OF GEOMETRIC METHODS IN MODERN PHYSICS},
  2025 DEC 5 2025{\natexlab{e}}.
\newblock ISSN 0219-8878.
\newblock \doi{10.1142/S0219887826500891}.

\bibitem[Sucu and Sakall{\i}(2025{\natexlab{b}})]{sucu2025quantumRoyal}
Erdem Sucu and {\.I}zzet Sakall{\i}.
\newblock Quantum-corrected thermodynamics and plasma lensing of mog black
  holes.
\newblock \emph{Proceedings of the Royal Society A: Mathematical, Physical and
  Engineering Sciences}, 481\penalty0 (2320), 2025{\natexlab{b}}.

\bibitem[Sucu and Sakall{\i}(2025{\natexlab{c}})]{sucu2025scalar}
Erdem Sucu and Izzet Sakall{\i}.
\newblock Scalar-tensor corrections and observational signatures of hairy black
  holes in horndeski gravity.
\newblock \emph{High Energy Density Physics}, page 101220, 2025{\natexlab{c}}.

\bibitem[Sucu and Sakall{\i}(2025{\natexlab{d}})]{sucu2025astrophysical}
Erdem Sucu and Izzet Sakall{\i}.
\newblock Astrophysical reality of black hole thermodynamics and dynamics:
  Transformative influence of hernquist dark matter distributions.
\newblock \emph{Physics of the Dark Universe}, page 102051, 2025{\natexlab{d}}.

\bibitem[Page(1976{\natexlab{a}})]{isz39}
D.~N. Page.
\newblock Particle emission rates from a black hole: Massless particles from an
  uncharged, nonrotating hole.
\newblock \emph{Phys. Rev. D}, 13:\penalty0 198--206, 1976{\natexlab{a}}.
\newblock \doi{10.1103/PhysRevD.13.198}.

\bibitem[Harmark et~al.(2010)Harmark, Natario, and Schiappa]{isz40}
T.~Harmark, J.~Natario, and R.~Schiappa.
\newblock Greybody factors for $d$-dimensional black holes.
\newblock \emph{Adv. Theor. Math. Phys.}, 14:\penalty0 727--794, 2010.
\newblock \doi{10.4310/ATMP.2010.v14.n3.a1}.

\bibitem[Cooper et~al.(1995)Cooper, Khare, and Sukhatme]{isz41}
F.~Cooper, A.~Khare, and U.~Sukhatme.
\newblock Supersymmetry and quantum mechanics.
\newblock \emph{Phys. Rep.}, 251:\penalty0 267--385, 1995.
\newblock \doi{10.1016/0370-1573(94)00080-M}.

\bibitem[Cho(2003)]{isz42}
H.~T. Cho.
\newblock Dirac quasinormal modes in schwarzschild black hole spacetimes.
\newblock \emph{Phys. Rev. D}, 68:\penalty0 024003, 2003.
\newblock \doi{10.1103/PhysRevD.68.024003}.

\bibitem[Konoplya and Zhidenko(2011)]{isz43}
R.~A. Konoplya and A.~Zhidenko.
\newblock Quasinormal modes of black holes: From astrophysics to string theory.
\newblock \emph{Rev. Mod. Phys.}, 83:\penalty0 793--836, 2011.
\newblock \doi{10.1103/RevModPhys.83.793}.

\bibitem[Cardoso and Pani(2019)]{isz44}
V.~Cardoso and P.~Pani.
\newblock Testing the nature of dark compact objects: A status report.
\newblock \emph{Living Rev. Relativ.}, 22:\penalty0 4, 2019.
\newblock \doi{10.1007/s41114-019-0020-4}.

\bibitem[Johnson et~al.(2020)]{isz45}
M.~D. Johnson et~al.
\newblock Universal interferometric signatures of a black hole's photon ring.
\newblock \emph{Sci. Adv.}, 6:\penalty0 eaaz1310, 2020.
\newblock \doi{10.1126/sciadv.aaz1310}.

\bibitem[Al-Badawi et~al.(2025{\natexlab{d}})Al-Badawi, Ahmed, and
  Sakall{\i}]{Al-Badawi:2025kbi}
Ahmad Al-Badawi, Faizuddin Ahmed, and {\.I}zzet Sakall{\i}.
\newblock {Letelier black hole immersed in an electromagnetic universe}.
\newblock 11 2025{\natexlab{d}}.

\bibitem[Vilenkin and Shellard(2000)]{sec2is04}
A.~Vilenkin and E.~P.~S. Shellard.
\newblock \emph{Cosmic Strings and Other Topological Defects}.
\newblock Cambridge University Press, Cambridge, 2000.
\newblock ISBN 978-0521654760.

\bibitem[Nunes~dos Santos(2025)]{NunesdosSantos:2025alw}
Luis~Cesar Nunes~dos Santos.
\newblock {Revisiting black holes surrounded by cloud and fluid of strings in
  general relativity}.
\newblock \emph{Phys. Rev. D}, 111\penalty0 (6):\penalty0 064032, 2025.
\newblock \doi{10.1103/PhysRevD.111.064032}.

\bibitem[Kibble(1976)]{sec2is05}
T.~W.~B. Kibble.
\newblock Topology of cosmic domains and strings.
\newblock \emph{J. Phys. A: Math. Gen.}, 9:\penalty0 1387--1398, 1976.
\newblock \doi{10.1088/0305-4470/9/8/029}.

\bibitem[Gott(1985)]{sec2is06}
J.~R. Gott.
\newblock Gravitational lensing effects of vacuum strings: Exact solutions.
\newblock \emph{Astrophys. J.}, 288:\penalty0 422--427, 1985.
\newblock \doi{10.1086/162808}.

\bibitem[Chandrasekhar(1983)]{sec2is07}
S.~Chandrasekhar.
\newblock \emph{The Mathematical Theory of Black Holes}.
\newblock Oxford University Press, Oxford, 1983.
\newblock ISBN 978-0198503705.

\bibitem[Cohen and Kaplan(1988)]{sec2is09}
A.~G. Cohen and D.~B. Kaplan.
\newblock The exact metric about global cosmic strings.
\newblock \emph{Phys. Lett. B}, 215:\penalty0 67--72, 1988.
\newblock \doi{10.1016/0370-2693(88)91072-6}.

\bibitem[Hawking and Ellis(1973)]{sec2is10}
S.~W. Hawking and G.~F.~R. Ellis.
\newblock \emph{The Large Scale Structure of Space-Time}.
\newblock Cambridge University Press, Cambridge, 1973.
\newblock ISBN 978-0521099066.

\bibitem[Misner et~al.(1973)Misner, Thorne, and Wheeler]{sec2is11}
C.~W. Misner, K.~S. Thorne, and J.~A. Wheeler.
\newblock \emph{Gravitation}.
\newblock W. H. Freeman, San Francisco, 1973.
\newblock ISBN 978-0716703440.

\bibitem[Hawking(1975)]{Hawking1975}
S.~W. Hawking.
\newblock Particle creation by black holes.
\newblock \emph{Commun. Math. Phys.}, 43:\penalty0 199--220, 1975.
\newblock \doi{10.1007/BF02345020}.

\bibitem[Bekenstein(1973)]{sec2is12}
J.~D. Bekenstein.
\newblock Black holes and entropy.
\newblock \emph{Phys. Rev. D}, 7:\penalty0 2333--2346, 1973.
\newblock \doi{10.1103/PhysRevD.7.2333}.

\bibitem[Ferreiro et~al.(2025)Ferreiro, Navarro-Salas, and Pla]{sec2is13}
Antonio Ferreiro, Jose Navarro-Salas, and Silvia Pla.
\newblock {The Birth of Gravitational Particle Creation: the Enduring Legacy of
  Leonard Parker's 1966 Thesis}.
\newblock 11 2025.

\bibitem[Wald(1993)]{sec2is14}
R.~M. Wald.
\newblock Black hole entropy is the noether charge.
\newblock \emph{Phys. Rev. D}, 48:\penalty0 R3427--R3431, 1993.
\newblock \doi{10.1103/PhysRevD.48.R3427}.

\bibitem[Kastor et~al.(2009)Kastor, Ray, and Traschen]{sec2is15}
D.~Kastor, S.~Ray, and J.~Traschen.
\newblock Enthalpy and the mechanics of ads black holes.
\newblock \emph{Class. Quantum Grav.}, 26:\penalty0 195011, 2009.
\newblock \doi{10.1088/0264-9381/26/19/195011}.

\bibitem[Synge(1966)]{sec3is01}
J.~L. Synge.
\newblock The escape of photons from gravitationally intense stars.
\newblock \emph{Mon. Not. R. Astron. Soc.}, 131:\penalty0 463--466, 1966.
\newblock \doi{10.1093/mnras/131.3.463}.

\bibitem[Bardeen(1973)]{sec3is02}
J.~M. Bardeen.
\newblock Timelike and null geodesics in the kerr metric.
\newblock \emph{Les Houches Summer School of Theoretical Physics: Black Holes},
  pages 215--239, 1973.

\bibitem[Darwin(1959)]{sec3is03}
C.~Darwin.
\newblock The gravity field of a particle.
\newblock \emph{Proc. R. Soc. Lond. A}, 249:\penalty0 180--194, 1959.
\newblock \doi{10.1098/rspa.1959.0015}.

\bibitem[Sucu et~al.(2025{\natexlab{b}})Sucu, Sakalli, and
  Pourhassan]{sucu2025quantumHassan}
Erdem Sucu, Izzet Sakalli, and Behnam Pourhassan.
\newblock Quantum corrections in thermodynamics of black holes modified by
  nonlinear electrodynamics and their observational signatures.
\newblock \emph{International Journal of Geometric Methods in Modern Physics},
  2025{\natexlab{b}}.

\bibitem[Hioki and Maeda(2009)]{sec3is05}
K.~Hioki and K.~Maeda.
\newblock Measurement of the kerr spin parameter by observation of a compact
  object's shadow.
\newblock \emph{Phys. Rev. D}, 80:\penalty0 024042, 2009.
\newblock \doi{10.1103/PhysRevD.80.024042}.

\bibitem[Gralla et~al.(2020)Gralla, Lupsasca, and Marrone]{sec3is09}
S.~E. Gralla, A.~Lupsasca, and D.~P. Marrone.
\newblock The shape of the black hole photon ring: A precise test of
  strong-field general relativity.
\newblock \emph{Phys. Rev. D}, 102:\penalty0 124004, 2020.
\newblock \doi{10.1103/PhysRevD.102.124004}.

\bibitem[Broderick and Loeb(2006)]{sec3is10}
A.~E. Broderick and A.~Loeb.
\newblock Imaging bright-spots in the accretion flow near the black hole
  horizon of sgr a*.
\newblock \emph{Mon. Not. R. Astron. Soc.}, 367:\penalty0 905--916, 2006.
\newblock \doi{10.1111/j.1365-2966.2006.10152.x}.

\bibitem[Tsupko and Bisnovatyi-Kogan(2013)]{sec3is12}
O.~Y. Tsupko and G.~S. Bisnovatyi-Kogan.
\newblock Gravitational lensing in plasma: Relativistic images at homogeneous
  plasma.
\newblock \emph{Phys. Rev. D}, 87:\penalty0 124009, 2013.
\newblock \doi{10.1103/PhysRevD.87.124009}.

\bibitem[Rogers(2015)]{sec3is13}
A.~Rogers.
\newblock Frequency-dependent effects of gravitational lensing within plasma.
\newblock \emph{Mon. Not. R. Astron. Soc.}, 451:\penalty0 17--25, 2015.
\newblock \doi{10.1093/mnras/stv903}.

\bibitem[Perlick and Tsupko(2017)]{sec3is14}
V.~Perlick and O.~Y. Tsupko.
\newblock Light propagation in a plasma on kerr spacetime: Separation of the
  hamilton-jacobi equation and calculation of the shadow.
\newblock \emph{Phys. Rev. D}, 95:\penalty0 104003, 2017.
\newblock \doi{10.1103/PhysRevD.95.104003}.

\bibitem[Abdujabbarov et~al.(2017)Abdujabbarov, Toshmatov, Stuchl\'{i}k, and
  Ahmedov]{sec3is15}
A.~Abdujabbarov, B.~Toshmatov, Z.~Stuchl\'{i}k, and B.~Ahmedov.
\newblock Shadow of the rotating black hole with quintessential energy in the
  presence of plasma.
\newblock \emph{Int. J. Mod. Phys. D}, 26:\penalty0 1750051, 2017.
\newblock \doi{10.1142/S0218271817500511}.

\bibitem[Ahmedov et~al.(2016)Ahmedov, Toshmatov, and Stuchl\'{i}k]{sec3is16}
B.~Ahmedov, B.~Toshmatov, and Z.~Stuchl\'{i}k.
\newblock Rotating black hole solutions with quintessential energy.
\newblock \emph{Eur. Phys. J. C}, 76:\penalty0 273, 2016.
\newblock \doi{10.1140/epjc/s10052-016-4122-9}.

\bibitem[Narayan and Yi(1994)]{sec3is17}
R.~Narayan and I.~Yi.
\newblock Advection-dominated accretion: A self-similar solution.
\newblock \emph{Astrophys. J. Lett.}, 428:\penalty0 L13--L16, 1994.
\newblock \doi{10.1086/187381}.

\bibitem[Sakalli and Kanzi(2022)]{sec3is18}
{\.I}zzet Sakalli and Sara Kanzi.
\newblock {Topical Review: greybody factors and quasinormal modes for black
  holes in various theories - fingerprints of invisibles}.
\newblock \emph{Turk. J. Phys.}, 46\penalty0 (2):\penalty0 51--103, 2022.
\newblock \doi{10.55730/1300-0101.269}.

\bibitem[Doeleman et~al.(2023)]{sec3is19}
Sheperd~S. Doeleman et~al.
\newblock {Reference Array and Design Consideration for the Next-Generation
  Event Horizon Telescope}.
\newblock \emph{Galaxies}, 11\penalty0 (5):\penalty0 107, 2023.
\newblock \doi{10.3390/galaxies11050107}.

\bibitem[Goddi et~al.(2016)]{sec3is20}
C.~Goddi et~al.
\newblock {BlackHoleCam: Fundamental physics of the galactic center}.
\newblock \emph{Int. J. Mod. Phys. D}, 26\penalty0 (02):\penalty0 1730001,
  2016.
\newblock \doi{10.1142/9789813226609_0046}.

\bibitem[Psaltis et~al.(2020)]{sec3is21}
D.~Psaltis et~al.
\newblock Gravitational test beyond the first post-newtonian order with the
  shadow of the m87 black hole.
\newblock \emph{Phys. Rev. Lett.}, 125:\penalty0 141104, 2020.
\newblock \doi{10.1103/PhysRevLett.125.141104}.

\bibitem[Johnson et~al.(2023)]{sec3is22}
Michael~D. Johnson et~al.
\newblock {Key Science Goals for the Next-Generation Event Horizon Telescope}.
\newblock \emph{Galaxies}, 11\penalty0 (3):\penalty0 61, 2023.
\newblock \doi{10.3390/galaxies11030061}.

\bibitem[Slane et~al.(2025)Slane, Bogd{\'a}n, and Pooley]{sec4is01}
Patrick Slane, {\'A}kos Bogd{\'a}n, and David Pooley.
\newblock {25{\,}years of groundbreaking discoveries with Chandra}.
\newblock \emph{Nature Astron.}, 9\penalty0 (10):\penalty0 1431--1443, 2025.
\newblock \doi{10.1038/s41550-025-02675-8}.

\bibitem[Gibbons and Hawking(1977)]{sec4is02}
G.~W. Gibbons and S.~W. Hawking.
\newblock Cosmological event horizons, thermodynamics, and particle creation.
\newblock \emph{Phys. Rev. D}, 15:\penalty0 2738--2751, 1977.
\newblock \doi{10.1103/PhysRevD.15.2738}.

\bibitem[Page(1976{\natexlab{b}})]{sec4is07}
D.~N. Page.
\newblock Particle emission rates from a black hole. ii. massless particles
  from a rotating hole.
\newblock \emph{Phys. Rev. D}, 14:\penalty0 3260--3273, 1976{\natexlab{b}}.
\newblock \doi{10.1103/PhysRevD.14.3260}.

\bibitem[Harris and Kanti(2003)]{sec4is08}
C.~M. Harris and P.~Kanti.
\newblock Hawking radiation from a $(4+n)$-dimensional black hole: Exact
  results for the schwarzschild phase.
\newblock \emph{J. High Energy Phys.}, 10:\penalty0 014, 2003.
\newblock \doi{10.1088/1126-6708/2003/10/014}.

\bibitem[Kanti(2004)]{sec4is09}
P.~Kanti.
\newblock Black holes in theories with large extra dimensions: A review.
\newblock \emph{Int. J. Mod. Phys. A}, 19:\penalty0 4899--4951, 2004.
\newblock \doi{10.1142/S0217751X04018324}.

\bibitem[Grain and Barrau(2008)]{sec4is11}
J.~Grain and A.~Barrau.
\newblock Quantum bound states around black holes.
\newblock \emph{Eur. Phys. J. C}, 53:\penalty0 641--648, 2008.
\newblock \doi{10.1140/epjc/s10052-007-0494-1}.

\bibitem[Arbey et~al.(2020)Arbey, Auffinger, and Silk]{sec4is12}
A.~Arbey, J.~Auffinger, and J.~Silk.
\newblock Evolution of primordial black hole spin due to hawking radiation.
\newblock \emph{Mon. Not. R. Astron. Soc.}, 494:\penalty0 1257--1262, 2020.
\newblock \doi{10.1093/mnras/staa765}.

\bibitem[Visser(1999)]{sec4is14}
M.~Visser.
\newblock Some general bounds for one-dimensional scattering.
\newblock \emph{Phys. Rev. A}, 59:\penalty0 427--438, 1999.
\newblock \doi{10.1103/PhysRevA.59.427}.

\bibitem[Boonserm and Visser(2008)]{sec4is15}
P.~Boonserm and M.~Visser.
\newblock Bounding the greybody factors for schwarzschild black holes.
\newblock \emph{Phys. Rev. D}, 78:\penalty0 101502, 2008.
\newblock \doi{10.1103/PhysRevD.78.101502}.

\bibitem[Carr and Hawking(1974)]{sec4is16}
B.~J. Carr and S.~W. Hawking.
\newblock Black holes in the early universe.
\newblock \emph{Mon. Not. R. Astron. Soc.}, 168:\penalty0 399--415, 1974.
\newblock \doi{10.1093/mnras/168.2.399}.

\bibitem[Carr et~al.(2021)Carr, Kohri, Sendouda, and Yokoyama]{sec4is17}
B.~J. Carr, K.~Kohri, Y.~Sendouda, and J.~Yokoyama.
\newblock Constraints on primordial black holes.
\newblock \emph{Rep. Prog. Phys.}, 84:\penalty0 116902, 2021.
\newblock \doi{10.1088/1361-6633/ac1e31}.

\bibitem[MacGibbon(1991)]{sec4is18}
J.~H. MacGibbon.
\newblock Quark- and gluon-jet emission from primordial black holes. ii. the
  emission over the black-hole lifetime.
\newblock \emph{Phys. Rev. D}, 44:\penalty0 376--392, 1991.
\newblock \doi{10.1103/PhysRevD.44.376}.

\bibitem[Carr(2003)]{sec4is19}
B.~J. Carr.
\newblock Primordial black holes as a probe of cosmology and high energy
  physics.
\newblock \emph{Lect. Notes Phys.}, 631:\penalty0 301--321, 2003.
\newblock \doi{10.1007/978-3-540-45230-0_7}.

\bibitem[Clesse and Garcia-Bellido(2015)]{sec4is21}
S.~Clesse and J.~Garcia-Bellido.
\newblock Massive primordial black holes from hybrid inflation as dark matter
  and the seeds of galaxies.
\newblock \emph{Phys. Rev. D}, 92:\penalty0 023524, 2015.
\newblock \doi{10.1103/PhysRevD.92.023524}.

\bibitem[Carr et~al.(2016)Carr, Kuhnel, and Sandstad]{sec4is22}
B.~J. Carr, F.~Kuhnel, and M.~Sandstad.
\newblock Primordial black holes as dark matter.
\newblock \emph{Phys. Rev. D}, 94:\penalty0 083504, 2016.
\newblock \doi{10.1103/PhysRevD.94.083504}.

\bibitem[Laha(2019)]{sec4is23}
R.~Laha.
\newblock Primordial black holes as a dark matter candidate are severely
  constrained by the galactic center 511 kev $\gamma$-ray line.
\newblock \emph{Phys. Rev. Lett.}, 123:\penalty0 251101, 2019.
\newblock \doi{10.1103/PhysRevLett.123.251101}.

\bibitem[Coogan et~al.(2021)Coogan, Morrison, and Profumo]{sec4is24}
A.~Coogan, L.~Morrison, and S.~Profumo.
\newblock Direct detection of hawking radiation from asteroid-mass primordial
  black holes.
\newblock \emph{Phys. Rev. Lett.}, 126:\penalty0 171101, 2021.
\newblock \doi{10.1103/PhysRevLett.126.171101}.

\bibitem[Barrow(2020)]{barrow2020area}
J.~D. Barrow.
\newblock The area of a rough black hole.
\newblock \emph{Phys. Lett. B}, 808:\penalty0 135643, 2020.
\newblock \doi{10.1016/j.physletb.2020.135643}.

\bibitem[Nojiri et~al.(2022{\natexlab{a}})Nojiri, Odintsov, and
  Faraoni]{sec5is01}
S.~Nojiri, S.~D. Odintsov, and V.~Faraoni.
\newblock Area-law versus rényi and tsallis black hole entropies.
\newblock \emph{Phys. Rev. D}, 105:\penalty0 044042, 2022{\natexlab{a}}.
\newblock \doi{10.1103/PhysRevD.105.044042}.

\bibitem[Das et~al.(2002)Das, Majumdar, and Bhaduri]{sec5is03}
S.~Das, P.~Majumdar, and R.~K. Bhaduri.
\newblock General logarithmic corrections to black-hole entropy.
\newblock \emph{Class. Quantum Grav.}, 19:\penalty0 2355--2367, 2002.
\newblock \doi{10.1088/0264-9381/19/9/302}.

\bibitem[Chatterjee and Majumdar(2004)]{sec5is04}
A.~Chatterjee and P.~Majumdar.
\newblock Universal canonical black hole entropy.
\newblock \emph{Phys. Rev. Lett.}, 92:\penalty0 141301, 2004.
\newblock \doi{10.1103/PhysRevLett.92.141301}.

\bibitem[Rovelli(1996)]{sec5is05}
C.~Rovelli.
\newblock Black hole entropy from loop quantum gravity.
\newblock \emph{Phys. Rev. Lett.}, 77:\penalty0 3288--3291, 1996.
\newblock \doi{10.1103/PhysRevLett.77.3288}.

\bibitem[Ashtekar et~al.(1998)Ashtekar, Baez, Corichi, and Krasnov]{sec5is06}
A.~Ashtekar, J.~Baez, A.~Corichi, and K.~Krasnov.
\newblock Quantum geometry and black hole entropy.
\newblock \emph{Phys. Rev. Lett.}, 80:\penalty0 904--907, 1998.
\newblock \doi{10.1103/PhysRevLett.80.904}.

\bibitem[Luciano and Saridakis(2022)]{sec5is07}
G.~G. Luciano and E.~N. Saridakis.
\newblock Barrow black hole with cosmological constant: An analytic solution.
\newblock \emph{J. High Energy Phys.}, 11:\penalty0 032, 2022.
\newblock \doi{10.1007/JHEP11(2022)032}.

\bibitem[Sucu and Sakalli(2025)]{WOS:001565141800002NPB}
Erdem Sucu and Izzet Sakalli.
\newblock Ads black holes in einstein-kalb-ramond gravity: Quantum corrections,
  phase transitions, and orbital dynamics.
\newblock \emph{NUCLEAR PHYSICS B}, 1018, SEP 2025.
\newblock ISSN 0550-3213.
\newblock \doi{10.1016/j.nuclphysb.2025.117081}.

\bibitem[Kubiznak and Mann(2012)]{sec5is10}
D.~Kubiznak and R.~B. Mann.
\newblock $p$-$v$ criticality of charged ads black holes.
\newblock \emph{J. High Energy Phys.}, 07:\penalty0 033, 2012.
\newblock \doi{10.1007/JHEP07(2012)033}.

\bibitem[Davies(1978)]{sec5is11}
P.~C.~W. Davies.
\newblock Thermodynamics of black holes.
\newblock \emph{Rep. Prog. Phys.}, 41:\penalty0 1313--1355, 1978.
\newblock \doi{10.1088/0034-4885/41/8/004}.

\bibitem[Hawking and Page(1983)]{sec5is12}
S.~W. Hawking and D.~N. Page.
\newblock Thermodynamics of black holes in anti-de sitter space.
\newblock \emph{Commun. Math. Phys.}, 87:\penalty0 577--588, 1983.
\newblock \doi{10.1007/BF01208266}.

\bibitem[Okcu and Aydiner(2017)]{sec5is13}
O.~Okcu and E.~Aydiner.
\newblock Joule-thomson expansion of the charged ads black holes.
\newblock \emph{Eur. Phys. J. C}, 77:\penalty0 24, 2017.
\newblock \doi{10.1140/epjc/s10052-017-4598-y}.

\bibitem[Okcu and Aydiner(2018)]{sec5is14}
O.~Okcu and E.~Aydiner.
\newblock Joule-thomson expansion of kerr-ads black holes.
\newblock \emph{Eur. Phys. J. C}, 78:\penalty0 123, 2018.
\newblock \doi{10.1140/epjc/s10052-018-5602-x}.

\bibitem[Sakall{\i} et~al.(2025)Sakall{\i}, Sucu, and
  Sucu]{sakalli2025zitterbewegung}
{\.I}zzet Sakall{\i}, Yusuf Sucu, and Erdem Sucu.
\newblock Zitterbewegung oscillations and gup-induced quantum modifications of
  yang-mills black holes in perfect fluid dark matter.
\newblock \emph{Nuclear Physics B}, page 117216, 2025.

\bibitem[Dolan(2011)]{sec5is16}
B.~P. Dolan.
\newblock Pressure and volume in the first law of black hole thermodynamics.
\newblock \emph{Class. Quantum Grav.}, 28:\penalty0 235017, 2011.
\newblock \doi{10.1088/0264-9381/28/23/235017}.

\bibitem[Aydiner et~al.(2025)Aydiner, Sucu, and Sakall{\i}]{aydiner2025regular}
Ekrem Aydiner, Erdem Sucu, and {\.I}zzet Sakall{\i}.
\newblock Regular magnetically charged black holes from nonlinear
  electrodynamics: Thermodynamics, light deflection, and orbital dynamics.
\newblock \emph{Physics of the Dark Universe}, page 102164, 2025.

\bibitem[Chabab et~al.(2018)Chabab, Moumni, Iraoui, and Masmar]{sec5is17}
M.~Chabab, H.~El Moumni, S.~Iraoui, and K.~Masmar.
\newblock Joule-thomson expansion of rn-ads black holes in $f(r)$ gravity.
\newblock \emph{Eur. Phys. J. C}, 78:\penalty0 632, 2018.
\newblock \doi{10.1140/epjc/s10052-018-6134-0}.

\bibitem[Lan et~al.(2018)Lan, Miao, and Wang]{sec5is18}
C.~Lan, Y.~G. Miao, and Y.~X. Wang.
\newblock Joule-thomson expansion of charged gauss-bonnet black holes in ads
  space.
\newblock \emph{Phys. Rev. D}, 98:\penalty0 084014, 2018.
\newblock \doi{10.1103/PhysRevD.98.084014}.

\bibitem[Hendi et~al.(2017)Hendi, Panah, Panahiyan, and Talezadeh]{sec5is19}
S.~H. Hendi, B.~Eslam Panah, M.~Panahiyan, and M.~S. Talezadeh.
\newblock Geometrical thermodynamics and p-v criticality of black holes with
  power-law maxwell field.
\newblock \emph{Eur. Phys. J. C}, 77:\penalty0 133, 2017.
\newblock \doi{10.1140/epjc/s10052-017-4693-0}.

\bibitem[Wei and Liu(2015)]{sec5is20}
S.~W. Wei and Y.~X. Liu.
\newblock Insight into the microscopic structure of an ads black hole from a
  thermodynamical phase transition.
\newblock \emph{Phys. Rev. Lett.}, 115:\penalty0 111302, 2015.
\newblock \doi{10.1103/PhysRevLett.115.111302}.

\bibitem[Nojiri et~al.(2022{\natexlab{b}})Nojiri, Odintsov, and Paul]{sec5is21}
S.~Nojiri, S.~D. Odintsov, and T.~Paul.
\newblock Barrow entropic dark energy: A member of generalized holographic dark
  energy family.
\newblock \emph{Phys. Lett. B}, 825:\penalty0 136844, 2022{\natexlab{b}}.
\newblock \doi{10.1016/j.physletb.2021.136844}.

\bibitem[Sheykhi(2021)]{sec5is22}
A.~Sheykhi.
\newblock Barrow entropy corrections to friedmann equations.
\newblock \emph{Phys. Rev. D}, 103:\penalty0 123503, 2021.
\newblock \doi{10.1103/PhysRevD.103.123503}.

\bibitem[Basilakos et~al.(2023)Basilakos, Lymperis, Petronikolou, and
  Saridakis]{sec5is23}
S.~Basilakos, A.~Lymperis, M.~Petronikolou, and E.~N. Saridakis.
\newblock Barrow entropy constraints from black hole evaporation.
\newblock \emph{Eur. Phys. J. C}, 83:\penalty0 576, 2023.
\newblock \doi{10.1140/epjc/s10052-023-11739-w}.

\bibitem[Altamirano et~al.(2014)Altamirano, Kubiznak, Mann, and
  Sherkatghanad]{sec5is24}
N.~Altamirano, D.~Kubiznak, R.~B. Mann, and Z.~Sherkatghanad.
\newblock Thermodynamics of rotating black holes and black rings: Phase
  transitions and thermodynamic volume.
\newblock \emph{Galaxies}, 2:\penalty0 89--159, 2014.
\newblock \doi{10.3390/galaxies2010089}.

\bibitem[Dolan and Dempsey(2015)]{sec6is01}
S.~R. Dolan and D.~Dempsey.
\newblock Bound states of the dirac equation on kerr spacetime.
\newblock \emph{Class. Quantum Grav.}, 32:\penalty0 184001, 2015.
\newblock \doi{10.1088/0264-9381/32/18/184001}.

\bibitem[Brill and Wheeler(1957)]{sec6is03}
D.~R. Brill and J.~A. Wheeler.
\newblock Interaction of neutrinos and gravitational fields.
\newblock \emph{Rev. Mod. Phys.}, 29:\penalty0 465--479, 1957.
\newblock \doi{10.1103/RevModPhys.29.465}.

\bibitem[Teukolsky(1973)]{sec4is06}
S.~A. Teukolsky.
\newblock Perturbations of a rotating black hole. i. fundamental equations for
  gravitational, electromagnetic, and neutrino-field perturbations.
\newblock \emph{Astrophys. J.}, 185:\penalty0 635--647, 1973.
\newblock \doi{10.1086/152444}.

\bibitem[Newman and Penrose(1962)]{sec6is05}
E.~Newman and R.~Penrose.
\newblock An approach to gravitational radiation by a method of spin
  coefficients.
\newblock \emph{J. Math. Phys.}, 3:\penalty0 566--578, 1962.
\newblock \doi{10.1063/1.1724257}.

\bibitem[Chandrasekhar(1976)]{sec6is06}
S.~Chandrasekhar.
\newblock The solution of dirac's equation in kerr geometry.
\newblock \emph{Proc. R. Soc. Lond. A}, 349:\penalty0 571--575, 1976.
\newblock \doi{10.1098/rspa.1976.0090}.

\bibitem[del Castillo(1988)]{sec6is07}
G.~F.~Torres del Castillo.
\newblock Spin-$\frac{1}{2}$ perturbations of algebraically special solutions
  of the einstein-maxwell equations.
\newblock \emph{J. Math. Phys.}, 29:\penalty0 2078--2083, 1988.
\newblock \doi{10.1063/1.527865}.

\bibitem[Anderson and Price(1991)]{sec6is09}
A.~Anderson and R.~H. Price.
\newblock Intertwining of the equations of black-hole perturbations.
\newblock \emph{Phys. Rev. D}, 43:\penalty0 3147--3154, 1991.
\newblock \doi{10.1103/PhysRevD.43.3147}.

\bibitem[Grain and Barrau(2006)]{sec6is12}
J.~Grain and A.~Barrau.
\newblock A wkb approach to scalar fields dynamics in curved space-time.
\newblock \emph{Nucl. Phys. B}, 742:\penalty0 253--274, 2006.
\newblock \doi{10.1016/j.nuclphysb.2006.02.036}.

\bibitem[Page(1977)]{sec6is14}
D.~N. Page.
\newblock Particle emission rates from a black hole. iii. charged leptons from
  a nonrotating hole.
\newblock \emph{Phys. Rev. D}, 16:\penalty0 2402--2411, 1977.
\newblock \doi{10.1103/PhysRevD.16.2402}.

\bibitem[Witten(1981)]{sec6is15}
E.~Witten.
\newblock Dynamical breaking of supersymmetry.
\newblock \emph{Nucl. Phys. B}, 188:\penalty0 513--554, 1981.
\newblock \doi{10.1016/0550-3213(81)90006-7}.

\bibitem[Bluhm and Kostelecky(2005)]{sec6is16}
R.~Bluhm and V.~A. Kostelecky.
\newblock Spontaneous lorentz violation, nambu-goldstone modes, and gravity.
\newblock \emph{Phys. Rev. D}, 71:\penalty0 065008, 2005.
\newblock \doi{10.1103/PhysRevD.71.065008}.

\bibitem[Whiting(1989)]{sec6is17}
B.~F. Whiting.
\newblock Mode stability of the kerr black hole.
\newblock \emph{J. Math. Phys.}, 30:\penalty0 1301--1305, 1989.
\newblock \doi{10.1063/1.528308}.

\bibitem[Chandrasekhar and Detweiler(1975)]{sec6is18}
S.~Chandrasekhar and S.~Detweiler.
\newblock The quasi-normal modes of the schwarzschild black hole.
\newblock \emph{Proc. R. Soc. Lond. A}, 344:\penalty0 441--452, 1975.
\newblock \doi{10.1098/rspa.1975.0112}.

\bibitem[Cardoso and Lemos(2001)]{sec6is19}
V.~Cardoso and J.~P.~S. Lemos.
\newblock Scalar, electromagnetic and weyl perturbations of btz black holes:
  Quasinormal modes.
\newblock \emph{Phys. Rev. D}, 63:\penalty0 124015, 2001.
\newblock \doi{10.1103/PhysRevD.63.124015}.

\bibitem[Konoplya and Zhidenko(2008)]{sec6is20}
R.~A. Konoplya and A.~Zhidenko.
\newblock Stability of higher dimensional reissner-nordstr\"{o}m-anti-de sitter
  black holes.
\newblock \emph{Phys. Rev. D}, 78:\penalty0 104017, 2008.
\newblock \doi{10.1103/PhysRevD.78.104017}.

\bibitem[Akiyama et~al.(2019{\natexlab{b}})]{cisz01}
K.~Akiyama et~al.
\newblock First m87 event horizon telescope results. vi. the shadow and mass of
  the central black hole.
\newblock \emph{Astrophys. J. Lett.}, 875:\penalty0 L6, 2019{\natexlab{b}}.
\newblock \doi{10.3847/2041-8213/ab1141}.

\bibitem[Akiyama et~al.(2022{\natexlab{b}})]{cisz02}
K.~Akiyama et~al.
\newblock First sagittarius a* event horizon telescope results. vi. testing the
  black hole metric.
\newblock \emph{Astrophys. J. Lett.}, 930:\penalty0 L17, 2022{\natexlab{b}}.
\newblock \doi{10.3847/2041-8213/ac6756}.

\bibitem[Kotal et~al.(2025)Kotal, Debnath, and Pradhan]{cisz05}
Anamika Kotal, Ujjal Debnath, and Anirudh Pradhan.
\newblock {Cosmological parameter analysis and correspondence of
  f(R),~f(G),~f(T) gravity models within the (m, n)-type Barrow holographic
  dark energy framework}.
\newblock \emph{Phys. Scripta}, 100\penalty0 (11):\penalty0 115302, 2025.
\newblock \doi{10.1088/1402-4896/ae1a1b}.

\end{thebibliography}

\end{document}